\documentclass[%
 reprint,
 amsmath,amssymb,
 aps,
]{revtex4-2}
\mathchardef\mhyphen="2D
\usepackage{graphicx}
\usepackage{dcolumn}
\usepackage{bm}
\usepackage{float}
\usepackage[dvipsnames]{xcolor}

\begin{document}
\preprint{APS/123-QED}

\title{SARM: Sparse Autoregressive Models for Scalable Generation of Sparse Images in Particle Physics}

\author{Yadong Lu$^{a}$}
\author{Julian Collado$^{b}$}
\author{Daniel Whiteson$^{c}$}
\author{Pierre Baldi$^{b}$}
 \email{Corresponding author: pfbaldi@ics.uci.edu}

\affiliation{$^{a}$Department of Statistics, University of California, Irvine, CA, USA 92627}
\affiliation{$^{b}$Department of Computer Science, University of California, Irvine, CA, USA 92627}
\affiliation{$^{c}$Department of Physics and Astronomy, University of California, Irvine, CA, USA 92627}



\date{\today}
\begin{abstract}
Generation of simulated data is essential for data analysis in particle physics, but current Monte Carlo methods are very computationally expensive.
Deep-learning-based generative models have  successfully  generated simulated data at lower cost, but struggle when the data are very sparse. 
We introduce a novel deep sparse autoregressive model (SARM) that explicitly learns the sparseness of the data with a tractable likelihood, making it more stable and  interpretable when compared to Generative Adversarial Networks (GANs) and other methods. 
In two case studies, we compare SARM to a GAN model and a non-sparse autoregressive model.
As a quantitative measure of performance, we compute the Wasserstein distance ($W_p$) between the distributions of physical quantities calculated on the generated images and on the training images. In the first study, featuring images of jets in which $90\%$ of the pixels are zero-valued, SARM produces images with $W_p$ scores that are $24\mhyphen52\%$ better than the scores obtained with other state-of-the-art generative models. In the second study, on calorimeter images in the vicinity of muons  where $98\%$ of the pixels are zero-valued, SARM  produces images with $W_p$ scores that are  $66\mhyphen68\%$ better.  Similar observations made with other metrics confirm the usefulness of SARM for sparse data in particle physics. Original data and software will be made available upon acceptance of the manuscript from the UCI Machine Learning in Physics web portal at: http://mlphysics.ics.uci.edu/.
\end{abstract}

\maketitle
\section{Introduction}
\label{sec:intro}

Experiments in particle physics seek to uncover the building blocks of matter and their interactions, which determine the structure of the Universe from subatomic to cosmic distances.
Analyses of the data produced by these experiments make extensive use of simulations to predict the experimental signature of particle interactions under various theoretical hypothesis. These simulations are used in likelihood-free inference as well as in the development of data selection and analysis strategies which optimize the statistical power of the data.
Current state-of-the-art simulators apply Monte Carlo techniques to the microphysical processes governing individual particles' propagation and interaction \cite{GEANT}, making them computationally expensive \cite{Aad:2010ah,Rahmat_2012}.
 
\begin{figure*}
    \centering
    \includegraphics[width=0.75\linewidth]{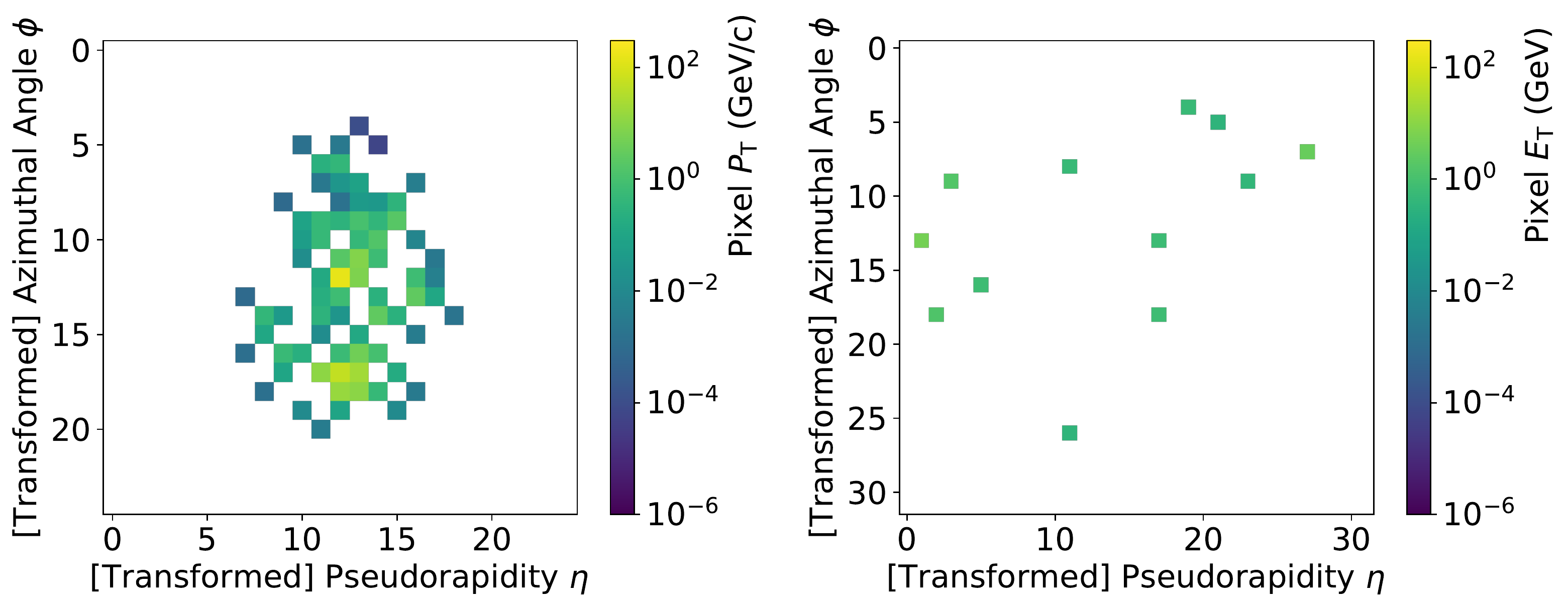}
    \caption{Calorimeter images in particle physics are often very sparse, where most of their pixels have very small values. \textbf{Left}: Typical signal image of a hadronic jet from \cite{LAGAN} \textbf{Right}: Typical signal image of the vicinity of a muon from \cite{muons}.}
\label{sparse_images}
\end{figure*}

Detectors in particle physics experiments have a multi-layer architecture which produces highly structured data. 
One essential layer, the calorimeter, measures the energy of passing particles, and is subdivided into small cells to ensure spatial resolution. In collider experiments, the calorimeter is typically cylindrical \cite{Nikiforou:2013nba}, while in fixed-target experiments it may be a surface \cite{LHCB:2000ab}. 
In both cases, the data can be represented as an image, allowing for the application of image-processing methods initially developed for natural images.
However, in contrast to natural images, pixels in calorimeter images (figure~\ref{sparse_images}) are very sparse, where usually 90\% or more of the pixel values are zero.
In addition, these images are not as uniform as natural images, featuring clusters in the center and noise in the periphery.

Recently, deep generative models \cite{GAN, VAE, pmlr-v48-oord16}
have produced high-quality artificial  natural images \cite{cyclegan, biggan, glow}
at a relatively low computational cost.
The successful application of machine learning in high energy physics \cite{Baldi:NatureHEP, DLHEP:IgorOstrovsky, Baldi:DecorrelateAdversarialHEP, NOvABaldi, Collado:JetFlavorClassification, Collado:DarkKnowledge, Collado:DUNE, BaldiDLBook2020}, 
and generative models in natural images has inspired the use of these models for generating image-like data in 
physical sciences applications \cite{LAGAN, CosmoGAN, PhysicsGAN1, PhysicsGAN2, PhysicsGAN3, PhysicsGAN4, PhysicsGAN5, PhysicsGAN6, PhysicsGAN7, PhysicsGAN8, PhysicsGAN9, PhysicsGAN10}, often employing Generative Adversarial Networks (GAN)  \cite{GAN} or, less frequently, Variational Auto-encoders (VAE) \cite{VAE}.
However, the extreme sparsity of the images in particle physics and other areas of the physical sciences \cite{sparsityInHEP} presents unique challenges for generative models.

The leading applications of GAN-based generative models for sparse image synthesis in high-energy physics, LAGAN \cite{LAGAN} and CALOGAN \cite{CALOGAN}, make use of the ReLU activation function in the final layer to induce sparsity in the output image. The flat portion of the ReLU activation function can lead to many error gradients being zero at the output layer, creating challenges \cite{sparseGradientsGANWorkshopNIPS} for stochastic gradient descent \cite{sgd, adam} methods.
In addition, GANs are notoriously unstable during training \cite{convergence1} and can suffer from mode collapse, which restricts the diversity of events in the generated data \cite{modecollapse, Radford2015UnsupervisedRL}. Despite these difficulties, GANs have been one of the most popular deep generative models in particle physics.

However, other generative models may be better suited for sparse data. 
For example, deep autoregressive models (ARMs) have also demonstrated impressive performance for generating natural images among likelihood based generative models \cite{pixelcnn++, pmlr-v48-oord16}. In this paper, we  develop {\it sparse} autoregressive models (SARM), a class of ARMs specifically tuned to produce sparse images. We present a systematic approach for designing SARMs and demonstrate their effectiveness through multiple experiments. SARMs are stable during training with respect to hyperparameter variations and weight initializations. SARMs are also interpretable in the sense that it is possible for these model to produce an analytic likelihood for any given sample.
We then evaluate SARMs on two benchmark data sets. Given their flexibility, SARMs may be applicable to  areas beyond particle physics where sparse images must be generated. 

\section{Datasets}
\label{Data}

An important statistical task in the analysis of particle physics data is identifying the particle source of a particular detector signature.   Below, we describe two datasets, one which  distinguishes between the detector signatures of single quarks and collimated pairs of quarks, and a second which distinguishes between muons produced in isolation and those produced as part of a shower of particles.

\subsection{Jet Substructure Study}

Quarks or gluons produced in collisions leave a particular detector signature: a {\it jet}, or shower of collimated particles, which deposit most of their energy in a tight core.  In many applications, it is important to distinguish the signatures of a single quark or gluon from that of a collimated pair of quarks, which may leave two potentially overlapping cores. This task is a natural setting for image-recognition algorithms, and has been the focus of many deep learning studies \cite{sparsityInHEP,Almeida:2015jua,de_Oliveira_jet_images_deep_learning_edition_2016,Barnard:2016qma,Komiske:2016rsd} which rely on simplified calorimeter simulations due to the cost of generating realistic samples.
Thus, an inexpensive generation of realistic datasets would be very valuable as a classification training sample. 

We use a set of benchmark jet images from Ref.~\cite{LAGAN}, where a full  description of this dataset can be found as well as the code to generate it. In this dataset, quark pairs from $W$-boson decay are labeled as signal and single quark or gluon jets are labeled as background images.
The intensity of each pixel value represents the sum of the momenta transverse to the beam ($P_{\textrm{T}}$) over the particles which strike a particular cell. 
The images are generated using PYTHIA 8.219 \cite{pythia} simulations of proton collisions at a center-of-mass energy $\sqrt{s}=14$~TeV,  selecting jets with $250 < P_{\textrm{T}} < 300$~GeV. 
Instead of a realistic detector simulation, the calorimeter response is mimicked via a regular $0.1 \times 0.1$ grid in the $\eta$ and $\phi$ coordinates. 
The jet images are constructed and preprocessed as described in \cite{de_Oliveira_jet_images_deep_learning_edition_2016}, including the centering and rotations of the images. 
The resulting images are $25 \times 25$ pixels, with intensity values in the [0,276] range. 
We divide them into a training set containing 400,000 images for the signal and 400,000 images for the background, and a testing set containing 36,000 images for the signal and 36,000 images for the background.
A typical image from this dataset is shown in figure~\ref{sparse_images}.
This dataset has a high degree of sparseness: more than 90\% of its pixels are zero valued.

\subsection{Muon Isolation Study}
Muons leave a very clear detector signature which is difficult to mimic. However, physicists must distinguish between two modes of muon production: a rare mode in which muons are produced from the decay of a heavy boson and are isolated in the detector, and a second prolific mode in which muons are produced inside a jet, surrounded by other particles.  Fluctuations in the jet can occasionally produce apparently-isolated muons.

We use a set of benchmark calorimeter images from \cite{muons}, where  muons from heavy bosons are labeled as signal and muons produced within jets are labeled as background.
The signal muons are generated with the process $p p \rightarrow Z'\rightarrow \mu^+ \mu^-$ with a $Z'$ mass of $20$~GeV/$c^2$. 
Background muons are generated with the process $p p \rightarrow b \bar{b}$. 
Both signal and background datasets are generated at a center of mass energy $\sqrt{s}= 13$~TeV. 
The collisions and immediate decays are simulated with {\sc madgraph5 2.3.3 \cite{madgraph}, 
showering and hadronization with {\sc pythia} 6.428 \cite{pythia}, and detector response with {\sc delphes} 3.4.0 \cite{delphes} using the {\sc delphes} ATLAS detector model. }
Additional proton interactions are overlaid on top of the primary process, at a rate of 50 additional interactions per event.
This dataset only considers muons with $P_{\textrm{T}}$ in the range: $P_{\textrm{T}} \in [10,15]$~GeV$/c$.
The signal events are weighted to match the transverse muon momentum distribution of the background events.
The calorimeter images in the vicinity of the muon are created from the calorimeter deposits within $\eta-\phi$ radius of $R<0.4$, where each pixel represents the momentum transverse to the beam axis. The deposits are preprocessed by centering the image on the coordinates of the identified muon propagated to the calorimeter.
The images are pixelated using a 32x32 grid to roughly match the granularity of the calorimeters of ATLAS and CMS, and the pixels have values in the range [0, 172]. 
The training set contains 41250 signal images and 41246 background images, and the testing set contains 41344 signal images and 41151 background images.
A typical image from this dataset is shown in figure~\ref{sparse_images}.
This dataset has an even greater level of sparsity: more than 98\% of its pixels have zero-value.

\section{Autoregressive Models (ARMs)}
\label{intro arm}
\begin{figure}[ht]
    \centering
    \includegraphics[width=\linewidth]{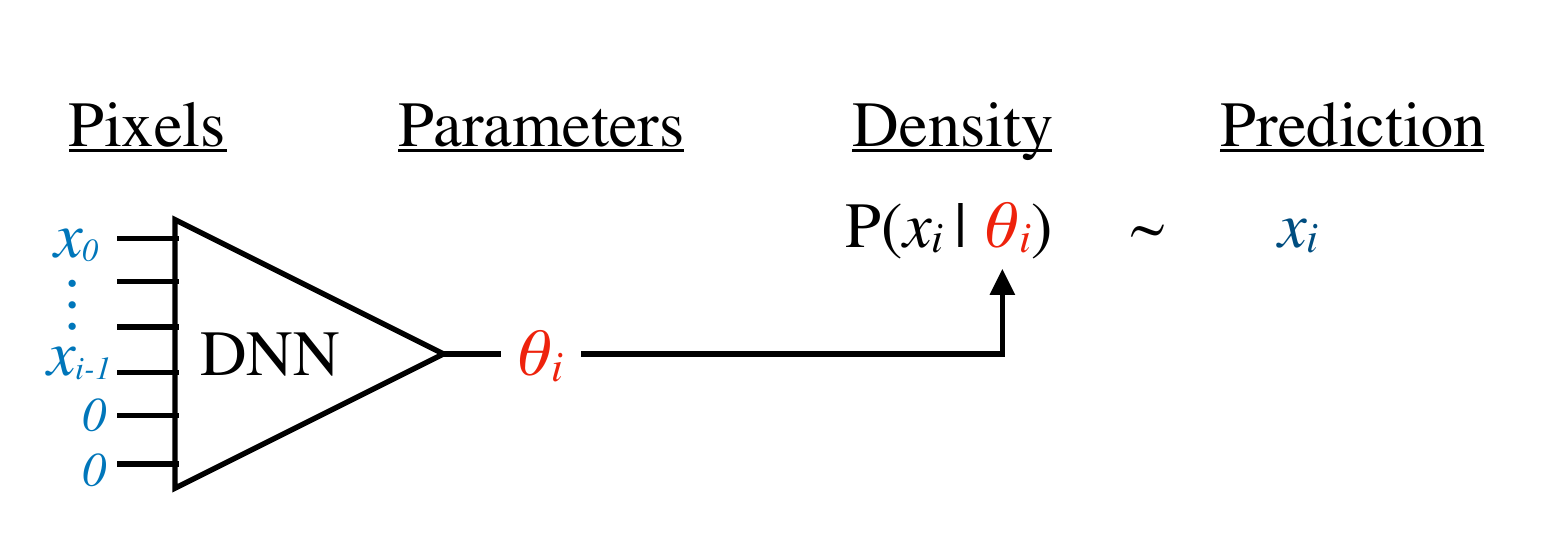}
    \caption{Pixel generation process by a deep ARM to create an image with $D$ pixels. 
    For the pixel $x_i$, a deep neural network (DNN) is evaluated on a vector with values $x_0, \ldots ,x_{i-1}$, zero-padded to length $D$. 
    The output of the network are the parameters $\theta_i$ of a parametric probability density $P(x_i|\theta_i)$, from which $x_i$ is sampled.}
    \label{fig:ARM_scheme}
\end{figure}

Autoregressive models (ARMs) approximate a high dimensional data distribution $P_{{\textrm{data}}}(\mathbf{x})$ with $P(\mathbf{x})$, the distribution induced by the model where $\mathbf{x} \in \mathbb{R}^{D}$. 
For example, when working with images, $P_{\textrm{data}}(\mathbf{x})$  represents the distribution of the values of $D$ pixels in the image.   ARMs are generative models that create outputs sequentially, where each new output is conditioned on the previous output \cite{NADE}.
Formally, ARMs transform the problem of learning the joint distribution $P_{\textrm{data}}(\mathbf{x})$ into learning a sequence of tractable conditional distributions $P(x_{i} | x_{j<i})$.
The ordering of the pixels can influence the model's performance and will be discussed later in the paper. 
ARMs rely on the basic factorization:
\begin{align}
 P(\mathbf{x}) &=P(x_0,x_1,\ldots,x_D) \nonumber\\ 
               &=P(x_{0})P(x_{1}|x_{0})P(x_{2}|x_{0}, x_{1})\ldots P(x_{D-1}|x_{0} \ldots x_{D-2})  \label{conditional}
\end{align}
The conditional densities $P(x_{i} | x_{j<i})$ can be parameterized by deep neural networks \cite{MADE, pmlr-v48-oord16, pixelcnn++, RNADE}
so that: (1)  $P(x_{i} | x_{j<i})=
P(x_{i} | \theta_i)$, where $\theta_i$ represents the parameters of a distribution (e.g. mean and standard deviation); (2) 
$\theta_i=f_i(x_0,\ldots,x_{i-1})$, such that $\theta_i$ depends on previous output; and (3) 
the function $f_i$ is implemented by a neural network. At generation time, the pixel values
 $x_i$ are generated sequentially by sampling in order from the distributions $P(x_i | \theta_i )$. 
A simplified implementation of this process using a single 
neural network is depicted
in figure~\ref{fig:ARM_scheme}. 
The weights of the neural networks that compute the $\theta_{i}$'s are shared across different values of $i$, for regularization \cite{RNADE} purposes and to reduce computational costs, hence the zero-padding of the input vector.

A common concern with ARMs is that by generating pixels in sequence, conditioning only on previously visited pixels, the model may not be able to take into account the dependence of a current pixel on subsequent pixels. However, this
is not the case because the weights are trained using all the data (i.e. ``past'' and ``future'' pixels) and the model always learns to generate the joint marginal distribution of previous and current pixels. 
This idea is further illustrated with a toy example in Appendix \ref{toy example}.

\begin{figure}
    \centering
    \includegraphics[width=\linewidth]{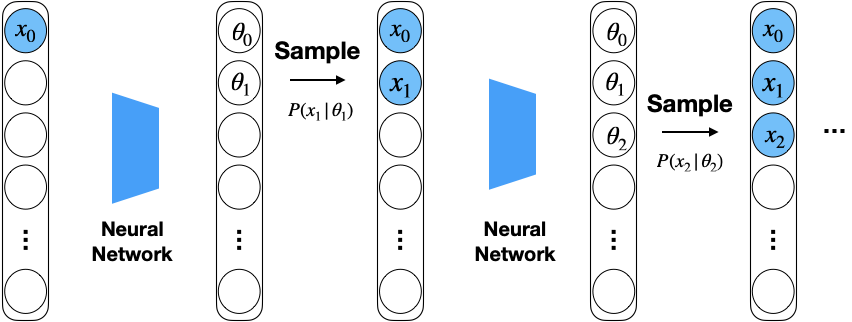}
    \caption{
    Generation process of a deep autoregressive model.
    During generation, the first pixel $x_{0}$ is sampled from $x_{0} \sim P(x_{0}|\theta_{0})$.
    Next, the pixel $x_{0}$ is zero-padded to a $D$ dimensional vector
    and passed to the neural network  ARM model, which evaluates the parameters $\bm{\theta} = \{\theta_{0},  \ldots , \theta_{D-1}\}$, though only 
     $\theta_{1}$ is needed to sample the next pixel $x_{1} \sim P(x_{1}|\theta_{1})$. 
    The pixels $x_{0}$ and $x_{1}$ are again zero-padded to create a $D$ dimensional vector which is passed into the neural network to generate the next pixel. 
    This process is repeated until all pixels are generated. 
    Note that the same neural network is used at each generation step, and part of its weight connections are disabled to preserve the autoregressive structure.}
    \label{arm}
\end{figure}

Learning in ARMs is different from learning in other generative models such as GANs and VAEs. 
ARMs directly minimize the discrepancy, in terms of KL divergence, between the
data distribution  $P_{\textrm{data}}(\mathbf{x})$
and the 
model distribution
$P(\mathbf{x})$ which is produced explicitly. 
In contrast, neither GANs nor VAEs produce a tractable marginal likelihood model $P(\mathbf{x})$ and, as a result,
they have to resort to approximations for minimizing the KL divergence between the data and model distributions.
ARMs avoid this issue by sequentially modeling each conditional probability distribution, allowing them to minimize the KL divergence directly with a tractable likelihood $P(\mathbf{x})$. 
Leveraging the flexibility of deep neural networks to learn each conditional probability, ARMs are able to approximate a large family of continuous distributions in $\mathbb{R}^{D}$ \cite{Neural_Autoregressive_Flows}. 

The implementation of ARMs for images can follow several approaches \cite{MADE, DARNs, pmlr-v48-oord16, pixelcnn++}. 
For scalability during training and generation, we use a single neural network to model the parameters of the conditional probabilities at each step, where some connections are intentionally disabled to preserve the autoregressive structure (see Appendix \ref{mask}), similar to the structure used in \cite{MADE}. 
Given a training image, this makes it possible to calculate {\it all} the parameters $\theta_{0},  \ldots , \theta_{D-1}$ in parallel, instead of calculating each $\theta_{i}$ sequentially.
During generation, the model generates the output elements one-by-one as illustrated in figure~\ref{arm}.

\section{Sparse Autoregressive Models (SARMs)}
\label{sparse_arm}

To deal with sparsity in images, we introduce sparse ARMs (SARMs) in which each conditional distribution is a mixture comprising a Dirac delta distribution at the zero pixel value, as one of its components.
The probability associated with the zero-pixel value is learnable by 
gradient descent, providing a flexible and efficient way of modeling and fitting highly sparse datasets. The other components of the mixture can be modeled in different ways, as described below.

\subsection{Sparse Images Likelihood Models}
In SARM, the likelihood function for the $i$-th pixel $x_{i}$ is formulated as: 
\begin{equation}
    p(x_{i}|\theta_{i}) = \gamma_{i} \cdot \delta_{x_{i}=0} +  (1-\gamma_{i}) \cdot \delta_{x_{i}\neq 0} \cdot p(x_{i}| \phi_{i} )
\end{equation}
where the parameters $\theta_{i}=\{\gamma_{i}, \phi_{i}\}$ are predicted by the underlying neural network taking $x_{0},  \ldots , x_{i-1}$ as its inputs.
Since the pixel values in the calorimeter images represent the physical deposition of energy, they must be non-negative, i.e. $p(x_{i}| \phi_{i} ) > 0$ only when $x_{i} > 0$. 
To satisfy this constraint, we explore two options. 
First, we use a mixture of a Dirac delta distribution at zero with a discrete distribution for the non-zero pixels (D+D). Second, we  use a mixture of Dirac delta distribution at zero with a continuous distribution for the non-zero pixels (D+C).

\textbf{Discrete Mixture Model (D+D):} 
We discretize each pixel value $x_{i}$ by rounding it to the nearest value in a pre-determined grid with points $\{0, g_{1},  \ldots , g_{N}\}$, where $g_{j}>0$ for $j$ from 1 to $N$, and $g_{N}$ corresponds to the largest pixel value after rounding.  
The model learns the probability of each discrete value as a categorical distribution: 
\begin{align}
\label{sm}
    p(x_{i}|\theta_{i}) = \gamma_{i,0}\cdot \delta_{x_{i}=0} + \sum^{N}_{j=1} \gamma_{i,j} \cdot \delta_{x_{i} = g_{j}}
\end{align}

where each $\gamma_{i,j}$ is predicted by the parameter $\theta_{i}=(\theta_{i0},  \ldots , \theta_{iN})$ using a softmax function.
When the grid is uniform, this likelihood is the same as the discretized softmax likelihood used by Pixel RNN \cite{pmlr-v48-oord16}, which has achieved state-of-the-art results on benchmark datasets of natural images. \cite{imagenet_cvpr09}.
However, in particle physics the distribution of pixel values 
is typically far from uniform. In many typical cases, there is a large number of pixels with small values, and a few pixels with large values, as seen in  figure~\ref{pixel_center_outer}. 
To  better represent the pixel distribution and minimize the error due to quantization, we assign more grid points to the region of low pixel values.
We achieve this by using a power transformation $\hat{x} = x^{1/p}$ on the pixel values, where 
$p$ is a hyperparameter such that $p \geq 1$.

\textbf{Discrete and Continuous Mixture Model (D+C):} The pixel values of natural images are usually represented by unsigned integer values between 0 and 255. 
However, in particle physics images, the pixel values are typically real-valued.
To avoid explicit rounding, SARM (D+C) is built with a truncated logistic distribution that models the non-zero distribution component of each pixel.
To generate the D+C mixture,  we reparameterize each pixel as $ x_{i}  = \tilde{x}_{i} \cdot z_{i}$, where $\tilde{x}_{i}$ follows a truncated logistic distribution $TL(\mu_{i}, s_{i})$ with mean $\mu_{i}$ and scale parameter $s_{i}$. 
Here $z_{i} \sim \text{Bern}(\gamma_{i})$ is a Bernoulli random variable with probability $p(z_{i}=1) = \gamma_{i}$, which controls the sparsity level.  
By assuming  independence of $\tilde{x}_{i}$ and $z_{i}$, the likelihood function of $x_{i}$ becomes:

\begin{equation}
\label{MTL likelihood}
    p(x_{i}|\theta_{i}) = \gamma_{i} \cdot \delta_{z_{i}=0} + (1-\gamma_{i}) \cdot  \delta_{z_{i}\neq 0} \cdot p(\tilde{x}_{i}| \mu_{i}, s_{i} ) 
\end{equation}
    
\noindent
where $\theta_{i} = \{\mu_{i}, s_{i}, \gamma_{i}\}$ are functions of the previous pixel values $x_{0:i-1}$, to ensure the autoregressive structure.
In order to allow for unconstrained optimization,  we treat $\log(s_{i})$ 
as the learning parameter and
take its exponential in the likelihood equation \ref{MTL likelihood}. 
Since the pixel distribution could be multi-modal,  we use a mixture of truncated logistic (MTL) distributions for $\tilde{x}_{i}$ which is more flexible. 

The mixture of truncated logistic likelihood differs from the discretized logistic mixture used in Pixel CNN++ \cite{pixelcnn++} in the way it handles continuous pixel values.  Pixel CNN++ requires discretizing $x_{i}$ and then maximizing the probability on the discretized grid. 
In contrast, SARM can directly maximize the probability density function of $x_{i}$, allowing it to handle continuous pixel values without incurring quantization errors.

\begin{figure*}
    \centering
    \includegraphics[width=0.78\textwidth]{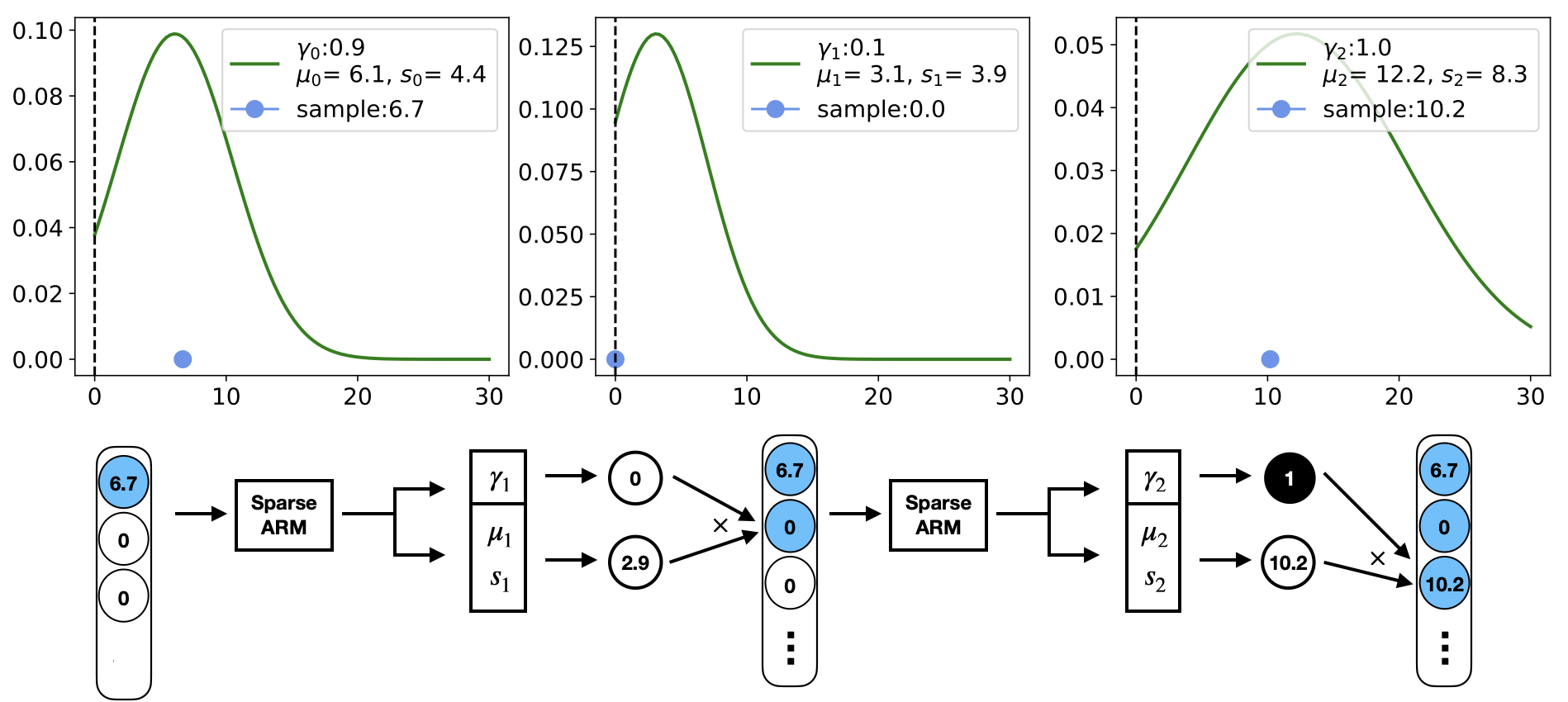}
    \caption{Generation process for the D+C model. The blue circle dots represent the value sampled for each pixel.
    For example, given the first pixel value of 6.7, sampled from the empirical distribution of the dataset, the neural network outputs the distribution parameters $\gamma_{1}=0.1, \mu_{1}=3.1, s_{1}=3.9$ to generate the second pixel.
   Then a Bernoulli random variable is sampled from $z_{1} \sim \text{Bern}(\gamma_{i})$ and a logistic random variable is sampled from $\tilde{x}_{i} \sim \text{Logistic}(\mu_{i}, s_{i})$. The value of the second pixel $x_{i}$ is produced by the product of these two variables as: $x_{i} = z_{i} \cdot \tilde{x}_{i} = 0 \cdot 2.9 = 0$.
     This sequential process is repeated until every pixel is generated.
    }
\label{flow_TN}
\end{figure*}

There are several differences between the D+D and the D+C models. 
The D+D model allows enough flexibility to represent multi-modal distributions, as each grid point has its own learnable probability. 
However, there is a price for this flexibility.
It is significantly more time-consuming to generate an ($N+1$)-way softmax vector and sample from a discrete mixture (D+D) than it is to generate the parameters of  $\gamma, \mu, s$ and then sample from a discrete and continuous mixture (D+C).
Other constrained domain distributions such as the exponential and the gamma distributions were also considered but led to inferior results. The
exponential distribution suffers from a lack of  flexibility due to having only one learnable parameter.

\subsection{Multi-Stage Generation for Heterogeneous Areas}
\label{multiscale}
\begin{figure*}
\centering
\begin{minipage}{.5\textwidth}
  \centering
  \includegraphics[width=0.95\linewidth]{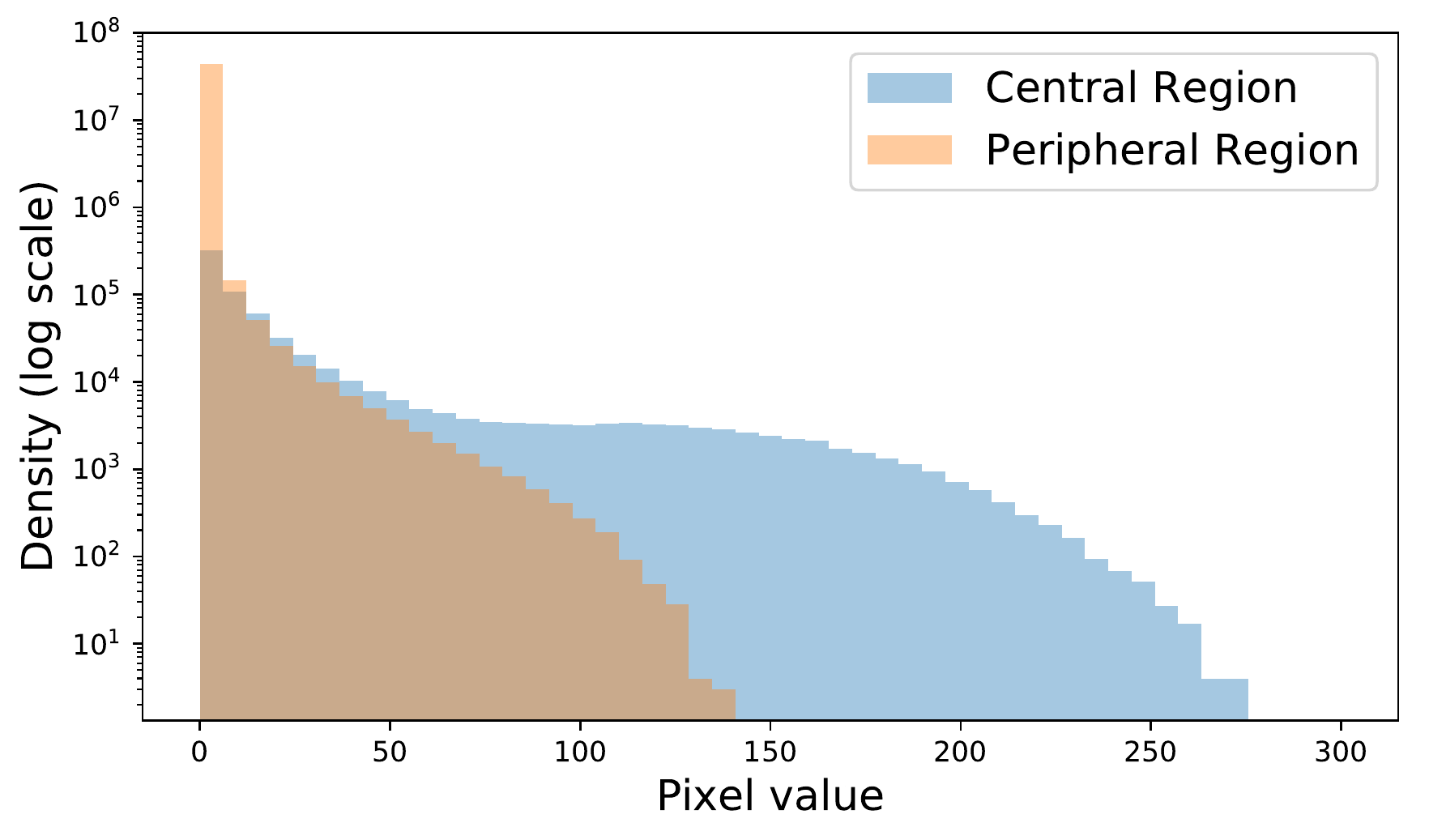}
\end{minipage}%
\begin{minipage}{.5\textwidth}
  \centering
  \includegraphics[width=0.95\linewidth]{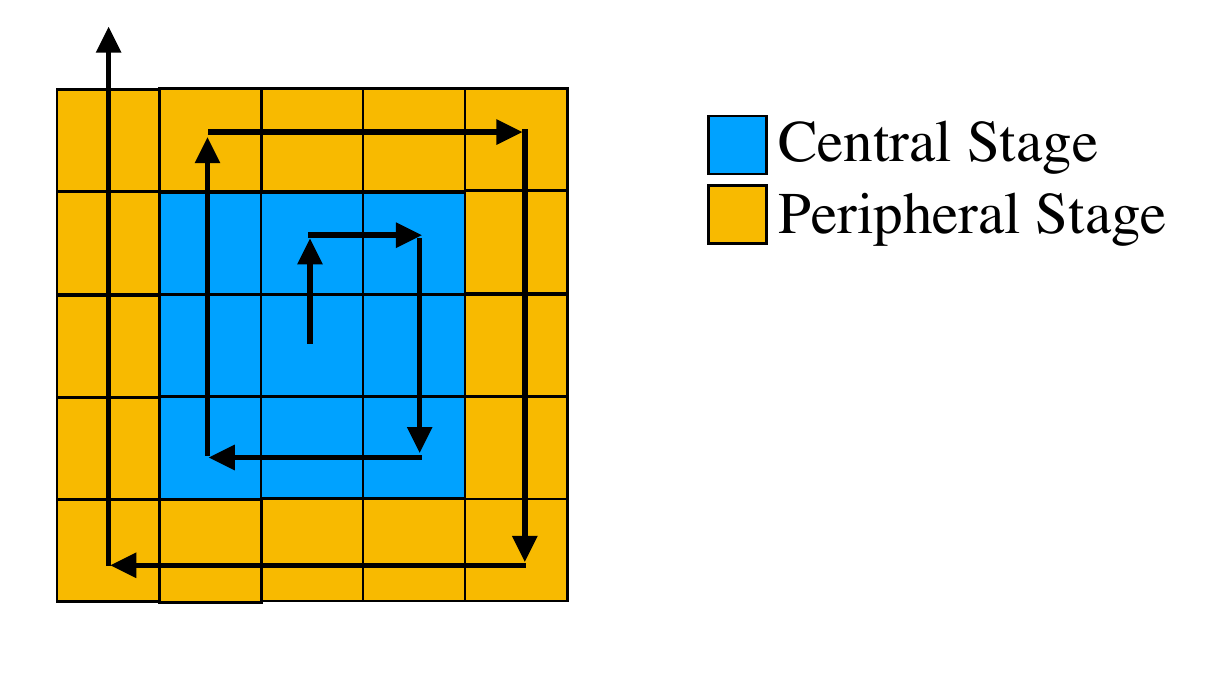}
\end{minipage}
\caption{
\textbf{(Left)} Distribution of pixel values in the jet substructure dataset for the 9 pixels in the center of the images (central region) and the rest of the pixels (peripheral region). Note that the majority of the pixels in the peripheral region are zero-valued and in general have lower variance than pixels in the central region.
\textbf{(Right)} Two-stage generation for the central and peripheral regions using a spiral path and two different SARM modules. Using different networks for each region improves performance.
}
\label{pixel_center_outer}
\end{figure*}

In many ARM applications, a single network is used to predict the parameters $\theta_i$ of the conditional probability distribution $P(x_{i}|\theta_{i})$. 
This approach works well if the distribution of pixel values is similar across pixels, as is often the case in natural images.
However, as shown in figure~\ref{pixel_center_outer} (left), the pixel value distribution in the central square of a calorimeter image containing a jet is very different from the distribution in the rest of the image (see also \cite{de_Oliveira_jet_images_deep_learning_edition_2016}).
In order to handle these heterogeneous regions, we use a two-stage approach by stacking two distinct deep SARM modules, one for the center and one for the periphery. When the model generates the image from the inside out, the outer module generates pixels conditioned on the outputs of the center module, as illustrated in figure~\ref{pixel_center_outer} (right).
We refer to this model as SARM-2 while the single stage model is SARM-1. Since the center may not have a clear border, we treat the size of the center relative to the periphery as a hyperparameter during training. Note that in general the number of stages depends on the structure of the data and 
is not limited to two. Furthermore, it is possible to learn the SARMs associated with each region in any order.


Thus, in summary, through the experiments to be presented, we show that a good heuristic approach for SARM design is to: (1) decompose the images into relevant regions (e.g. center vs background); (2) use a different SARM for each region type; and (3) within each region type, preferably choose a systematic and congruent order for generating the pixels, as these compare favorably to random generation orders. By systematic and congruent orders we mean orders that have some kind of continuity for the location of the pixels being generated--subsequent generated pixels should be close in the image--while respecting the geometry  of the highly activated region (e.g. a spiral order for a globular region, a linear order for a linear region).


\section{Evaluation Methods}
\label{methods}

The goal is to train generative models which create images indistinguishable from images created by the slower Monte Carlo methods.  We compare the performance of our models, both in terms of image quality and generation time, against two other generative models: LAGAN \cite{LAGAN}, the current state-of-the-art generative model for sparse images in particle physics; and Pixel CNN++ \cite{pixelcnn++}, a widely used autoregressive model for natural images not tuned for sparse images.  We evaluate all models on both datasets described above; note that LAGAN was designed to handle images typically found in the jet substructure dataset, while the muon dataset features extreme sparsity in comparison. We measure the quality of the generated images both qualitatively and quantitatively. 

{\bf Qualitative Evaluation:} We examine typical images generated by each model, as well as the pixel-wise average intensity of the generated images, using the images produced by the Monte Carlo methods, which in the jet substructure study are referred to as the Pythia images. 
Additional qualitative comparisons are described in the Appendix \ref{further_analysis_jet} and \ref{further_analysis_muon}.

{\bf Quantitative Evaluation:} Comparisons of distributions in high-dimensional datasets should focus on the scientific context and potential applications.  In particle physics, the calorimeter information is typically used to calculate physical quantities, such as invariant mass or transverse momentum ($P_{\textrm{T}}$), which are especially revealing as metrics because they have not been explicitly optimized by the models. In addition, calorimeter images are used to train classifiers which can identify particles from their patterns of depositions.

One-dimensional distributions of mass and $P_{\textrm{T}}$ can be evaluated in comparison to the distributions from Monte Carlo generators using the Wasserstein distance, the minimum cost to transform one distribution into the other one, expressed by:
\begin{equation}
W_{p}(P, Q)=\left(\inf _{J \in \mathcal{J}(P, Q)} \int\|x-y\|^{p} d J(x, y)\right)^{1 / p}
\end{equation}
where $\mathcal{J}(x,y)$ is the family of joint probability distribution of $x$ and $y$; $P$ and $Q$ are marginal distributions, and $p \geq 1$. When $p=1$, this metric is also known as the Earth Mover's Distance \cite{Wasserstein_distance}. 
To match the results in \cite{LAGAN}, we computed $W_{1}(P, Q)$, where $P$ represents one of jet observable distributions from the Pythia images, and $Q$ represents the corresponding jet observable distribution from the generated images. 

An important motivation for developing generative models for fast simulations is to provide a computationally inexpensive method to augment existing datasets in classification task \cite{de_Oliveira_jet_images_deep_learning_edition_2016, Baldi:2016fql}.
The jet substructure dataset was generated to train classifiers to distinguish between jets from $W$ boson decays (signal) and those from single quarks and gluons, a well-known classification task \cite{de_Oliveira_jet_images_deep_learning_edition_2016, Baldi:2016fql}.
The muon isolation  dataset was generated to train classifiers to distinguish isolated muons from those due to heavy-flavor jet production.
Therefore, an essential test for the quality
of the generated images is whether they can be used in these classification tasks. 
To quantify this, the generated images were used as training sets to develop a classifier whose performance was assessed using the Monte Carlo images.
The same convolutional neural network architecture was trained with the same hyperparameters on five different data sets: Monte Carlo images,
images generated by SARM-2 (D+C) images generated by SARM-2 (D+D),
images generated by LAGAN, and images generated by Pixel CNN++.
Because higher quality images should lead to improved classification of the Monte Carlo images, we used the classification performance as the 
evaluation metric.

{\bf Speed:} Each generative model was used to generate batches of images multiple times to measure the average speed of image generation. 

\section{Results}
\label{exp}

\subsection{Jet Substructure Study}
\label{Jet substructure study}

\subsubsection{Qualitative Analysis}
\label{Qualitative analysis}
\begin{figure*}
    \centering
    \includegraphics[width=0.8\linewidth]{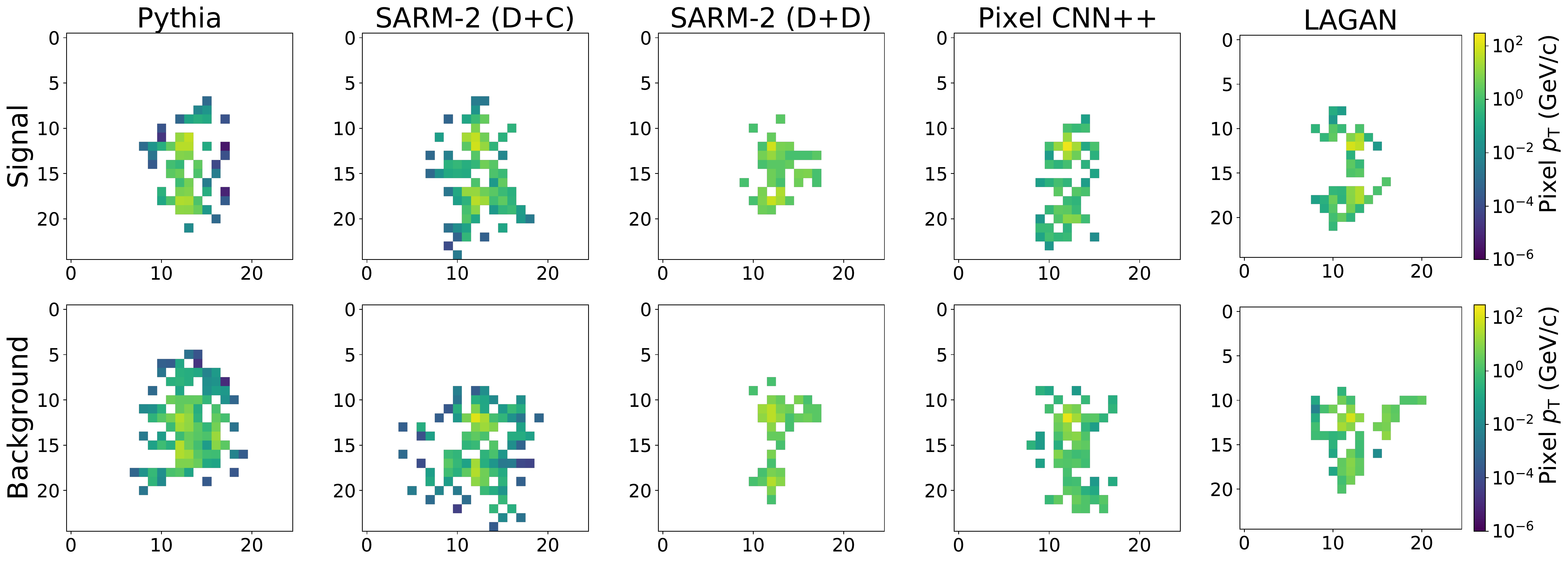}
    \caption{ 
    Example jet images generated from each model. Notice that SARM-2 (D+C) is able to produce small value pixels in the periphery of the images. 
    The intensity of each pixel is shown on a $\log$ scale, where the white space represents pixels with value zero.
    }
    \label{fig:example_images}
\end{figure*}

\begin{figure*}
    \centering
    \includegraphics[width=0.8\linewidth]{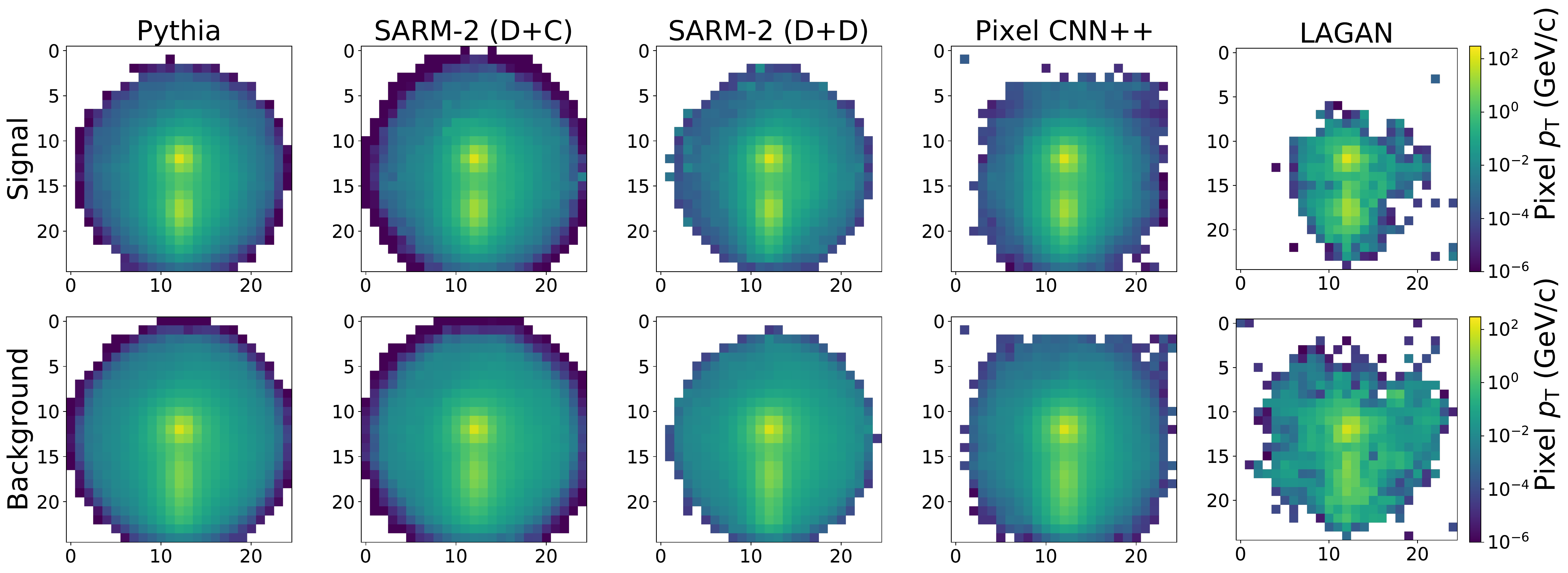}
    \caption{
    Pixel-wise average of the images generated by each model. Notice that LAGAN struggles to capture the distribution of low value pixels in the periphery of the images and has a non-smooth radial transition compared to the autoregressive models.
    The intensity of each pixel is shown on a $\log$ scale, where the white space represents pixels with value zero.
    }
    \label{mean_comparison}
\end{figure*}

An example image from each generative model and from the Pythia Monte Carlo generator is shown in figure~\ref{fig:example_images}. 
It is clear that SARM-2 (D+C) excels at generating pixels with small values around the periphery in comparison to the other models. Additional samples for each model can be seen in Appendix \ref{sample_images_appendix}.
To assess the overall quality of the generated images, figure~\ref{mean_comparison} shows the pixel-wise average of each dataset. 
The autoregressive models, SARMs and Pixel CNN++, are able to model the peripheral radial region around the center more accurately.
This region has higher degree of sparseness than the center region, making it more challenging for the generative models to accurately capture. We note that the images from the SARM-2 (D+C) model appear to be most similar to the Pythia images, while the other models are less able to generate the peripheral region faithfully.
In addition, Pixel CNN++ struggles to achieve the radial structure present in the Pythia images and creates a square-like structure instead.
In general, the images from figure~\ref{mean_comparison} generated by the autoregressive models show a smooth transition from the highly activated center to the sparse border, similar to that seen in the Pythia dataset.
In contrast, the border of the LAGAN images is irregular, which could be due to its reliance on the ReLU activation function to induce the sparsity, making the model unable to estimate the sparseness level directly.

\subsubsection{Quantitative Analysis: Jet Observables as Metrics for Quality}
\begin{figure*}
    \centering
    \includegraphics[width=0.8\textwidth]{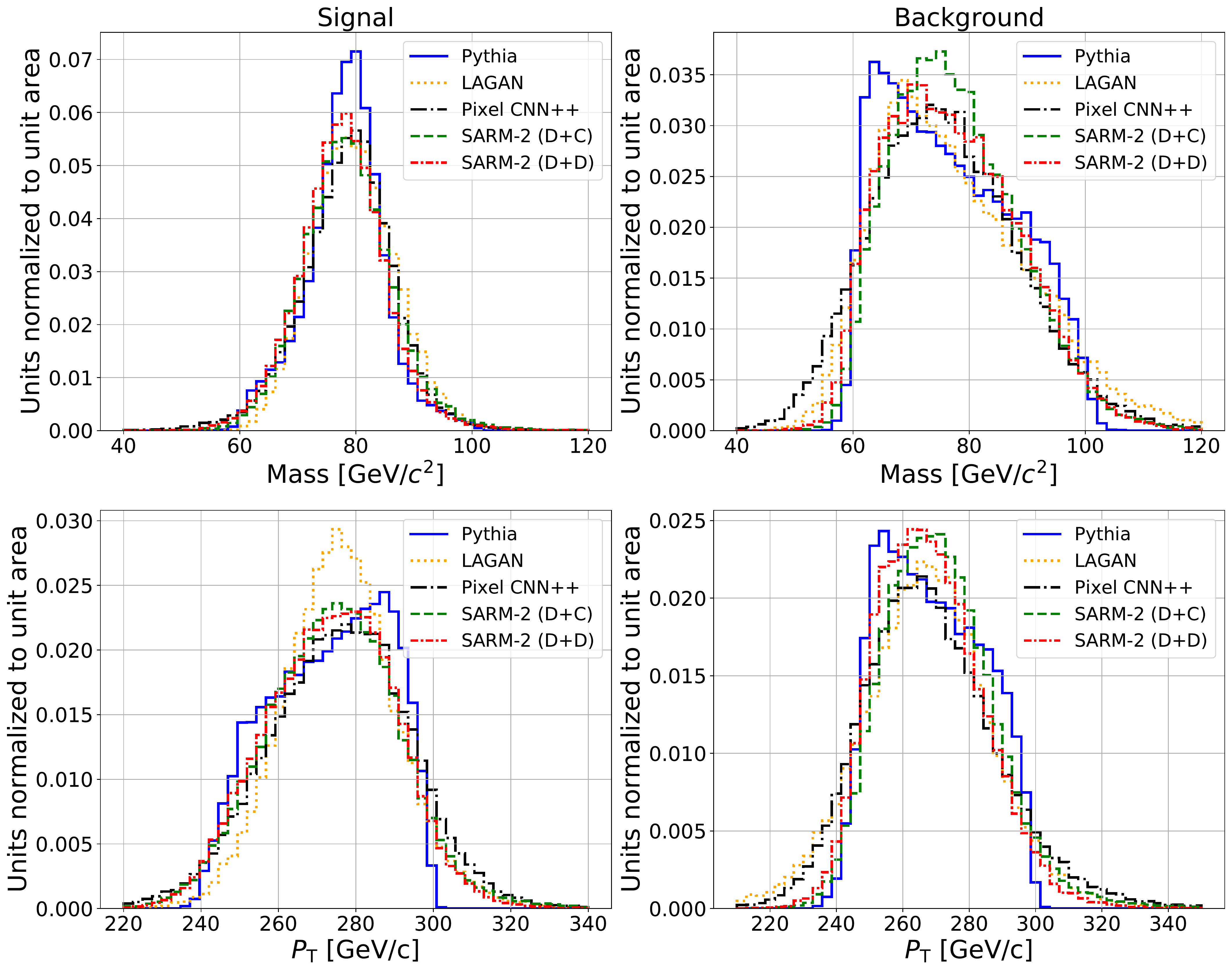}
    \caption{ Distributions of jet observables (\textbf{Top}:  Mass, \textbf{Bottom}: $P_{\textrm{T}}$) calculated from images generated by several generative models and from the original images generated by Pythia.  Signal images, with two collimated quarks, are on the left; background images, with a single quark or gluon, are on the right.}
    \label{mass_pt}
\end{figure*}

To quantify the fidelity of the images generated by each model as compared with the original samples, we insert them into typical applications in particle physics.  In the context of collisions that produce jets, it is common to calculate the invariant mass of the jet, and the transverse momentum.  Distributions of jet mass and $P_{\textrm{T}}$ are shown in figure~\ref{mass_pt} for all models, which all succeed in matching the general shape, though discrepancies are visible, and Wasserstein distances are shown in table \ref{divergence table}.

\begin{table}
\caption{ Comparison of images created by various generative models with original images from Pythia, evaluated using the Wasserstein distance (with $p=1$) between one-dimensional distributions of physical quantities calculated from the images: jet $P_{\textrm{T}}$ and invariant mass, also shown in figure~\ref{mass_pt}.  Smaller values indicate a closer match to the Pythia images. Four SARMs are evaluated, those with either one-stage (SARM-1) or two-stage (SARM-2) models, and those with 
either discrete and continuous distributions (D+C) or a mixture of discrete distributions (D+D).
}
\label{divergence table}
\centering
\vspace{0.1in}
\begin{tabular}{lllll}
\hline \hline
                              & \multicolumn{2}{c}{$P_{\textrm{T}}$} & \multicolumn{2}{c}{Mass} 
                              \\
\cline{2-5} 
Model                              & Signal   & Background & Signal    & Background 
                              \\ 
\hline                            
LAGAN        &   3.15 & 3.29 & 1.45 & 1.39
\\
Pixel CNN++         & 3.46 & 3.59 & 1.09 & 1.56
\\
SARM-1 (D+C)      & 2.33 & 2.46 & 1.07  &  1.54  
\\
SARM-2 (D+C)    & 2.32 & 2.71 & 1.06 & 1.39
\\
SARM-1 (D+D)  & 1.95 & 2.52 & 1.34  &   2.45
\\
\textbf{SARM-2 (D+D)}  &  \textbf{1.44} &  \textbf{1.66}     & \textbf{0.94}   & \textbf{0.92}   
\\
\hline \hline
\end{tabular}
\end{table}

All SARM variants achieve lower distances in the $P_{\textrm{T}}$ distributions than LAGAN and Pixel CNN+, and comparable or better distances in jet mass.  The best results in all categories are obtained by the SARM-2 (D+D).
Compared to the best of Pixel CNN++ and LAGAN, SARM-2 (D+D) provides a $51.92$\%  improvement for $P_{\textrm{T}}$, and a 23.79\% improvement for mass, averaged over the signal and background sets. 
These results demonstrate the effectiveness of taking sparseness into account during learning and generation. 
Secondly, the SARM-2 models clearly outperform the SARM-1 models for both the (D+D) and (D+C) likelihoods, which shows the effectiveness of the multi-stage approach in modeling heterogeneous areas in the images. 

\subsubsection{Classification of Generated Images}
\begin{figure}
    \centering
    \includegraphics[width=.9\linewidth]{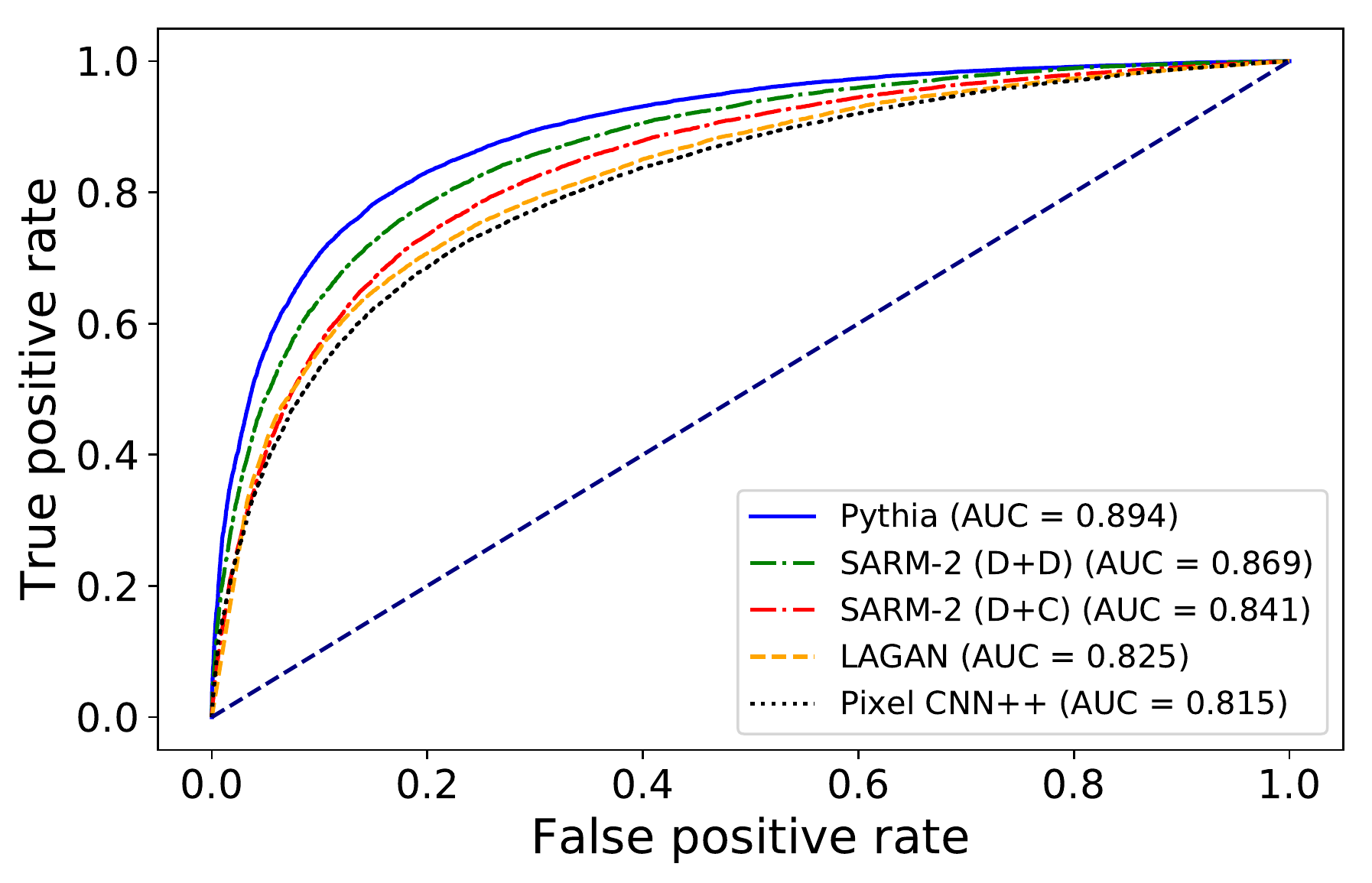}
    \caption{ Evaluation of the fidelity of images generated by several models in the context of a classification task. Images generated by the model are used to train a network to discriminate between signal and background, but performance is measured using the original Pythia images.}
\label{classifier_train}
\end{figure}
An important application of generated calorimeter images is to augment training sets for networks learning to perform vital signal-background classification tasks.  As a high-level test of the image quality, we train networks using images generated by each model (200k signal, 200k background), and evaluate the performance on the original images from Pythia (20k signal, 20k background). Training sets which best mimic the original Pythia images should lead to networks which most closely match the performance of a network trained on Pythia images. Detailed information about the classifier and training procedure are given in Appendix \ref{classifier}. The receiver operating characteristic (ROC) curves for networks trained on images from Pythia , SARM-2 (D+C), SARM-2 (D+D), Pixel CNN++ and LAGAN are shown in figure~\ref{classifier_train}. Classifiers trained on both SARM generated datasets have higher AUC (area under the ROC curve) scores than  the classifiers trained on the LAGAN images and Pixel CNN++ images.

\subsubsection{Generation Order}
\label{lagan_order}
SARMs generate images pixel by pixel, conditioning each step on the  previously generated pixels.  The order of the pixel generation corresponds to a dependency decomposition in Equation \ref{conditional}, which may impact training performance. The traversal path is especially important for images containing heterogeneous areas. For natural images, \cite{MADE} uses an ensemble of models with random paths, while Pixel CNN++ and other models~\cite{pixelcnn++, pmlr-v48-oord16} use the row-by-row pixel ordering. 

\begin{table}[ht]
\caption{
Quality of jet substructure signal images generated by SARM-1 (D+D) with various pixel-generation orderings. The quality is measured by the Wasserstein distance for the  physical observables ($P_{\textrm{T}}$ and mass) between the generated images and the original Pythia images. Spiral-in clockwise/counterclockwise (CW/CCW), spiral-out clockwise/counterclockwise (CW/CCW), column-wise, row-wise, and two random approaches are compared. The outward spiral orders show good performance due to the radial structure of the images.
}
\vspace{0.1in}
\label{order_performance_lagan}
\begin{tabular}{lll}
\hline \hline
               & $P_{\textrm{T}}$ (std)  & Mass (std) \\
               \hline
Spiral-out CCW  & 1.94 (0.09) & 1.38 (0.10) \\
Spiral-out CW   & 2.47 (0.23) & 1.53 (0.22) \\
Spiral-in CCW & 3.64 (0.32) & 1.62 (0.14) \\
Spiral-in CW  & 3.20 (0.22) & 1.45 (0.16) \\
Row-wise       & 3.06 (0.30) & 2.01 (0.11)\\
Column-wise    & 3.38 (0.39) & 1.90 (0.08)\\
Random I       & 4.05 (0.51) & 1.74 (0.53)\\
Random II       & 3.41 (0.33)& 1.25 (0.26) \\
\hline \hline
\end{tabular}
\end{table}


The average performance of various pixel orderings  for SARM-1 (D+D) over 10 repeated runs is shown in table \ref{order_performance_lagan}. Each order is evaluated by using the Wasserstein distance between the distributions of the generated signal images and the Pythia signal images for the jet $P_{\textrm{T}}$ and invariant mass. 

The spiral paths, clockwise (CW) and counterclockwise (CCW), achieve the stronger results.
This could be understood in terms of mutual information between neighboring pixels. Unlike the other orderings, the spiral ordering is continuous, i.e. it
always generates a pixel adjacent to the previously generated pixel. Furthermore, the spiral order is congruent with the globular shape of the highly activated region in the jet images, e.g. Fig. \ref{mean_comparison}.  
Starting the spiral from the center outperforms inward spirals,  indicating that it may be easier to learn the correlations between the pixels starting with pixels that are more active (more non-zero pixel values). 
The difference between CW and CCW is likely due to asymmetries generated by the rotation and centering steps in the preprocessing of the data. We use this asymmetric version of the data in order to enable direct comparison to the LAGAN model.
These results confirm that non-random, systematic, generation orders that have good continuity and congruence properties perform well (and outperform random orders).
A full exploration of the ordering dependency is beyond the scope of this work and computationally challenging due to the factorial number of possible orderings.

\subsubsection{Computational Costs}
\label{costs_lagan}
Table \ref{speed} shows the speed of the generative models in comparison to the Monte Carlo method (Pythia). 
The SARM-2 models are five times slower than LAGAN, which is mainly due to
the extra computational cost of the autoregressive structure.
On the other hand, the SARM-2 models are two orders of magnitude faster than Pythia and Pixel CNN++. The forward pass of the Pixel CNN++ model is computationally expensive due to the ResNet blocks with convolutional layers and skip connections \cite{pixelcnn++, resnet}.  In contrast, SARMs use a simple feed forward network with disabled connections to preserve autoregressive structure.
The speed of the generative models is measured on a machine with 4 TITANX GPU cards each with 12G of memory.
The speed of Pythia was assessed in \cite{LAGAN} using Amazon Web Services (AWS) and an IntelR XeonR E5-2686 v4 at 2.30GHz CPU.

\begin{table}[ht]
\centering
\caption{Comparison of image generation speed between the Monte Carlo approach (Pythia) and various generative models. The SARM-2 models are slower than LAGAN, but still considerably faster than Pythia and Pixel CNN++.}
\label{speed}
\vspace{0.1in}
\begin{tabular}{ll}
\hline \hline
Model                    & Speed (images/sec) \\
\hline
Pythia \cite{LAGAN}     & 34                \\
Pixel CNN++              & 50                  \\
SARM-2 (D+D)  & 1612                \\ 
SARM-2 (D+C)          & 2480             \\
LAGAN                    & 10176            \\
\hline\hline
\end{tabular}
\end{table}

There is room to further optimize the speed of the SARM models. For instance, we find that reducing the size of the intermediate upsampling layer of the SARM (D+D) drastically reduces the memory requirements and improves the generation speed. Another direction is to explore model pruning and compression.

\subsection{Muon Isolation Study}

\subsubsection{Qualitative Analysis: Average Generated Images}

\begin{figure*}
    \centering
    \includegraphics[width=0.8\textwidth]{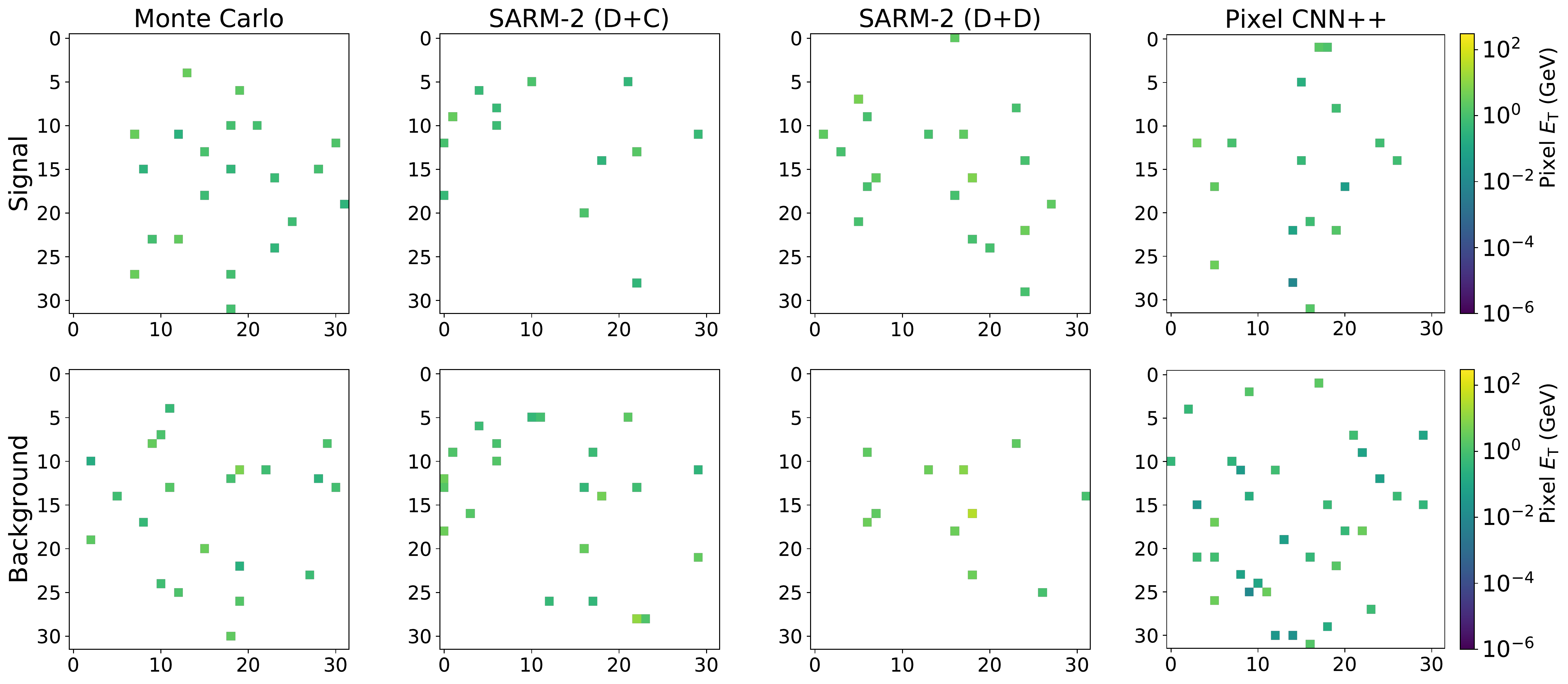}
    \caption{
    Example calorimeter images in the vicinity of a muon from the generative models as well as the original Monte Carlo generator. Top row shows isolated muons  (signal), while the bottom shows muons produced in association with a jet (background). The intensity of each pixel is shown on a log scale,  where the white space represents pixels with value zero. 
    }
    \label{muon_typical_image}
\end{figure*}

Typical calorimeter images in the vicinity of a muon generated by the standard Monte Carlo method, Pixel CNN++ as well as two SARMs are shown in figure~\ref{muon_typical_image}.  In this context, LAGAN suffered from mode collapse and failed to generate reasonable quality images (See figure~\ref{muon_mean_with_lagan} in the Appendix). This is a well known problem when training GANs \cite{convergence1, modecollapse, LAGAN}, especially with sparse data.

\begin{figure*}
    \centering
    \includegraphics[width=0.8\textwidth]{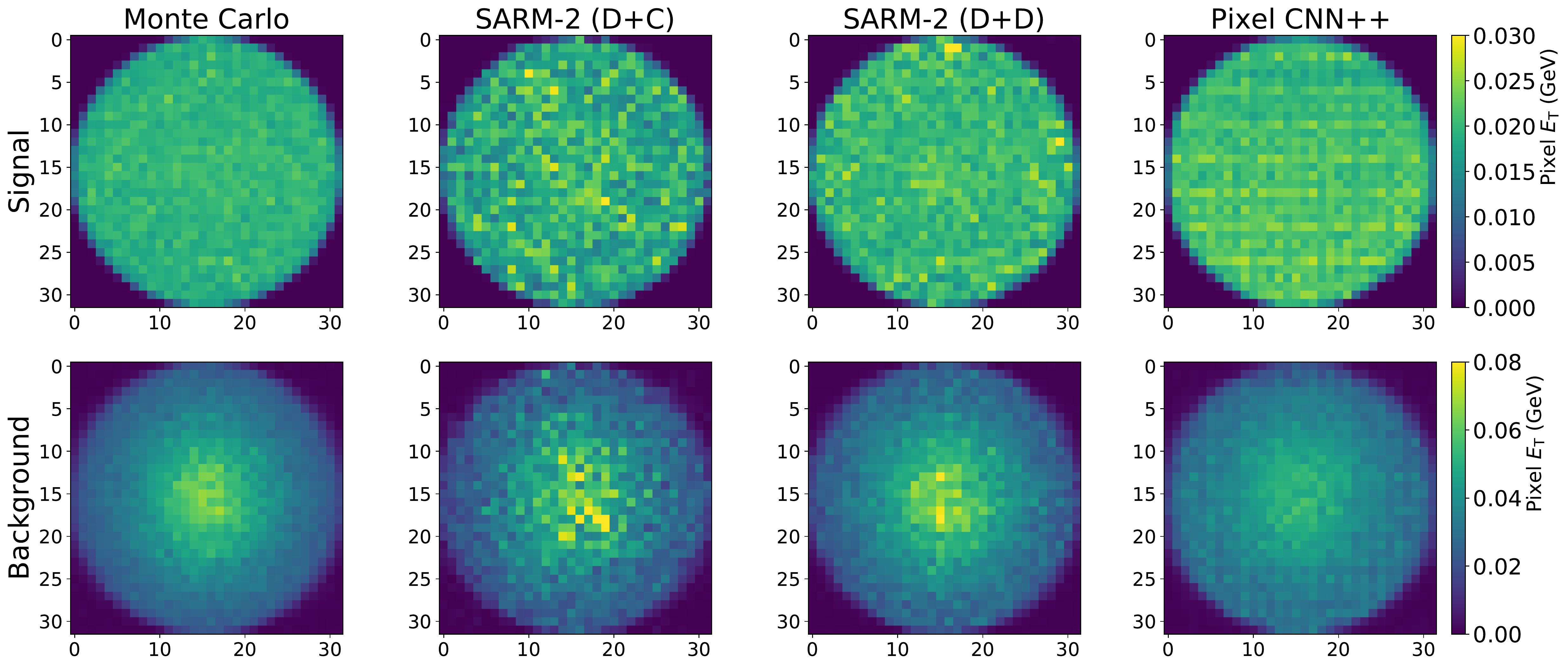}
    \caption{ Pixel-wise averages of calorimeter images in the vicinity of a muon from the generative models as well as the original Monte Carlo generator. Top row shows isolated muons (signal), where little calorimeter activity is expected. The bottom row shows muons produced in association with a jet (background), which deposits significant energy near the muon. A linear scale is used to reveal the differences between signal and background images.
    }
    \label{muon_mean}
\end{figure*}

Figure~\ref{muon_mean} shows the pixel-wise average images. The SARM-2 models and the Pixel CNN++ reproduce the radial symmetry seen in the original images. However, the average images produced by Pixel CNN++  contain noticeable artifacts, potentially due to the convolutional layers in the model \cite{checkerboard}.

\subsubsection{Quantitative Analysis: Calorimeter Observables as Metrics for Quality}
\begin{figure*}
    \centering
    \includegraphics[width=0.8\textwidth]{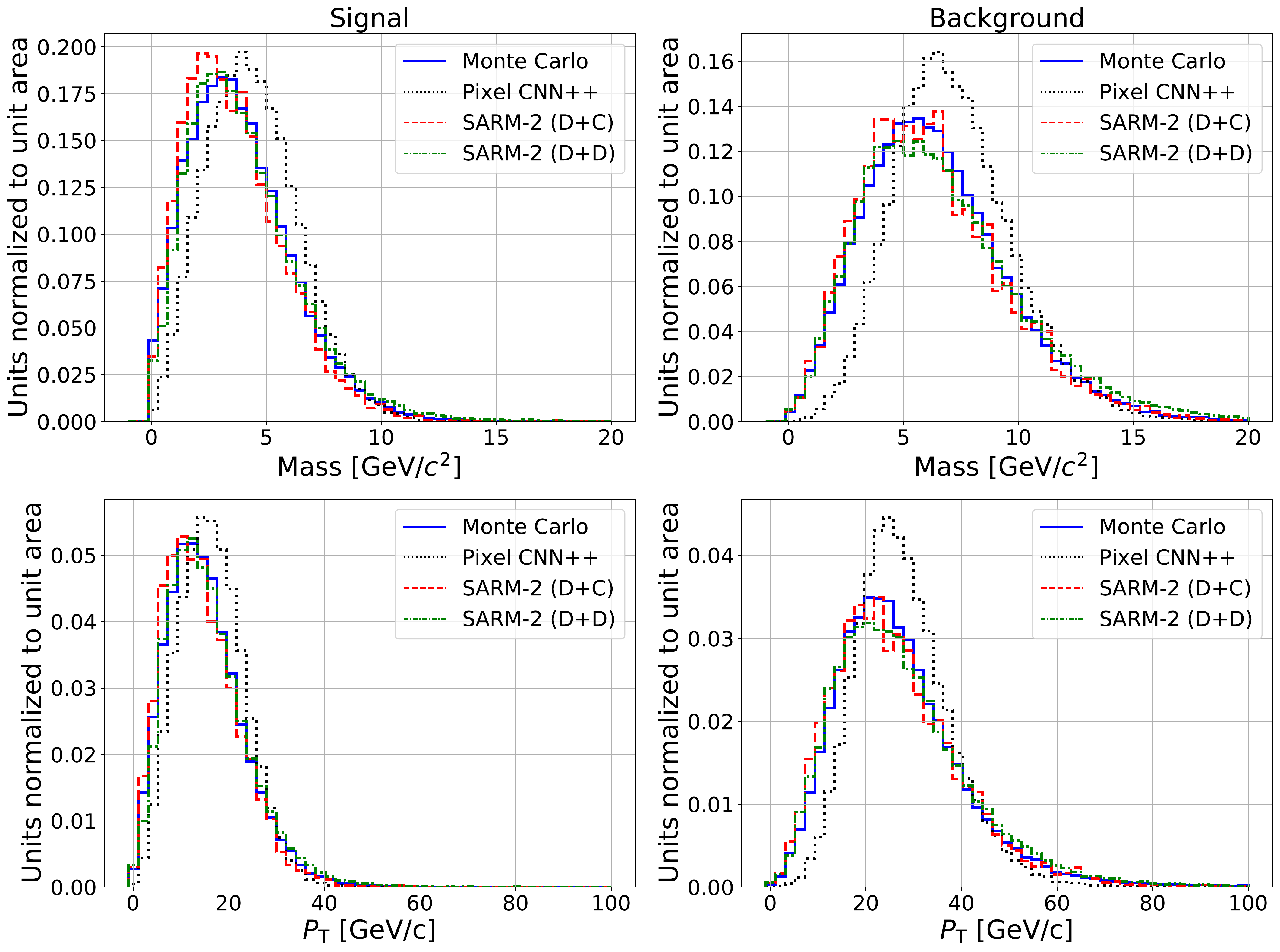}
    \caption{ Distributions of calorimeter observables (top: invariant mass, bottom: total $P_{\textrm{T}}$) calculated from images generated by several generative models and the originals generated by a Monte Carlo generator.  Signal images, in the vicinity of an isolated muon, are on the left. Background images, in the vicinity of a muon produced with an associated jet, are on the right. 
    }
    \label{fig:muon_mass_pt}
\end{figure*}

To assess the fidelity of the images quantitatively, we calculate physical quantities which summarize the content of the images and allow for comparison of one-dimensional distributions. While calorimeter images in the vicinity of a muon do not necessarily contain a clustered jet, the total $P_{\textrm{T}}$  and invariant mass of the entire image do have physical meaning.   Figure~\ref{fig:muon_mass_pt} shows the distributions of these quantities for the original Monte Carlo images, as well as for the generated images, and table \ref{muon_metrics} provides the corresponding Wasserstein distances.

\begin{table}
\caption{
Comparison of images created by various generative models to the original Monte Carlo images using the Wasserstein distance (with $p=1$) between one-dimensional distributions of physical quantities calculated from the images: $P_\textrm{T}$ and invariant mass, also shown in figure~\ref{fig:muon_mass_pt}.  Smaller values indicate a closer match to the Monte Carlo images. Two SARMs are evaluated, with either discrete and continuous distributions (D+C) or a mixture of discrete distributions (D+D).
}
\label{muon_metrics}
\centering
\vspace{0.1in}
\begin{tabular}{lllll}
\hline \hline
                              & \multicolumn{2}{c}{$P_{\textrm{T}}$} & \multicolumn{2}{c}{Mass} \\
\cline{2-5}                              
Model                              & Signal   & Background & Signal    & Background   \\
\hline                 
PixelCNN++                  & 1.75      &   2.92 &  0.58    & 0.82 \\
SARM-2 (D+C)            & 0.79 &  0.97  &  0.25   & \textbf{0.21}  \\
SARM-2 (D+D)             & \textbf{0.56} &  \textbf{0.93}  &  \textbf{0.17}   & 0.31  \\
\hline \hline
\end{tabular}
\end{table}

The datasets generated by both SARM-2 models have considerably smaller Wasserstein distances than the datasets generated by the Pixel CNN++ model for both signal and
background.  
The distributions of all the generated datasets approximate the shape of the Monte Carlo distributions quite well for $P_{\textrm{T}}$ and mass, but the distributions of the Pixel CNN++ dataset have a small shift towards higher values, for both the signal and the background. In addition, for the background they are more concentrated around the mean. 
This is potentially due to the fact that Pixel CNN++ fails to model the right tail of the pixel distribution, where the pixels have higher values but appear much less frequently in the data (figure~\ref{muon_pixel_intensity} in Appendix). 
The SARM-2 (D+D) has the best overall performance, with improvements of $68.08\%$ for $P_{\textrm{T}}$ and $66.44\%$ for mass, averaged over the signal and background datasets.

\subsubsection{Classification of Generated Images}
The fidelity of the images can be evaluated in the context of the data analysis task for which they were created, training a network to distinguish between signal (calorimeter images near isolated muons) and background (calorimeter images near non-isolated muons).

\begin{figure}[H]
    \centering
    \includegraphics[width=.9\linewidth]{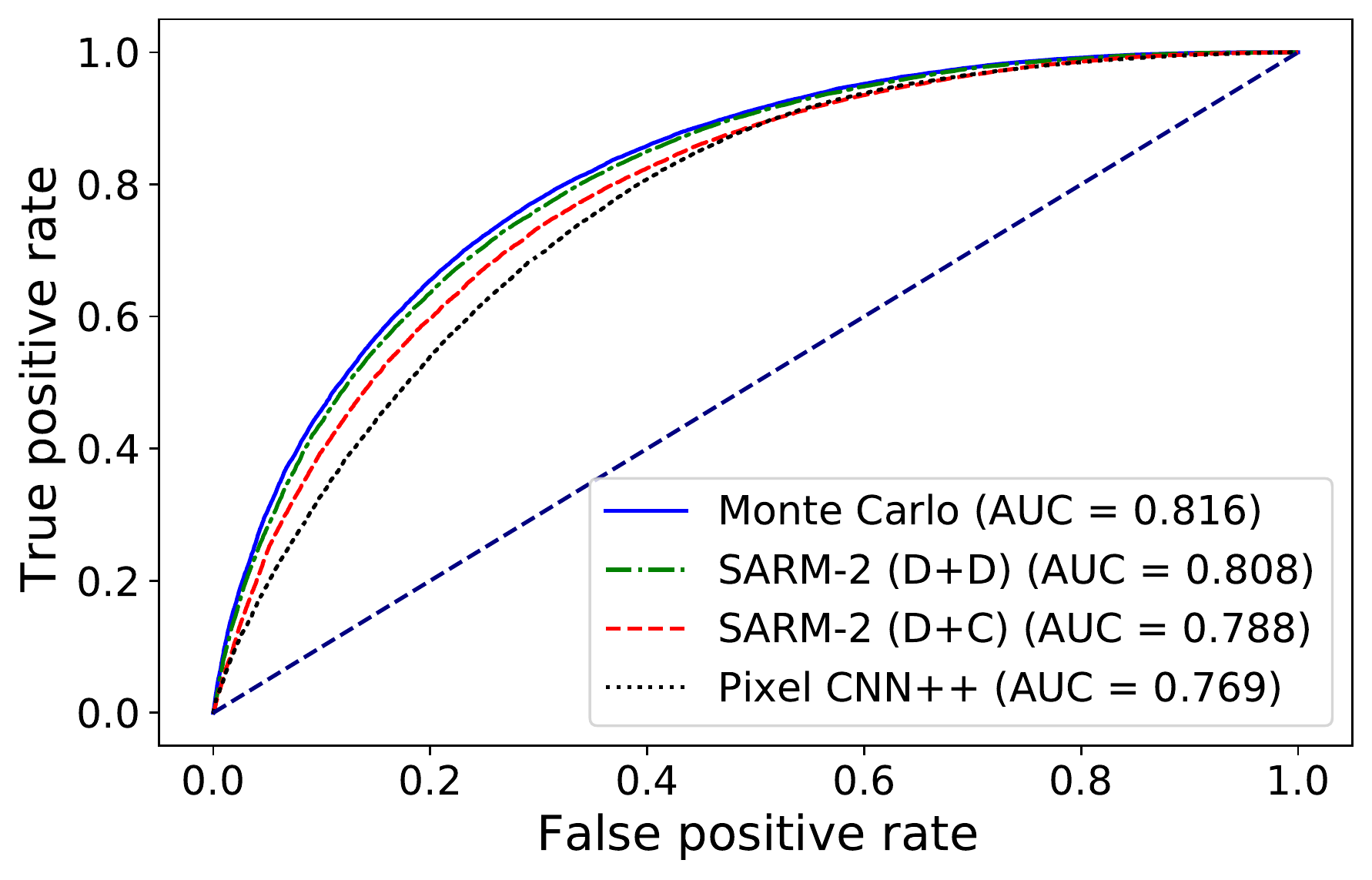}
    \caption{
    Evaluation of the fidelity of images generated by several models in the context of a classification task, distinguishing muons produced in isolation from those produced in association with a jet. Images generated by the model are used to train a network to discriminate between signal and background, but performance is measured using the original Monte Carlo images.}
    \label{classifier_train_muon}
\end{figure}

A convolutional neural network classifier was trained using images generated exclusively by each of the models (SARM-2 (D+C), SARM-2 (D+D), or Pixel CNN++); one additional network was trained using images  from the Monte Carlo generator.
The quality of the images is measured by comparing the classification performance of these networks on images from the Monte Carlo generator, see figure~\ref{classifier_train_muon}.  
The classifiers trained on each SARM dataset have higher AUC score than the classifier trained on the Pixel CNN++ dataset, providing additional evidence that the SARM datasets are more similar to the Monte Carlo images and thus better suited for downstream tasks such as data augmentation.

\subsubsection{Generation Order}
In this section, we discuss the impact of the pixel order for SARMs associated with the signal dataset of the muon isolation study. Similarly to section \ref{lagan_order}, we conducted 10 repeated experiments for each of the orders and summarized the results in Table \ref{order_performance_muon}.

In contrast to the jet substructure study, the muon isolation data is not rotated and the pixel value distribution is quite uniform. 
Therefore we see that different generation orders have similar performance in terms of mass and $P_{\textrm{T}}$ distances. 
In addition, all the models trained using systematic orders that have some continuity in the sequence of pixels slightly outperform the models trained using random orders. 
In combination, these results confirm the validity of the heuristic strategy outlined at the end of Section IV, providing general guidelines for SARM design and pixel generation when applying these models to other datasets.

\begin{table}[H]
\caption{ Quality of images generated by SARM-1 models with various pixel-generation orderings for the muon isolation signal dataset. The quality is measured by the Wasserstein distance for the  physical observables ($P_{\textrm{T}}$ and mass) between the generated images and the original Monte Carlo images. 
} 
\label{order_performance_muon}
\vspace{0.1in}
\centering
\begin{tabular}{lll}
\hline \hline
               & $P_{\textrm{T}}$ (std)   & Mass (std) \\
               \hline
Spiral-out CCW  & 0.99 (0.37) & 0.27 (0.10) \\
Spiral-out CW   & 0.92 (0.33) & 0.26 (0.09) \\
Spiral-in CCW & 0.81 (0.23) & 0.20 (0.05) \\
Spiral-in CW  & 0.95 (0.24) & 0.24 (0.07) \\
Row-wise       & 0.99 (0.28) & 0.20 (0.05) \\
Column-wise    & 0.90 (0.26) & 0.22 (0.05) \\
Random I       & 1.17 (0.30) & 0.32 (0.08) \\
Random II       & 1.34 (0.41) & 0.37 (0.11)\\
\hline \hline
\end{tabular}
\end{table}

\subsubsection{Computational Costs}
\label{costs_muon}
Calorimeter image generation speeds in the context of the muon isolation study are shown in table \ref{speed_muon} for the SARM models, Pixel CNN++ and the Monte Carlo generator. The SARM models are one to two orders of magnitude faster than Pixel CNN++, similar to the observation of the jet substructure study.
The generation speed of each generative model is measured with the same hardware as described in Section \ref{costs_lagan}. The speed for the Monte Carlo generator is measured on an Intel(R) Xeon(R) E5-2680 at 2.70GHz CPU.

\begin{table}[H]
\centering
\caption{Comparison of image generation speed between the Monte Carlo approach and various generative models.  The SARM-2 models are 
considerably faster than Pixel CNN++ and the Monte Carlo generator.}
\vspace{0.1in}
\label{speed_muon}
\begin{tabular}{ll}
\hline \hline
Model                    & Speed (images/sec) \\
\hline
Monte Carlo    &   5         \\
Pixel CNN++              & 10               \\
SARM-2 (D+D)  &  625         \\  
SARM-2 (D+C)          & 1136           \\
\hline\hline
\end{tabular}
\end{table}

\section{Conclusion}
\label{Conclusion}
Sparse images, prevalent in particle physics datasets, present unique challenges for generative models. We have developed and applied a new class of models, deep sparse autoregressive generative models (SARMs), specifically designed to handle extreme sparseness. These compositional models are also able to take advantage of the structure present in particle physics images by using a multi-stage generation approach.
Using several different metrics, we compared SARMs to other generative models, in particular  to Pixel CNN++, a popular autoregressive model not adapted for sparsity, and to LAGAN, a state-of-the-art GAN for sparse images. The comparisons were carried using two benchmark data sets.

In the first case study on jet substructure, the adaptation to sparseness enables SARMs to produce qualitatively and quantitatively higher quality images than Pixel CNN++ and LAGAN. 
SARM are also orders of magnitude faster than traditional Monte Carlo methods and Pixel CNN++, but slower than the non-autoregressive model LAGAN, showing a trade-off between speed and quality. The second case study features extremely sparse images corresponding to calorimeter images in the vicinity of muons.   While competing models produce artifacts or suffer from mode collapse, SARMs are able to
handle and model extreme degrees of sparseness. 

In sum, given the prevalence of sparse images in particle physics and beyond, SARMs can be expected to provide an important option for rapid, high-quality, image generation from training data. Because of their quality, the generated images in turn will be able to benefit a variety of downstream data analyses.

\acknowledgments
We  wish to acknowledge a hardware grant from NVIDIA. The work of YL, JC, and PB is in part supported by grants  NSF 1839429 and NSF NRT 1633631 to PB. DW is supported by the Department of Energy Office of Science. The authors would like to thank Benjamin Nachman for helpful feedback on an early draft. 

\clearpage
\newpage
\section*{Appendix}
\subsection{2D Toy Example}
\label{toy example}

\begin{figure*}
\centering
\begin{minipage}{.4\textwidth}
  \centering
  \includegraphics[width=.8\linewidth]{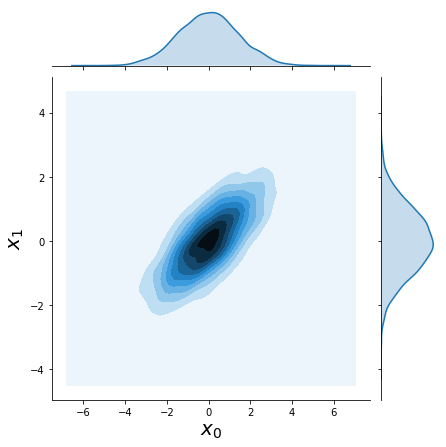}
\end{minipage}%
\begin{minipage}{.4\textwidth}
  \centering
  \includegraphics[width=.8\linewidth]{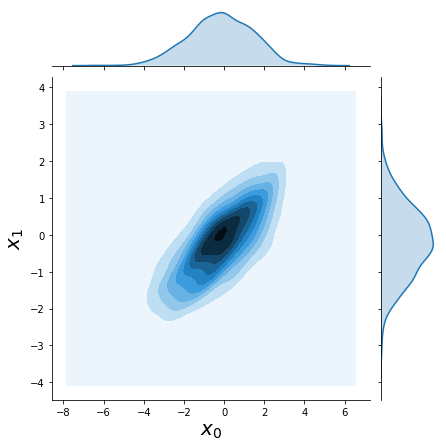}
\end{minipage}
\caption{
\textbf{Left}: 
Density plot of training data.
\textbf{Right}: 
Density plot of generated data.
The two distributions are very close, showing the ARM is able to learn the joint distribution of $x_0$ and $ x_1$ well.
}
\label{toy}
\end{figure*}

We simulate a dataset containing pairs of two variables $x_{0}$ and $x_{1}$, such that 
$x_{0} \sim p(x_{0}|x_{1})$ and $x_{1} \sim p(x_{1})$. In this toy example we show that the autoregressive model is
still able to learn to generate the joint distribution of $x_{0}$ and $x_{1}$, even though during training it is forced to learn $x_{0} \sim p(x_{0})$ first, and then to learn the dependency $p(x_{1}|x_{0})$.
The simulated training data contains 1000 pairs of $\{x_{0}, x_{1}\}$ according to $x_{1} \sim N(0, 1)$ and $x_{0}  = x_{1} + \epsilon$, where $\epsilon \sim N(0, 1)$, a standard normal distribution independent of $x_{1}$. The joint distribution of $x_{0}, x_{1}$ is shown in figure~\ref{toy}.
The toy autoregressive model learns to generate $x_{0}$ using two learnable parameters, $\mu_{0}$ and $\log(\sigma_{0})$, corresponding to the mean and log standard deviation of $x_{0}$.
It has a single linear layer for predicting $\mu_{0}$ and $\log(\sigma_{0})$, which corresponds to the mean and log standard deviation of $x_{1}$. The model is trained for 5000 iterations, by maximizing the likelihood  $p(x_{0}, x_{1})$.
During the generation stage,
the model generates $x_{0}$ without knowing $x_{1}$. Since the goal of the model is to generate the joint distribution of $(x_{0}, x_{1}) \sim P(x_{0}, x_{1})$, to do this it only needs to learn the marginal distribution, which is $x_{0} \sim N(0, 2)$ and the relationship $x_{1} = x_{0} - \epsilon$.
figure~\ref{toy} shows the result of training this model and we can see it correctly learns the means and variances of $\{x_{0}, x_{1}\}$ along with the data distribution despite the fact that it has to generate $x_0$ before generating $x_1$.
\\

\subsection{MADE Structure}
\label{mask}
The MADE structure enforces the auto-regressive property on fully connected layers by using a carefully selected binary mask on the weights of the layer.
The joint likelihood of the MADE structure can be evaluated in one forward pass of the network during training, which is not possible in other models like Pixel-RNN \cite{pmlr-v48-oord16} and Pixel CNN++ \cite{pixelcnn++}.  
This allows MADE to take advantage of the GPU acceleration.
In our SARM implementation, we consider a simple MADE structure with input $x$ and a stack of multiple hidden layers $\mathbf{h}(\mathbf{x})$, where each $\mathbf{h}(\mathbf{x})$ follows: 

\begin{equation}
\begin{aligned} \mathbf{h}(\mathbf{x}) &=\mathbf{f}\left(\mathbf{b}+\left(\mathbf{W} \odot \mathbf{M}^{\mathbf{W}}\right) \mathbf{x}\right) \\ 
\bm{\theta} &=\mathbf{f}\left(\mathbf{c}+\left(\mathbf{V} \odot \mathbf{M}^{\mathbf{V}}\right) \mathbf{h}(\mathbf{x})\right) \end{aligned}
\end{equation}

Here $\bm{\theta}$ is the output, and $\mathbf{f}$ is the activation function of the hidden layer.
In practice, we found Gaussian Error Linear Units (GeLU) \cite{GeLU}  works better in our experiments than other activations such as $Sigmoid$ and $tanh$. 
Both $\mathbf{W}$ and $\mathbf{V}$ are weight matrices, with corresponding masks: the hidden mask $\mathbf{M}^{\mathbf{W}}$, and the output mask $\mathbf{M}^{\mathbf{V}}$.
Each matrix is multiplied element-wise with each mask.

Suppose $\mathbf{x} \in R^{D}$, it can be shown that for the input mask:
\begin{equation}
\mathbf{M}_{k, d}^{\mathbf{W}}=1_{k\bmod D \leq d}=\left\{\begin{array}{ll}{1} & {\text { if } k\bmod D \leq d} \\ {0} & {\text { otherwise }}\end{array}\right.
\end{equation}
Likewise, suppose $\mathbf{h}(\mathbf{x}) \in R^{H}$, then for the output mask:
\begin{equation}
\mathbf{M}_{k, d}^{\mathbf{V}}=1_{k\bmod D < d}=\left\{\begin{array}{ll}{1} & {\text { if } k\bmod D < d} \\ {0} & {\text { otherwise }}\end{array}\right.
\end{equation}

Then the output $\bm{\theta}$ satisfies autoregressive structure: for any $i$, $\theta_{i}$ only depends on $x_{j<i}$. 
As shown in figure~\ref{arm}, the parameter $\theta_{i}$ is used to generate the $i$th pixel during generation. 
For example, if the likelihood is a logistic distribution, then $\theta_{i} = [\mu_{i}, s_{i}]$, where $\mu_{i}, s_{i}$ corresponds to the mean and scale of a logistic distribution.  

During generation, at step $i$ we take the previously generated $x_{0}, x_{1}, \ldots , x_{i-1}$ and pad the remaining $x_{i}, \ldots ,x_{D-1}$ with zeros.
Then we input this vector in the MADE structure
so that the output $\theta_{i}$
depends only on $x_{0}, \ldots ,x_{i-1}$. 
Finally,
 we sample the pixel $x_{i}$ conditioned on $\theta_{i}$ and repeat this process until every pixel is generated.

\subsection{Further Analysis of the Jet Structure Study}
\label{further_analysis_jet}

\begin{figure*}
    \centering
    \includegraphics[width=0.8\textwidth]{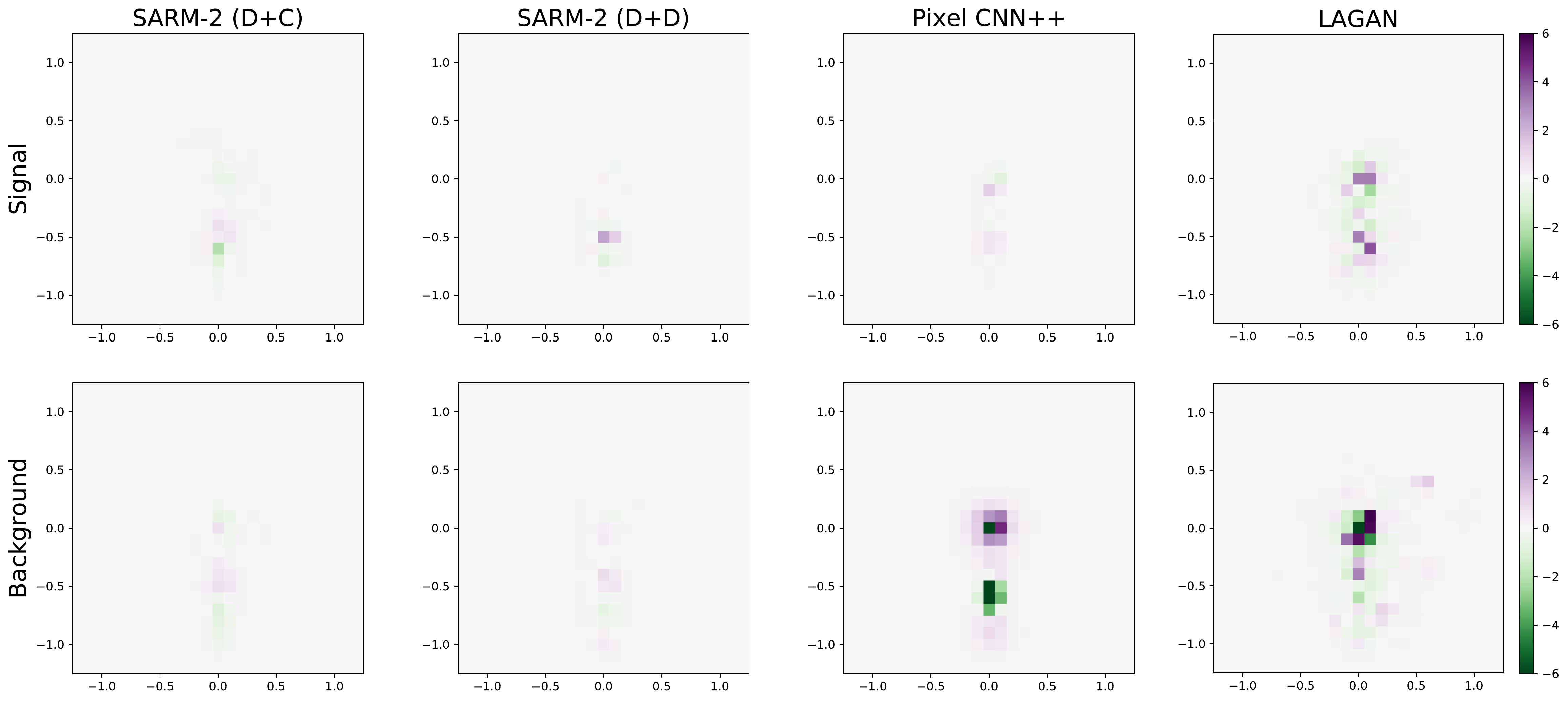}
    \caption{ 
    Error measured by subtracting of the pixel-wise average of the images created by each generative model and the pixel-wise average of the images generated with Pythia.
    The SARM models have lower error than both Pixel CNN++ and LAGAN with most of the errors are concentrated in the center of the image.
    }
    \label{diff}
\end{figure*}

Figure~\ref{diff} shows the subtraction between the pixel-wise average of the images from each generative model and the pixel-wise average from Pythia. 
Notice the differences are concentrated in the middle of the images where there are higher value pixels.
The images generated by both SARM models have small differences compared to the ones generated by LAGAN for both signal and background and by Pixel CNN++ for background. 
Also, Pixel CNN++ has higher errors in background images compared to signal images.

\begin{figure*}
    \centering
    \includegraphics[width=0.8\textwidth]{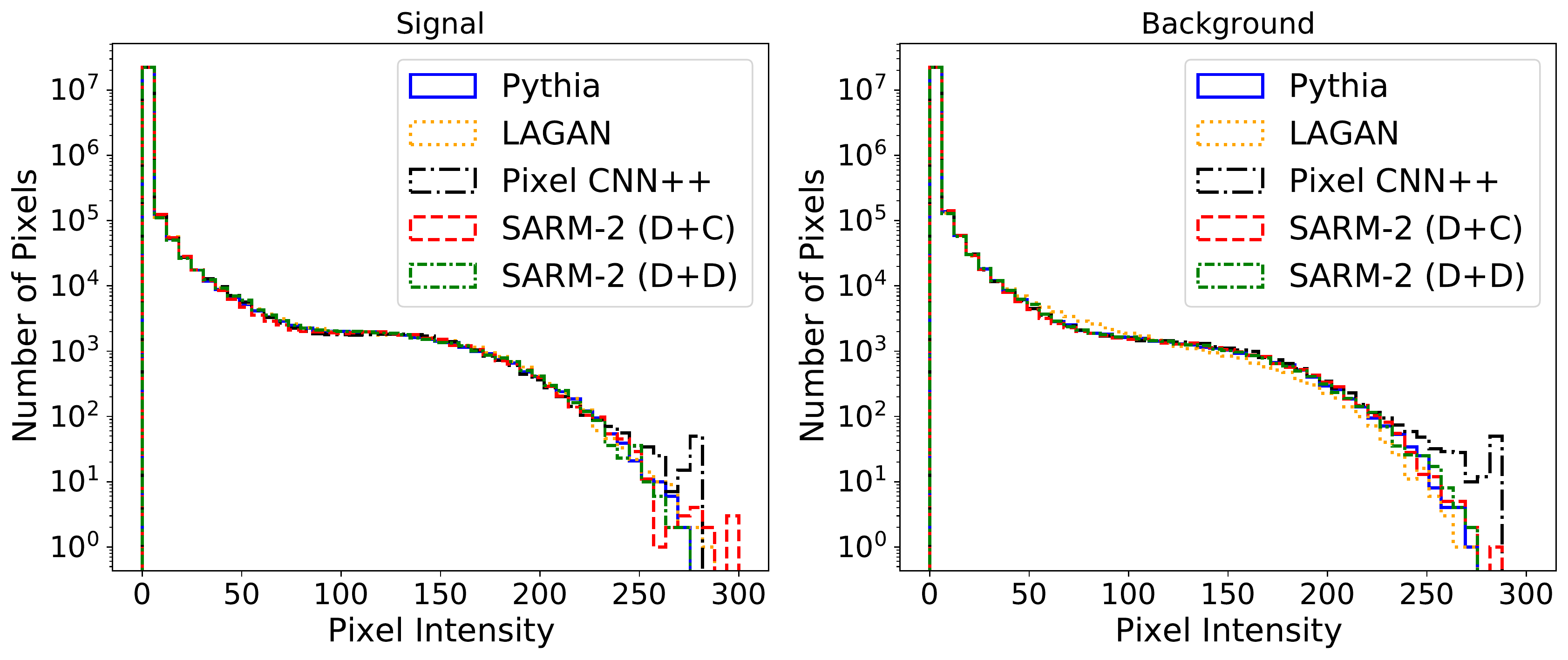}
    \caption{Distribution of aggregated pixel intensity in the generated images for jet substructure study. Notice most of the differences happen at high pixel values where there are fewer events. LAGAN also has a harder time replicating the distribution of background images across all pixel values compared to the other models.
     }
    \label{pixel_intensity}
\end{figure*}

Figure~\ref{pixel_intensity} shows the distribution of pixel values across all the generated images. 
For the signal images, all the models match the Pythia distribution for pixel values below 200 but the models have difficulties at higher values. SARM-2 (D+D) and LAGAN have the closest match at high pixel values while SARM-2 (D+C) and Pixel CNN++ overestimate them.
For the background images, most of the models accurately predict low value pixels, but LAGAN slightly overestimates pixels in the range 50 to 100 and underestimates them afterwards. 
For high pixel values, Pixel CNN++ strongly over-estimates pixels in the range 250-300 while the other models remain reasonably close to Pythia.
In both cases the models have difficulties learning the high value pixels, which is expected since there are very few pixels in this range in the Pythia distribution.

\subsection{Further Analysis of the Muon Isolation Study}
\label{further_analysis_muon}
\subsubsection{LAGAN}

\begin{figure*}
    \centering
    \includegraphics[width=0.9\textwidth]{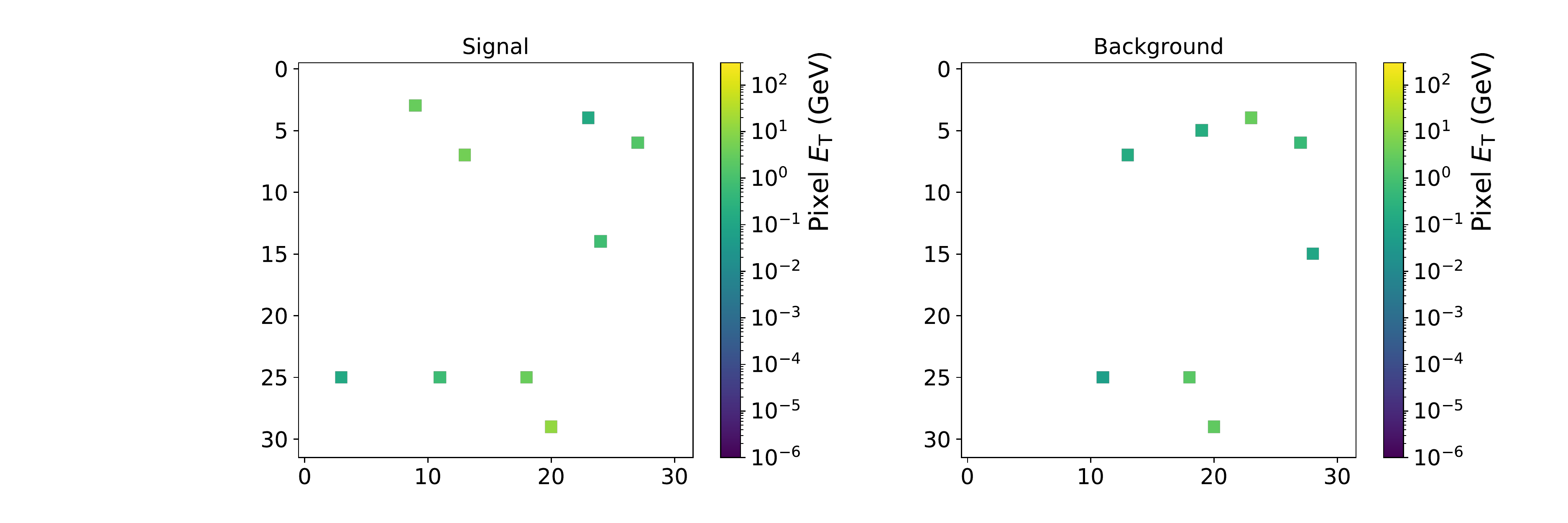}
    \caption{
    Typical muon images generated using LAGAN. The figures are plotted in log scale, where the white space represents pixels with value zero. 
    }
    \label{muon_typical_image_with_lagan}
\end{figure*}

\begin{figure*}
    \centering
    \includegraphics[width=0.9\textwidth]{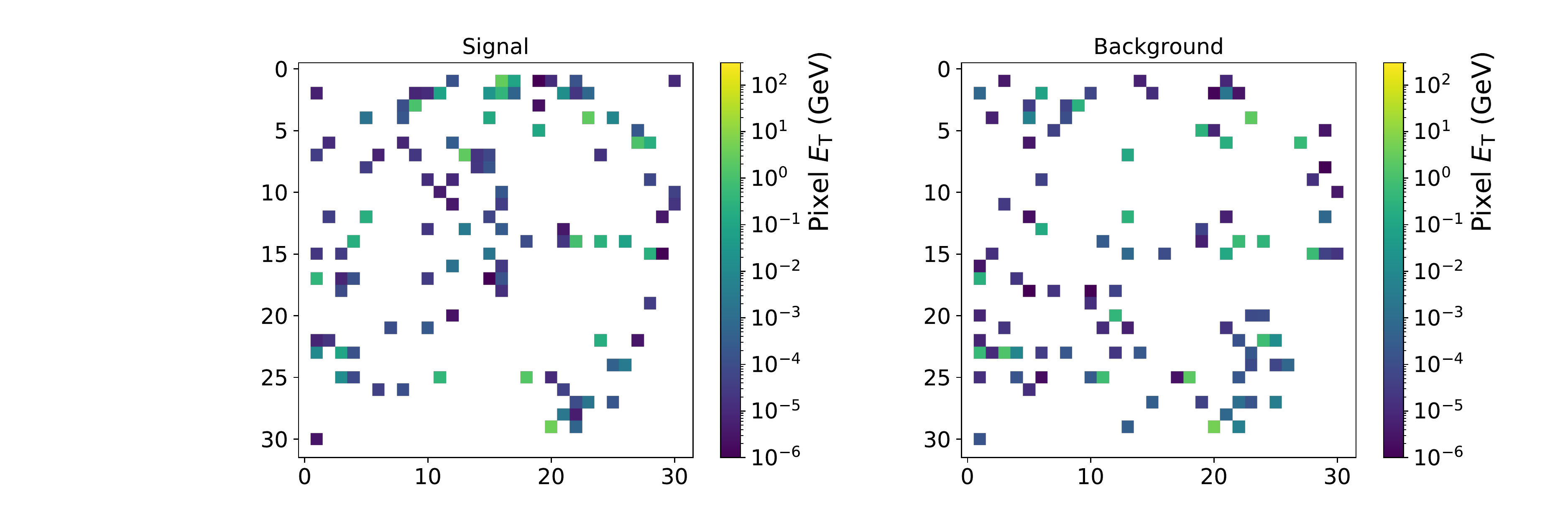}
    \caption{
       Pixel-wise average of muon images from LAGAN for signal and background. The average images generated by LAGAN fail to reproduce the radial structure present in the average Monte Carlo images (figure~\ref{muon_mean}).
    }
    \label{muon_mean_with_lagan}
\end{figure*}

\begin{figure*}
    \centering
    \includegraphics[width=0.8\textwidth]{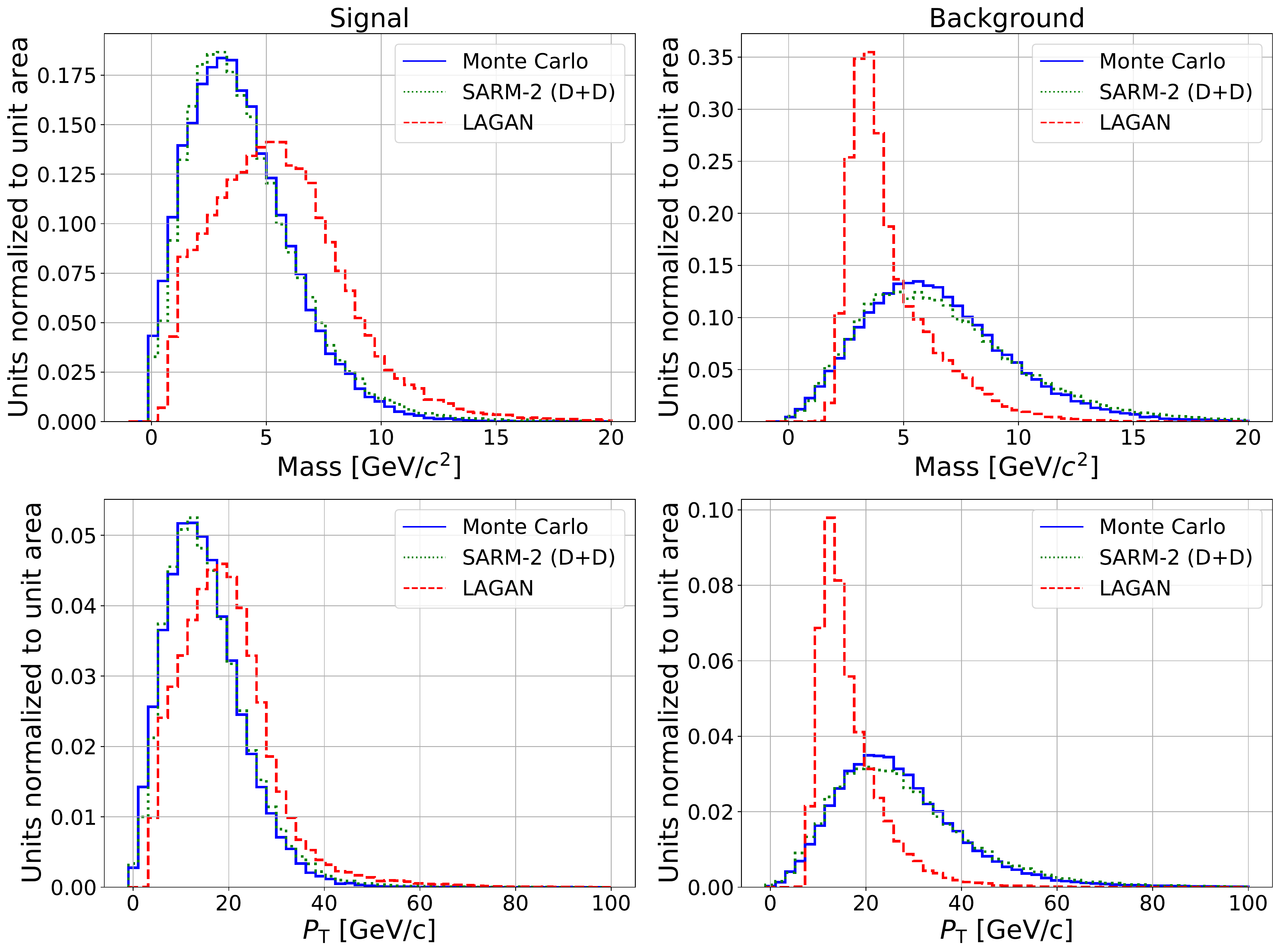}
    \caption{Comparison of the mass and $P_{\textrm{T}}$  distributions of the images generated by LAGAN, SARM-2 (D+D), and the Monte Carlo simulations for both signal and background muons.
    }
    
    \label{fig:muon_mass_pt_with_lagan}
\end{figure*}

Despite our best efforts, the LAGAN model performed poorly every time it was trained on the muon isolation dataset. As seen in Figures \ref{muon_typical_image_with_lagan} and \ref{muon_mean_with_lagan} the pixel-wise average image doesn't capture the radial structure present in the dataset and some of the pixels with high values seem to be present in many of the images. This seems to be due to a low amount of variability in the generated images, typical of mode collapse in GANs. This performance is also reflected in the distributions of $P_{\textrm{T}}$ and mass (figure~\ref{fig:muon_mass_pt_with_lagan}) and the respective Wasserstein distances which are one order of magnitude worse than the values for the other models (table \ref{muon_metrics_with_lagan}). 

\begin{table}[h]
\caption{
Wasserstein distance of the physical constituents jet $P_{\textrm{T}}$ and mass distributions between the original muon images from the Monte Carlo generator and the images created by the generative models. 
A small distance signifies a good agreement.
SARM-2 (D+D) is the two-stage SARM model with a discrete mixture.
}
\label{muon_metrics_with_lagan}
\centering
\vspace{0.2in}
\begin{tabular}{lllll}
\hline \hline
                              & \multicolumn{2}{c}{$P_{\textrm{T}}$} & \multicolumn{2}{c}{Mass} \\
\cline{2-5}                              
                              & Signal   & {Background} & Signal    & {Background}   \\
\hline           
LAGAN                  & 4.81      &   10.88 &  1.81    & 2.17 \\
SARM-2 (D+D)             & \textbf{0.56} &  \textbf{0.93}  &  \textbf{0.17}   & \textbf{0.31}  \\
\hline \hline
\end{tabular}
\end{table}

\subsubsection{SARM vs Pixel CNN++}
\begin{figure*}
    \centering
    \includegraphics[width=0.8\textwidth]{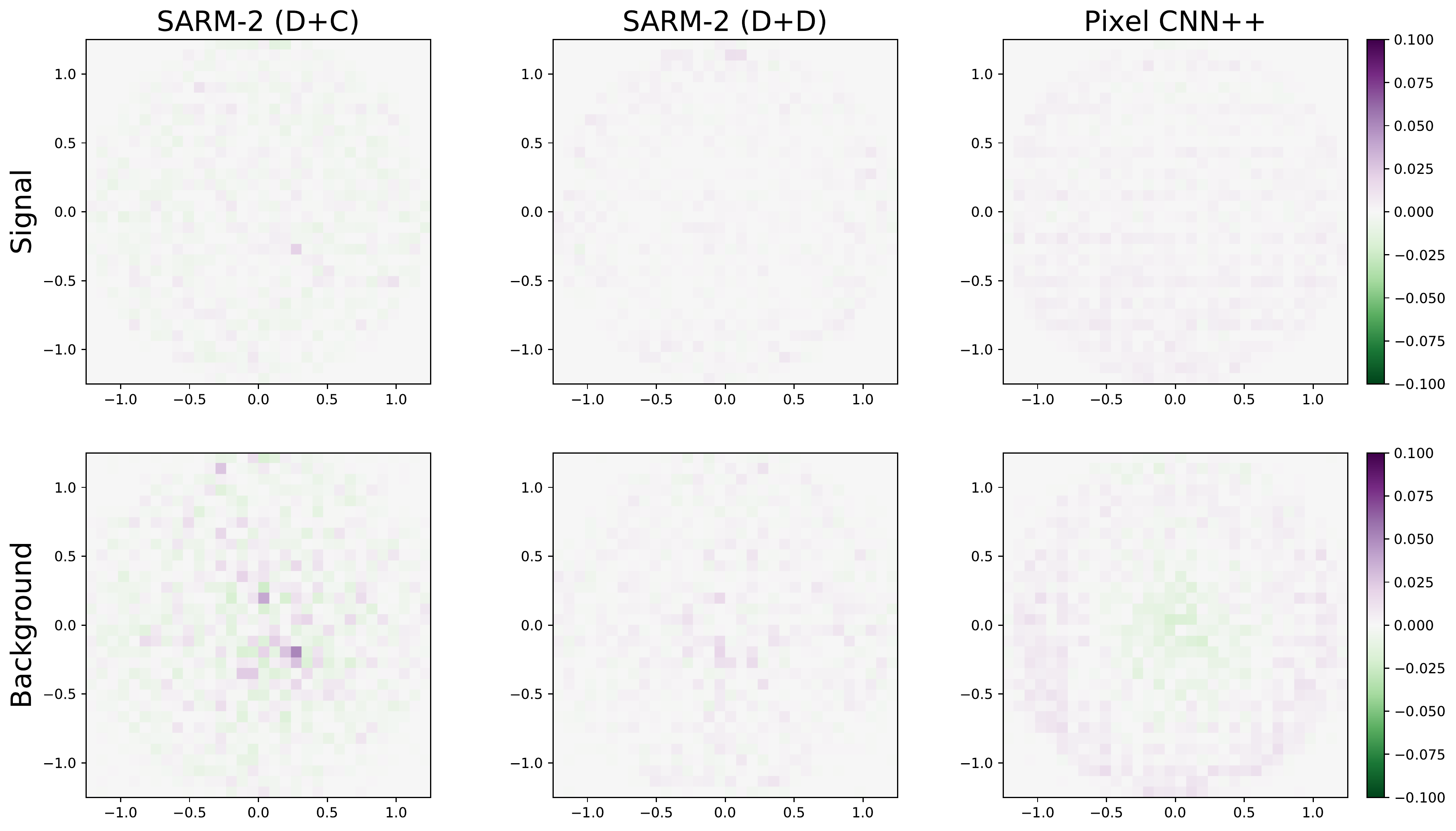}
    \caption{Subtraction between the pixel-wise average of generated images vs {Monte Carlo images}. 
    The errors are evenly distributed in the signal images, while they are concentrated in the center for the background images.
     In the center there is larger number of high intensity pixels. 
    }
    \label{muon_diff_comparison}
\end{figure*}

Figure~\ref{muon_diff_comparison} shows the subtraction between the pixel-wise average of the images from each generative model and the pixel-wise average from Pythia in the muon isolation dataset. 
For the signal data, all models show very small differences, evenly distributed across the radial structure of the images. 
In particular, Pixel CNN++ is over-representing most of the pixels in the artificial checkerboard pattern noted before. 
For the background data the errors are slightly higher for all models. The SARM models have more difficulties with the pixels in the center and tend to over-represent them while Pixel CNN++ under-represents the center and over-represents the periphery.

\begin{figure*}
    \centering
    \includegraphics[width=0.8\textwidth]{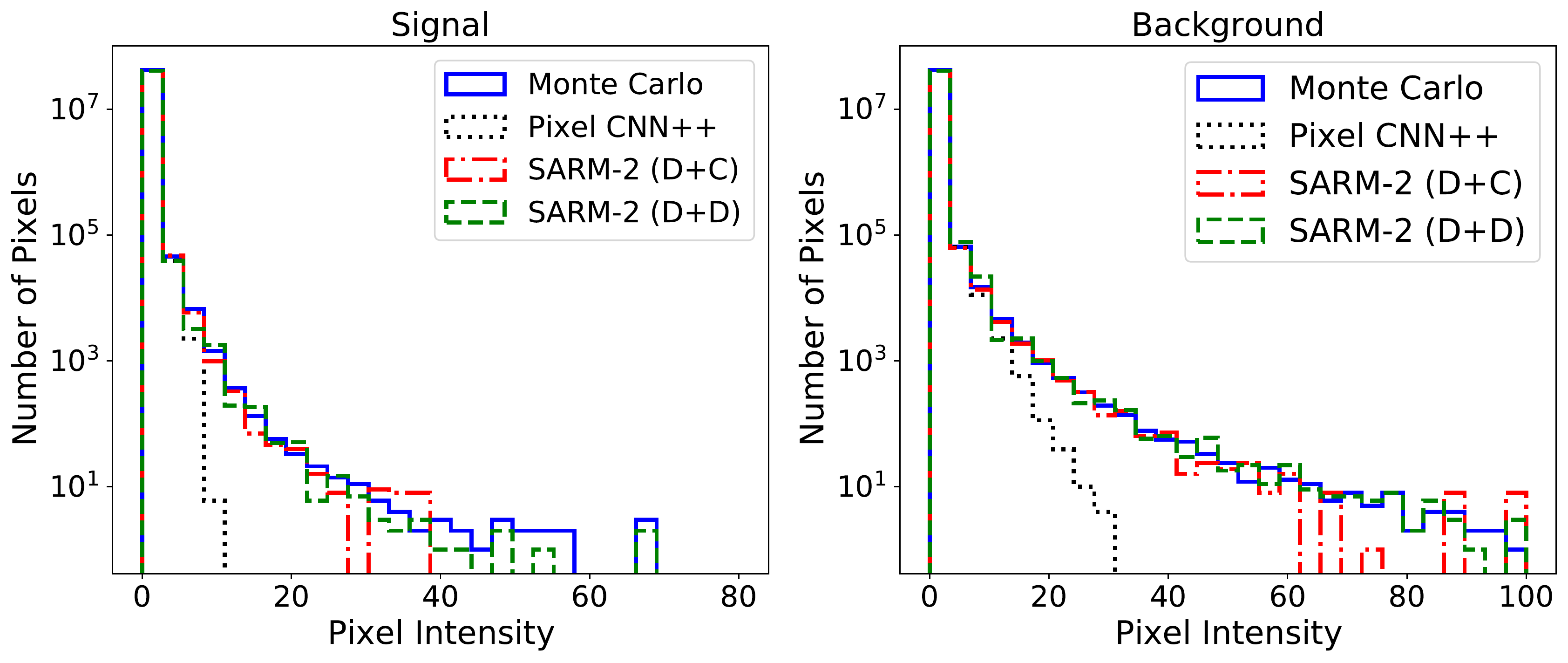}
    \caption{Distribution of pixel intensity for muon isolation study. Pixel CNN++ under-represents the distribution while the SARM models miss the high pixel values where there are fewer events.}
    \label{muon_pixel_intensity}
\end{figure*}

Figure~\ref{muon_pixel_intensity} shows the distribution of pixel values across all the generated images. 
For both signal and background the Pixel CNN++ model is under-representing pixels with high intensity, while the SARM models match the distribution quite well.
Like in the jet substructure study, most of the errors correspond to pixels with high intensity values, which is expected since these values are rare in the training data, making it difficult to correctly learn their distribution.

\subsection{Software Modifications}
\label{appendix_lagan_muon}

\subsubsection{LAGAN}
The code and weights of the original LAGAN model for the jet substructure study dataset are publicly available. 
This makes it possible to generate new images using the original model's weights for this dataset, but the model needs to be retrained to generate images of a different dataset. 
The model was retrained for the muon isolation study and it also had to be modified to adapt it to the larger images of $32\times32$ pixels since it has upsampling layers in the generator part of the GAN.

\subsubsection{PixelCNN++}
\label{Baseline_pcnn}
As a baseline for autoregressive models we used the Pixel CNN ++ \cite{pixelcnn++}.
Due to speed and memory restrictions, we had to modify the original model by reducing the number of filters in the masked convolutional layers and the number of residual blocks compared to the original model.
Both the number of filters and the number of residual blocks are optimized as hyperparameters using grid search with 5, 10 or 20 filters and 2 or 3 residual blocks. However, we found most hyperparameter combinations to have similar performance. The model with 20 filters and 3 blocks performs slightly better in the jet substructure study, and the model with 10 filters and 5 blocks performs slightly better in the muon isolation study.
Even though the models we used are smaller than the original model in \cite{pixelcnn++}, they are almost as slow as the traditional Monte Carlo methods (table \ref{speed} and \ref{speed_muon}).

\subsection{Architecture and Hyperparameter Optimization}
\label{classifier}

We performed a search over the architectures of the SARMs including the number of hidden layers structure, the size of the central area for the two-stage 
approach and the size of the intermediate upsampling layer using SHERPA \cite{hertel2020sherpa}. 
We also conducted search of the transformation parameter $p$ with values $[1, 1.1, 1.2, 1.3, 1.5, 2]$ for the D+D models. 
All models were implemented in Pytorch \cite{pytorch}, and were trained for 300 epochs with outward spiral (CCW) order using the Adam optimizer \cite{adam} with learning rate 3e-4, decreased by half every 100 epochs and mini-batch size 128. 

For the jet substructure study, the best SARM-2 configuration had a center area of side length 3.  For the D+D models, we used 5 hidden layers with an upsampling layer of size 10 and found that a power transformation with $p=1.0$ yields slightly better results. For the D+C models, we found that the model with 3 hidden layers and a mixture of 5 truncated logistic for the C component works well for both signal and background images. In the generation order experiments, similarly we used SARM-1 (D+D) models with 5 hidden layers, an upsampling layer of size 10 and a power transformation with $p=1.0$, effectively no transformation. And all models are trained with identical settings: learning rate of 3e-4, decreased by half every 100 epochs and mini-batch size 128. 
For the LAGAN model we used the publicly available version of LAGAN optimized by the original authors.

For the muon isolation study, the best model we found had 5 hidden layers, and a center area of side length 7 for both D+D and D+C  models.  For the SARM-2 (D+D), we used an upsampling layer of size 10 and found that a power transformation with $p=1.2$ for signal and $p=1.3$ for background provided the best results. 
And for the D+C models, we found again that a mixture of 5 truncated logistic for the C component works well for both signal and background images.

For the classification tasks, we trained five convolutional neural networks with the same structure on each of the datasets. We randomly split the data into a 90\% subset for training and a 10\% subset for validation. The validation set is used for early-stopping during training to avoid over-fitting.
The convolutional neural network model has 2 convolutional blocks, 2 fully connected layers with 100 rectified linear units, and a sigmoid unit at the end to predict the probability of the image being signal. 
Each convolutional block contains two convolutional layers with 3x3 kernels and 30 filters with rectified linear units followed by a maxpooling layer with 2x2 kernel.
All models were trained in PyTorch using the Adam optimizer, with a learning rate of $0.001$ and a batch size of 128.

\subsection{Complexity Analysis}

Next we compare the number of parameters for the different models in table \ref{para count}. Note that the original Pixel CNN++ model \cite{pixelcnn++} uses 160 convolutional filters. With all these filters, each forward pass takes more than 1 second on 4 NVIDIA TITANX GPU cards, resulting in a generation speed that is one order of magnitude slower than the traditional Monte Carlo methods, thus defeating the original purpose. Therefore, in our implementation of the
Pixel CNN++ model, we limit the number of its filters to 20 to speed up the generation process and reduce the memory requirements.
\begin{table}[H]
\centering
\caption{Model complexity comparison in terms of the number of parameters. }
\vspace{0.15in}
\label{para count}
\begin{tabular}{ll}
\hline \hline
Model                    & Num. of Parameters \\
\hline
Pythia \cite{LAGAN}         &     -  \\
Pixel CNN++                  &     0.7M     \\
SARM-2 (D+D)              &  21M  \\ 
SARM-2 (D+C)                 &     7M     \\
LAGAN                     &      5M           \\
\hline
\end{tabular}
\end{table}

\subsection{Sample Images}
\label{sample_images_appendix}
\begin{figure*}
    \centering
    \includegraphics[width=0.9\textwidth]{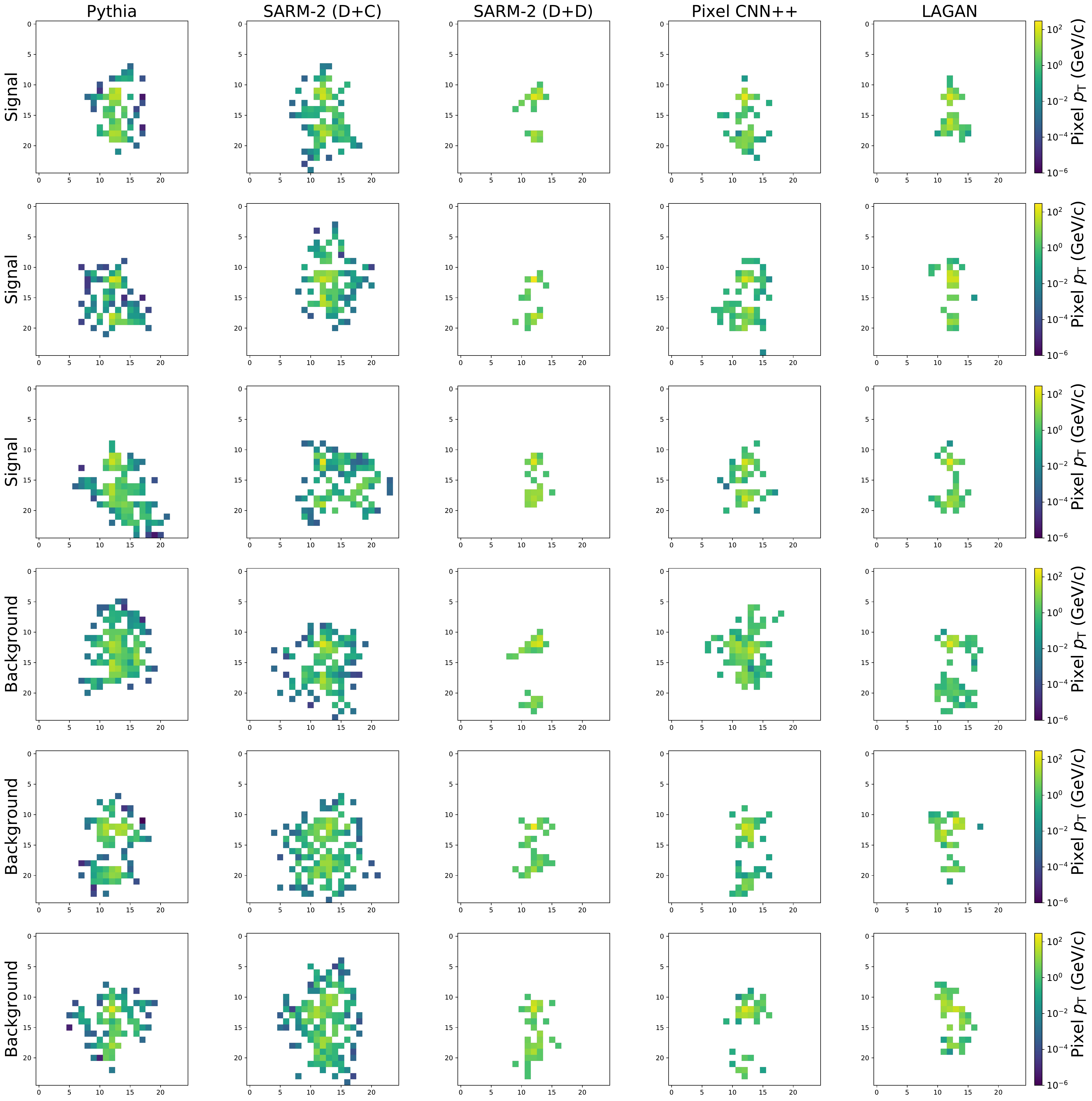}
    \caption{Additional typical images from the jet substructure study}
    \label{more_gen_jet}
\end{figure*}

\begin{figure*}
    \centering
    \includegraphics[width=0.9\textwidth]{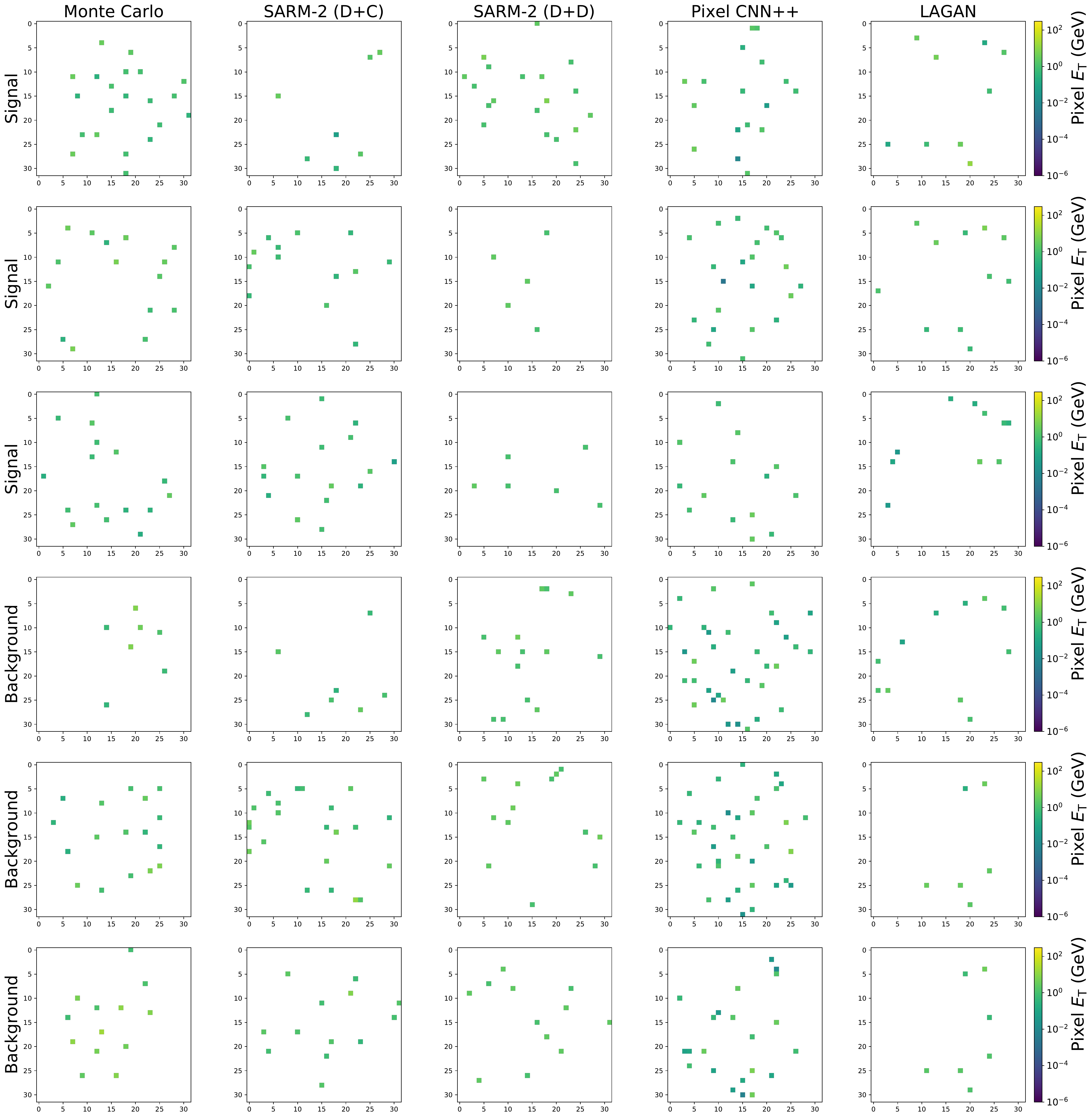}
    \caption{Additional typical images from the muon isolation study}
    \label{more_gen_muon}
\end{figure*}

In this section, we show more generated images from both the jet substructure study and the muon isolation study in figure~\ref{more_gen_jet} and figure~\ref{more_gen_muon}.  

\clearpage

\bibliography{apssamp}

\begin{thebibliography}{61}%
\makeatletter
\providecommand \@ifxundefined [1]{%
 \@ifx{#1\undefined}
}%
\providecommand \@ifnum [1]{%
 \ifnum #1\expandafter \@firstoftwo
 \else \expandafter \@secondoftwo
 \fi
}%
\providecommand \@ifx [1]{%
 \ifx #1\expandafter \@firstoftwo
 \else \expandafter \@secondoftwo
 \fi
}%
\providecommand \natexlab [1]{#1}%
\providecommand \enquote  [1]{``#1''}%
\providecommand \bibnamefont  [1]{#1}%
\providecommand \bibfnamefont [1]{#1}%
\providecommand \citenamefont [1]{#1}%
\providecommand \href@noop [0]{\@secondoftwo}%
\providecommand \href [0]{\begingroup \@sanitize@url \@href}%
\providecommand \@href[1]{\@@startlink{#1}\@@href}%
\providecommand \@@href[1]{\endgroup#1\@@endlink}%
\providecommand \@sanitize@url [0]{\catcode `\\12\catcode `\$12\catcode
  `\&12\catcode `\#12\catcode `\^12\catcode `\_12\catcode `\%12\relax}%
\providecommand \@@startlink[1]{}%
\providecommand \@@endlink[0]{}%
\providecommand \url  [0]{\begingroup\@sanitize@url \@url }%
\providecommand \@url [1]{\endgroup\@href {#1}{\urlprefix }}%
\providecommand \urlprefix  [0]{URL }%
\providecommand \Eprint [0]{\href }%
\providecommand \doibase [0]{https://doi.org/}%
\providecommand \selectlanguage [0]{\@gobble}%
\providecommand \bibinfo  [0]{\@secondoftwo}%
\providecommand \bibfield  [0]{\@secondoftwo}%
\providecommand \translation [1]{[#1]}%
\providecommand \BibitemOpen [0]{}%
\providecommand \bibitemStop [0]{}%
\providecommand \bibitemNoStop [0]{.\EOS\space}%
\providecommand \EOS [0]{\spacefactor3000\relax}%
\providecommand \BibitemShut  [1]{\csname bibitem#1\endcsname}%
\let\auto@bib@innerbib\@empty
\bibitem [{\citenamefont {Agostinelli}\ \emph {et~al.}(2003)\citenamefont
  {Agostinelli} \emph {et~al.}}]{GEANT}%
  \BibitemOpen
  \bibfield  {author} {\bibinfo {author} {\bibfnamefont {S.}~\bibnamefont
  {Agostinelli}} \emph {et~al.} (\bibinfo {collaboration} {GEANT4}),\
  }\bibfield  {title} {\bibinfo {title} {{GEANT4: A Simulation toolkit}},\
  }\bibfield  {journal} {\bibinfo  {journal} {Nucl. Instrum. Meth. A}\ }\textbf
  {\bibinfo {volume} {506}},\ \href
  {https://doi.org/10.1016/S0168-9002(03)01368-8}
  {10.1016/S0168-9002(03)01368-8} (\bibinfo {year} {2003})\BibitemShut
  {NoStop}%
\bibitem [{\citenamefont {Aad}\ \emph {et~al.}(2010)\citenamefont {Aad} \emph
  {et~al.}}]{Aad:2010ah}%
  \BibitemOpen
  \bibfield  {author} {\bibinfo {author} {\bibfnamefont {G.}~\bibnamefont
  {Aad}} \emph {et~al.} (\bibinfo {collaboration} {ATLAS}),\ }\bibfield
  {title} {\bibinfo {title} {{The ATLAS Simulation Infrastructure}},\
  }\bibfield  {journal} {\bibinfo  {journal} {Eur. Phys. J. C}\ }\textbf
  {\bibinfo {volume} {70}},\ \href
  {https://doi.org/10.1140/epjc/s10052-010-1429-9}
  {10.1140/epjc/s10052-010-1429-9} (\bibinfo {year} {2010})\BibitemShut
  {NoStop}%
\bibitem [{\citenamefont {Rahmat}\ \emph {et~al.}(2012)\citenamefont {Rahmat},
  \citenamefont {Kroeger},\ and\ \citenamefont {Giammanco}}]{Rahmat_2012}%
  \BibitemOpen
  \bibfield  {author} {\bibinfo {author} {\bibfnamefont {R.}~\bibnamefont
  {Rahmat}}, \bibinfo {author} {\bibfnamefont {R.}~\bibnamefont {Kroeger}},\
  and\ \bibinfo {author} {\bibfnamefont {A.}~\bibnamefont {Giammanco}},\
  }\bibfield  {title} {\bibinfo {title} {{The Fast Simulation of The {CMS}
  Experiment}},\ }\bibfield  {journal} {\bibinfo  {journal} {Journal of
  Physics: Conference Series}\ }\textbf {\bibinfo {volume} {396}},\ \href
  {https://doi.org/10.1088/1742-6596/396/6/062016}
  {10.1088/1742-6596/396/6/062016} (\bibinfo {year} {2012})\BibitemShut
  {NoStop}%
\bibitem [{\citenamefont {Nikiforou}(2013)}]{Nikiforou:2013nba}%
  \BibitemOpen
  \bibfield  {author} {\bibinfo {author} {\bibfnamefont {N.}~\bibnamefont
  {Nikiforou}} (\bibinfo {collaboration} {ATLAS}),\ }\bibfield  {title}
  {\bibinfo {title} {{Performance of the ATLAS Liquid Argon Calorimeter After
  Three Years of LHC Operation and Plans For a Future Upgrade}},\ }in\
  \href@noop {} {\emph {\bibinfo {booktitle} {3rd International Conference on
  Advancements in Nuclear Instrumentation Measurement Methods and their
  Applications}}}\ (\bibinfo {year} {(2013)})\ \Eprint
  {https://arxiv.org/abs/1306.6756} {arXiv:1306.6756} \BibitemShut {NoStop}%
\bibitem [{LHC(2000)}]{LHCB:2000ab}%
  \BibitemOpen
  \href@noop {} {\emph {\bibinfo {title} {{LHCb calorimeters: Technical Design
  Report}}}},\ Technical Design Report LHCb\ (\bibinfo  {publisher} {CERN},\
  \bibinfo {address} {Geneva},\ \bibinfo {year} {(2000)})\ \Eprint
  {https://arxiv.org/abs/https://cds.cern.ch/record/494264}
  {arXiv:https://cds.cern.ch/record/494264} \BibitemShut {NoStop}%
\bibitem [{\citenamefont {de~Oliveira}\ \emph {et~al.}(2017)\citenamefont
  {de~Oliveira}, \citenamefont {Paganini},\ and\ \citenamefont
  {Nachman}}]{LAGAN}%
  \BibitemOpen
  \bibfield  {author} {\bibinfo {author} {\bibfnamefont {L.}~\bibnamefont
  {de~Oliveira}}, \bibinfo {author} {\bibfnamefont {M.}~\bibnamefont
  {Paganini}},\ and\ \bibinfo {author} {\bibfnamefont {B.}~\bibnamefont
  {Nachman}},\ }\bibfield  {title} {\bibinfo {title} {{Learning Particle
  Physics by Example: Location-Aware Generative Adversarial Networks for
  Physics Synthesis}},\ }\href@noop {} {\bibfield  {journal} {\bibinfo
  {journal} {Comput Softw Big Sci}\ } (\bibinfo {year} {2017})},\ \Eprint
  {https://arxiv.org/abs/1701.05927} {arXiv:1701.05927} \BibitemShut {NoStop}%
\bibitem [{\citenamefont {Lu}\ \emph {et~al.}(2019)\citenamefont {Lu},
  \citenamefont {Collado}, \citenamefont {Bauer}, \citenamefont {Whiteson},\
  and\ \citenamefont {Baldi}}]{muons}%
  \BibitemOpen
  \bibfield  {author} {\bibinfo {author} {\bibfnamefont {Y.}~\bibnamefont
  {Lu}}, \bibinfo {author} {\bibfnamefont {J.}~\bibnamefont {Collado}},
  \bibinfo {author} {\bibfnamefont {K.}~\bibnamefont {Bauer}}, \bibinfo
  {author} {\bibfnamefont {D.}~\bibnamefont {Whiteson}},\ and\ \bibinfo
  {author} {\bibfnamefont {P.}~\bibnamefont {Baldi}},\ }\bibfield  {title}
  {\bibinfo {title} {{Sparse Image Generation with Decoupled Generative
  Models}},\ }in\ \href@noop {} {\emph {\bibinfo {booktitle} {Neural
  Information Processing Systems, Machine Learning and the Physical Sciences
  Workshop}}}\ (\bibinfo {year} {(2019)})\ \bibinfo {note}
  {\url{https://ml4physicalsciences.github.io/2019/files/NeurIPS_ML4PS_2019_161.pdf}}\BibitemShut
  {NoStop}%
\bibitem [{\citenamefont {Goodfellow}\ \emph {et~al.}(2014)\citenamefont
  {Goodfellow}, \citenamefont {Pouget-Abadie}, \citenamefont {Mirza},
  \citenamefont {Xu}, \citenamefont {Warde-Farley}, \citenamefont {Ozair},
  \citenamefont {Courville},\ and\ \citenamefont {Bengio}}]{GAN}%
  \BibitemOpen
  \bibfield  {author} {\bibinfo {author} {\bibfnamefont {I.}~\bibnamefont
  {Goodfellow}}, \bibinfo {author} {\bibfnamefont {J.}~\bibnamefont
  {Pouget-Abadie}}, \bibinfo {author} {\bibfnamefont {M.}~\bibnamefont
  {Mirza}}, \bibinfo {author} {\bibfnamefont {B.}~\bibnamefont {Xu}}, \bibinfo
  {author} {\bibfnamefont {D.}~\bibnamefont {Warde-Farley}}, \bibinfo {author}
  {\bibfnamefont {S.}~\bibnamefont {Ozair}}, \bibinfo {author} {\bibfnamefont
  {A.}~\bibnamefont {Courville}},\ and\ \bibinfo {author} {\bibfnamefont
  {Y.}~\bibnamefont {Bengio}},\ }\bibfield  {title} {\bibinfo {title}
  {{Generative Adversarial Networks}},\ }in\ \href
  {http://papers.nips.cc/paper/5423-generative-adversarial-nets.pdf} {\emph
  {\bibinfo {booktitle} {Advances in Neural Information Processing Systems
  27}}}\ (\bibinfo  {publisher} {Curran Associates, Inc.},\ \bibinfo {year}
  {(2014)})\ \Eprint {https://arxiv.org/abs/1406.2661} {arXiv:1406.2661}
  \BibitemShut {NoStop}%
\bibitem [{\citenamefont {Kingma}\ and\ \citenamefont {Welling}(2014)}]{VAE}%
  \BibitemOpen
  \bibfield  {author} {\bibinfo {author} {\bibfnamefont {D.~P.}\ \bibnamefont
  {Kingma}}\ and\ \bibinfo {author} {\bibfnamefont {M.}~\bibnamefont
  {Welling}},\ }\bibfield  {title} {\bibinfo {title} {{Auto-Encoding
  Variational Bayes}},\ }in\ \href@noop {} {\emph {\bibinfo {booktitle} {{2nd
  International Conference on Learning Representations, {ICLR}}}}}\ (\bibinfo
  {year} {(2014)})\ \Eprint {https://arxiv.org/abs/1312.6114} {arXiv:1312.6114}
  \BibitemShut {NoStop}%
\bibitem [{\citenamefont {Oord}\ \emph {et~al.}(2016)\citenamefont {Oord},
  \citenamefont {Kalchbrenner},\ and\ \citenamefont
  {Kavukcuoglu}}]{pmlr-v48-oord16}%
  \BibitemOpen
  \bibfield  {author} {\bibinfo {author} {\bibfnamefont {A.~V.}\ \bibnamefont
  {Oord}}, \bibinfo {author} {\bibfnamefont {N.}~\bibnamefont {Kalchbrenner}},\
  and\ \bibinfo {author} {\bibfnamefont {K.}~\bibnamefont {Kavukcuoglu}},\
  }\bibfield  {title} {\bibinfo {title} {{Pixel Recurrent Neural Networks}},\
  }in\ \href@noop {} {\emph {\bibinfo {booktitle} {Proceedings of The 33rd
  International Conference on Machine Learning {ICML}}}},\ \bibinfo {series}
  {Proceedings of Machine Learning Research}, Vol.~\bibinfo {volume} {48}\
  (\bibinfo  {publisher} {PMLR},\ \bibinfo {address} {New York, New York,
  USA},\ \bibinfo {year} {(2016)})\ pp.\ \bibinfo {pages} {1747--1756},\
  \Eprint {https://arxiv.org/abs/1601.06759} {arXiv:1601.06759} \BibitemShut
  {NoStop}%
\bibitem [{\citenamefont {Zhu}\ \emph {et~al.}(2017)\citenamefont {Zhu},
  \citenamefont {Park}, \citenamefont {Isola},\ and\ \citenamefont
  {Efros}}]{cyclegan}%
  \BibitemOpen
  \bibfield  {author} {\bibinfo {author} {\bibfnamefont {J.}~\bibnamefont
  {Zhu}}, \bibinfo {author} {\bibfnamefont {T.}~\bibnamefont {Park}}, \bibinfo
  {author} {\bibfnamefont {P.}~\bibnamefont {Isola}},\ and\ \bibinfo {author}
  {\bibfnamefont {A.~A.}\ \bibnamefont {Efros}},\ }\bibfield  {title} {\bibinfo
  {title} {{Unpaired Image-to-Image Translation using Cycle-Consistent
  Adversarial Networks}},\ }\href@noop {} {\bibfield  {journal} {\bibinfo
  {journal} {CoRR}\ } (\bibinfo {year} {2017})},\ \Eprint
  {https://arxiv.org/abs/1703.10593} {arXiv:1703.10593} \BibitemShut {NoStop}%
\bibitem [{\citenamefont {Brock}\ \emph {et~al.}(2019)\citenamefont {Brock},
  \citenamefont {Donahue},\ and\ \citenamefont {Simonyan}}]{biggan}%
  \BibitemOpen
  \bibfield  {author} {\bibinfo {author} {\bibfnamefont {A.}~\bibnamefont
  {Brock}}, \bibinfo {author} {\bibfnamefont {J.}~\bibnamefont {Donahue}},\
  and\ \bibinfo {author} {\bibfnamefont {K.}~\bibnamefont {Simonyan}},\
  }\bibfield  {title} {\bibinfo {title} {{Large Scale {GAN} Training for High
  Fidelity Natural Image Synthesis}},\ }in\ \href
  {https://openreview.net/forum?id=B1xsqj09Fm} {\emph {\bibinfo {booktitle}
  {International Conference on Learning Representations {ICLR}}}}\ (\bibinfo
  {year} {(2019)})\ \Eprint {https://arxiv.org/abs/1809.11096}
  {arXiv:1809.11096} \BibitemShut {NoStop}%
\bibitem [{\citenamefont {Kingma}\ and\ \citenamefont {Dhariwal}(2018)}]{glow}%
  \BibitemOpen
  \bibfield  {author} {\bibinfo {author} {\bibfnamefont {D.~P.}\ \bibnamefont
  {Kingma}}\ and\ \bibinfo {author} {\bibfnamefont {P.}~\bibnamefont
  {Dhariwal}},\ }\bibfield  {title} {\bibinfo {title} {Glow: Generative flow
  with invertible 1x1 convolutions},\ }in\ \href
  {http://papers.nips.cc/paper/8224-glow-generative-flow-with-invertible-1x1-convolutions.pdf}
  {\emph {\bibinfo {booktitle} {Advances in Neural Information Processing
  Systems 31}}}\ (\bibinfo  {publisher} {Curran Associates, Inc.},\ \bibinfo
  {year} {2018})\ pp.\ \bibinfo {pages} {10215--10224},\ \Eprint
  {https://arxiv.org/abs/1807.03039} {arXiv:1807.03039} \BibitemShut {NoStop}%
\bibitem [{\citenamefont {Baldi}\ \emph {et~al.}(2014)\citenamefont {Baldi},
  \citenamefont {Sadowski},\ and\ \citenamefont {Whiteson}}]{Baldi:NatureHEP}%
  \BibitemOpen
  \bibfield  {author} {\bibinfo {author} {\bibfnamefont {P.}~\bibnamefont
  {Baldi}}, \bibinfo {author} {\bibfnamefont {P.}~\bibnamefont {Sadowski}},\
  and\ \bibinfo {author} {\bibfnamefont {D.}~\bibnamefont {Whiteson}},\
  }\bibfield  {title} {\bibinfo {title} {{Searching for Exotic Particles in
  High-Energy Physics with Deep Learning}},\ }\href
  {https://doi.org/10.1038/ncomms5308} {\bibfield  {journal} {\bibinfo
  {journal} {Nature Communications}\ }\textbf {\bibinfo {volume} {5}},\
  \bibinfo {pages} {4308} (\bibinfo {year} {2014})},\ \Eprint
  {https://arxiv.org/abs/1402.4735} {arXiv:1402.4735} \BibitemShut {NoStop}%
\bibitem [{\citenamefont {Delaquis}\ \emph {et~al.}(2018)\citenamefont
  {Delaquis}, \citenamefont {Jewell}, \citenamefont {Ostrovskiy}, \citenamefont
  {Weber}, \citenamefont {Ziegler}, \citenamefont {Dalmasson}, \citenamefont
  {Kaufman}, \citenamefont {Richards}, \citenamefont {Albert}, \citenamefont
  {Anton}, \citenamefont {Badhrees}, \citenamefont {Barbeau}, \citenamefont
  {Bayerlein}, \citenamefont {Beck}, \citenamefont {Belov}, \citenamefont
  {Breidenbach}, \citenamefont {Brunner}, \citenamefont {Cao}, \citenamefont
  {Cen},\ and\ \citenamefont {Zeldovich}}]{DLHEP:IgorOstrovsky}%
  \BibitemOpen
  \bibfield  {author} {\bibinfo {author} {\bibfnamefont {S.}~\bibnamefont
  {Delaquis}}, \bibinfo {author} {\bibfnamefont {M.}~\bibnamefont {Jewell}},
  \bibinfo {author} {\bibfnamefont {I.}~\bibnamefont {Ostrovskiy}}, \bibinfo
  {author} {\bibfnamefont {M.}~\bibnamefont {Weber}}, \bibinfo {author}
  {\bibfnamefont {T.}~\bibnamefont {Ziegler}}, \bibinfo {author} {\bibfnamefont
  {J.}~\bibnamefont {Dalmasson}}, \bibinfo {author} {\bibfnamefont
  {L.}~\bibnamefont {Kaufman}}, \bibinfo {author} {\bibfnamefont
  {T.}~\bibnamefont {Richards}}, \bibinfo {author} {\bibfnamefont
  {J.}~\bibnamefont {Albert}}, \bibinfo {author} {\bibfnamefont
  {G.}~\bibnamefont {Anton}}, \bibinfo {author} {\bibfnamefont
  {I.}~\bibnamefont {Badhrees}}, \bibinfo {author} {\bibfnamefont
  {P.}~\bibnamefont {Barbeau}}, \bibinfo {author} {\bibfnamefont
  {R.}~\bibnamefont {Bayerlein}}, \bibinfo {author} {\bibfnamefont
  {D.}~\bibnamefont {Beck}}, \bibinfo {author} {\bibfnamefont {V.}~\bibnamefont
  {Belov}}, \bibinfo {author} {\bibfnamefont {M.}~\bibnamefont {Breidenbach}},
  \bibinfo {author} {\bibfnamefont {T.}~\bibnamefont {Brunner}}, \bibinfo
  {author} {\bibfnamefont {G.}~\bibnamefont {Cao}}, \bibinfo {author}
  {\bibfnamefont {W.}~\bibnamefont {Cen}},\ and\ \bibinfo {author}
  {\bibfnamefont {O.}~\bibnamefont {Zeldovich}},\ }\bibfield  {title} {\bibinfo
  {title} {{Deep Neural Networks for Energy and Position Reconstruction in
  EXO-200}},\ }\href {https://doi.org/10.1088/1748-0221/13/08/P08023}
  {\bibfield  {journal} {\bibinfo  {journal} {Journal of Instrumentation}\
  }\textbf {\bibinfo {volume} {13}}},\ \Eprint
  {https://arxiv.org/abs/1804.09641} {arXiv:1804.09641} \BibitemShut {NoStop}%
\bibitem [{\citenamefont {Shimmin}\ \emph {et~al.}(2017)\citenamefont
  {Shimmin}, \citenamefont {Sadowski}, \citenamefont {Baldi}, \citenamefont
  {Weik}, \citenamefont {Whiteson}, \citenamefont {Goul},\ and\ \citenamefont
  {Søgaard}}]{Baldi:DecorrelateAdversarialHEP}%
  \BibitemOpen
  \bibfield  {author} {\bibinfo {author} {\bibfnamefont {C.}~\bibnamefont
  {Shimmin}}, \bibinfo {author} {\bibfnamefont {P.}~\bibnamefont {Sadowski}},
  \bibinfo {author} {\bibfnamefont {P.}~\bibnamefont {Baldi}}, \bibinfo
  {author} {\bibfnamefont {E.}~\bibnamefont {Weik}}, \bibinfo {author}
  {\bibfnamefont {D.}~\bibnamefont {Whiteson}}, \bibinfo {author}
  {\bibfnamefont {E.}~\bibnamefont {Goul}},\ and\ \bibinfo {author}
  {\bibfnamefont {A.}~\bibnamefont {Søgaard}},\ }\bibfield  {title} {\bibinfo
  {title} {{Decorrelated Jet Substructure Tagging using Adversarial Neural
  Networks}},\ }\bibfield  {journal} {\bibinfo  {journal} {Physical Review D}\
  }\textbf {\bibinfo {volume} {96}},\ \href
  {https://doi.org/10.1103/PhysRevD.96.074034} {10.1103/PhysRevD.96.074034}
  (\bibinfo {year} {2017}),\ \Eprint {https://arxiv.org/abs/1703.03507}
  {arXiv:1703.03507} \BibitemShut {NoStop}%
\bibitem [{\citenamefont {Baldi}\ \emph {et~al.}(2019)\citenamefont {Baldi},
  \citenamefont {Bian}, \citenamefont {Hertel},\ and\ \citenamefont
  {Li}}]{NOvABaldi}%
  \BibitemOpen
  \bibfield  {author} {\bibinfo {author} {\bibfnamefont {P.}~\bibnamefont
  {Baldi}}, \bibinfo {author} {\bibfnamefont {J.}~\bibnamefont {Bian}},
  \bibinfo {author} {\bibfnamefont {L.}~\bibnamefont {Hertel}},\ and\ \bibinfo
  {author} {\bibfnamefont {L.}~\bibnamefont {Li}},\ }\bibfield  {title}
  {\bibinfo {title} {{Improved Energy Reconstruction in NOvA with Regression
  Convolutional Neural Networks}},\ }\bibfield  {journal} {\bibinfo  {journal}
  {Phys. Rev. D}\ }\textbf {\bibinfo {volume} {99}},\ \href
  {https://doi.org/10.1103/PhysRevD.99.012011} {10.1103/PhysRevD.99.012011}
  (\bibinfo {year} {2019}),\ \Eprint {https://arxiv.org/abs/1811.04557}
  {arXiv:1811.04557} \BibitemShut {NoStop}%
\bibitem [{\citenamefont {Guest}\ \emph {et~al.}(2016)\citenamefont {Guest},
  \citenamefont {Collado}, \citenamefont {Baldi}, \citenamefont {Hsu},
  \citenamefont {Urban},\ and\ \citenamefont
  {Whiteson}}]{Collado:JetFlavorClassification}%
  \BibitemOpen
  \bibfield  {author} {\bibinfo {author} {\bibfnamefont {D.}~\bibnamefont
  {Guest}}, \bibinfo {author} {\bibfnamefont {J.}~\bibnamefont {Collado}},
  \bibinfo {author} {\bibfnamefont {P.}~\bibnamefont {Baldi}}, \bibinfo
  {author} {\bibfnamefont {S.-C.}\ \bibnamefont {Hsu}}, \bibinfo {author}
  {\bibfnamefont {G.}~\bibnamefont {Urban}},\ and\ \bibinfo {author}
  {\bibfnamefont {D.}~\bibnamefont {Whiteson}},\ }\bibfield  {title} {\bibinfo
  {title} {{Jet Flavor Classification in High-Energy Physics with Deep Neural
  Networks}},\ }\bibfield  {journal} {\bibinfo  {journal} {Physical Review D}\
  }\textbf {\bibinfo {volume} {94}},\ \href
  {https://doi.org/10.1103/PhysRevD.94.112002} {10.1103/PhysRevD.94.112002}
  (\bibinfo {year} {2016}),\ \Eprint {https://arxiv.org/abs/1607.08633}
  {arXiv:1607.08633} \BibitemShut {NoStop}%
\bibitem [{\citenamefont {Sadowski}\ \emph {et~al.}(2015)\citenamefont
  {Sadowski}, \citenamefont {Collado}, \citenamefont {Whiteson},\ and\
  \citenamefont {Baldi}}]{Collado:DarkKnowledge}%
  \BibitemOpen
  \bibfield  {author} {\bibinfo {author} {\bibfnamefont {P.}~\bibnamefont
  {Sadowski}}, \bibinfo {author} {\bibfnamefont {J.}~\bibnamefont {Collado}},
  \bibinfo {author} {\bibfnamefont {D.}~\bibnamefont {Whiteson}},\ and\
  \bibinfo {author} {\bibfnamefont {P.}~\bibnamefont {Baldi}},\ }\bibfield
  {title} {\bibinfo {title} {{Deep Learning, Dark Knowledge, and Dark
  Matter}},\ }in\ \href@noop {} {\emph {\bibinfo {booktitle} {Proceedings of
  the NIPS 2014 Workshop on High-energy Physics and Machine Learning}}},\
  \bibinfo {series} {Proceedings of Machine Learning Research}, Vol.~\bibinfo
  {volume} {42}\ (\bibinfo  {publisher} {PMLR},\ \bibinfo {address} {Montreal,
  Canada},\ \bibinfo {year} {(2015)})\ pp.\ \bibinfo {pages} {81--87},\
  \bibinfo {note}
  {\url{http://proceedings.mlr.press/v42/sado14.html}}\BibitemShut {NoStop}%
\bibitem [{\citenamefont {Seong}\ \emph {et~al.}(2019)\citenamefont {Seong},
  \citenamefont {Hertel}, \citenamefont {Collado}, \citenamefont {Li},
  \citenamefont {Nayak}, \citenamefont {Bian},\ and\ \citenamefont
  {Baldi}}]{Collado:DUNE}%
  \BibitemOpen
  \bibfield  {author} {\bibinfo {author} {\bibfnamefont {I.}~\bibnamefont
  {Seong}}, \bibinfo {author} {\bibfnamefont {L.}~\bibnamefont {Hertel}},
  \bibinfo {author} {\bibfnamefont {J.}~\bibnamefont {Collado}}, \bibinfo
  {author} {\bibfnamefont {L.}~\bibnamefont {Li}}, \bibinfo {author}
  {\bibfnamefont {N.}~\bibnamefont {Nayak}}, \bibinfo {author} {\bibfnamefont
  {J.}~\bibnamefont {Bian}},\ and\ \bibinfo {author} {\bibfnamefont
  {P.}~\bibnamefont {Baldi}},\ }\bibfield  {title} {\bibinfo {title}
  {{Convolutional Neural Networks for Energy and Vertex Reconstruction in
  DUNE}},\ }in\ \href
  {https://ml4physicalsciences.github.io/files/NeurIPS_ML4PS\_2019\_77.pdf}
  {\emph {\bibinfo {booktitle} {33rd Conference on Neural Information
  Processing Systems (NeurIPS), Machine Learning and the Physical Sciences
  Workshop}}}\ (\bibinfo {year} {(2019)})\ \bibinfo {note}
  {\url{https://ml4physicalsciences.github.io/2019/files/NeurIPS\_ML4PS\_2019\_77.pdf}}\BibitemShut
  {NoStop}%
\bibitem [{\citenamefont {Baldi}(2020)}]{BaldiDLBook2020}%
  \BibitemOpen
  \bibfield  {author} {\bibinfo {author} {\bibfnamefont {P.}~\bibnamefont
  {Baldi}},\ }\href@noop {} {\emph {\bibinfo {title} {Deep Learning in Science:
  Theory, Algorithms, and Applications}}}\ (\bibinfo  {publisher} {Cambridge
  University Press},\ \bibinfo {address} {Cambridge, UK},\ \bibinfo {year}
  {(2020)})\ \bibinfo {note} {in press.}\BibitemShut {Stop}%
\bibitem [{\citenamefont {Mustafa}\ \emph {et~al.}(2019)\citenamefont
  {Mustafa}, \citenamefont {Bard}, \citenamefont {Bhimji}, \citenamefont
  {Luki{\'c}}, \citenamefont {Al-Rfou},\ and\ \citenamefont
  {Kratochvil}}]{CosmoGAN}%
  \BibitemOpen
  \bibfield  {author} {\bibinfo {author} {\bibfnamefont {M.}~\bibnamefont
  {Mustafa}}, \bibinfo {author} {\bibfnamefont {D.}~\bibnamefont {Bard}},
  \bibinfo {author} {\bibfnamefont {W.}~\bibnamefont {Bhimji}}, \bibinfo
  {author} {\bibfnamefont {Z.}~\bibnamefont {Luki{\'c}}}, \bibinfo {author}
  {\bibfnamefont {R.}~\bibnamefont {Al-Rfou}},\ and\ \bibinfo {author}
  {\bibfnamefont {J.~M.}\ \bibnamefont {Kratochvil}},\ }\bibfield  {title}
  {\bibinfo {title} {{CosmoGAN: Creating High-fidelity Weak Lensing Convergence
  Maps using Generative Adversarial Networks}},\ }\bibfield  {journal}
  {\bibinfo  {journal} {Computational Astrophysics and Cosmology}\ }\textbf
  {\bibinfo {volume} {6}},\ \href {https://doi.org/10.1186/s40668-019-0029-9}
  {10.1186/s40668-019-0029-9} (\bibinfo {year} {2019}),\ \Eprint
  {https://arxiv.org/abs/1706.02390} {arXiv:1706.02390} \BibitemShut {NoStop}%
\bibitem [{\citenamefont {Musella}\ and\ \citenamefont
  {Pandolfi}(2018)}]{PhysicsGAN1}%
  \BibitemOpen
  \bibfield  {author} {\bibinfo {author} {\bibfnamefont {P.}~\bibnamefont
  {Musella}}\ and\ \bibinfo {author} {\bibfnamefont {F.}~\bibnamefont
  {Pandolfi}},\ }\bibfield  {title} {\bibinfo {title} {{Fast and Accurate
  Simulation of Particle Detectors Using Generative Adversarial Networks}},\
  }\href@noop {} {\bibfield  {journal} {\bibinfo  {journal} {Computing and
  Software for Big Science}\ } (\bibinfo {year} {2018})},\ \Eprint
  {https://arxiv.org/abs/1805.00850} {arXiv:1805.00850} \BibitemShut {NoStop}%
\bibitem [{\citenamefont {Zhou}\ \emph {et~al.}(2019)\citenamefont {Zhou},
  \citenamefont {Endr\ifmmode~\mbox{\H{o}}\else \H{o}\fi{}di}, \citenamefont
  {Pang},\ and\ \citenamefont {St\"ocker}}]{PhysicsGAN2}%
  \BibitemOpen
  \bibfield  {author} {\bibinfo {author} {\bibfnamefont {K.}~\bibnamefont
  {Zhou}}, \bibinfo {author} {\bibfnamefont {G.}~\bibnamefont
  {Endr\ifmmode~\mbox{\H{o}}\else \H{o}\fi{}di}}, \bibinfo {author}
  {\bibfnamefont {L.-G.}\ \bibnamefont {Pang}},\ and\ \bibinfo {author}
  {\bibfnamefont {H.}~\bibnamefont {St\"ocker}},\ }\bibfield  {title} {\bibinfo
  {title} {{Regressive and Generative Neural Networks for Dcalar Field
  Theory}},\ }\href {https://doi.org/10.1103/PhysRevD.100.011501} {\bibfield
  {journal} {\bibinfo  {journal} {Phys. Rev. D}\ }\textbf {\bibinfo {volume}
  {100}},\ \bibinfo {pages} {011501} (\bibinfo {year} {2019})},\ \Eprint
  {https://arxiv.org/abs/1810.12879} {arXiv:1810.12879} \BibitemShut {NoStop}%
\bibitem [{\citenamefont {r.~{Khattak}}\ \emph {et~al.}(2018)\citenamefont
  {r.~{Khattak}}, \citenamefont {{Vallecorsa}},\ and\ \citenamefont
  {{Carminati}}}]{PhysicsGAN3}%
  \BibitemOpen
  \bibfield  {author} {\bibinfo {author} {\bibfnamefont {G.}~\bibnamefont
  {r.~{Khattak}}}, \bibinfo {author} {\bibfnamefont {S.}~\bibnamefont
  {{Vallecorsa}}},\ and\ \bibinfo {author} {\bibfnamefont {F.}~\bibnamefont
  {{Carminati}}},\ }\bibfield  {title} {\bibinfo {title} {{Three Dimensional
  Energy Parametrized Generative Adversarial Networks for Electromagnetic
  Shower Simulation}},\ }in\ \href {https://doi.org/10.1109/ICIP.2018.8451587}
  {\emph {\bibinfo {booktitle} {2018 25th IEEE International Conference on
  Image Processing (ICIP)}}}\ (\bibinfo {year} {2018})\ pp.\ \bibinfo {pages}
  {3913--3917},\ \bibinfo {note}
  {\url{https://ieeexplore.ieee.org/document/8451587}}\BibitemShut {NoStop}%
\bibitem [{\citenamefont {Alonso~Monsalve}\ and\ \citenamefont
  {Whitehead}(2020)}]{PhysicsGAN4}%
  \BibitemOpen
  \bibfield  {author} {\bibinfo {author} {\bibfnamefont {S.}~\bibnamefont
  {Alonso~Monsalve}}\ and\ \bibinfo {author} {\bibfnamefont {L.}~\bibnamefont
  {Whitehead}},\ }\bibfield  {title} {\bibinfo {title} {{Image-Based Model
  Parameter Optimization Using Model-Assisted Generative Adversarial
  Networks}},\ }\href {https://doi.org/10.1109/TNNLS.2020.2969327} {\bibfield
  {journal} {\bibinfo  {journal} {IEEE Transactions on Neural Networks and
  Learning Systems}\ }\textbf {\bibinfo {volume} {PP}},\ \bibinfo {pages} {1}
  (\bibinfo {year} {2020})},\ \Eprint {https://arxiv.org/abs/1812.00879}
  {arXiv:1812.00879} \BibitemShut {NoStop}%
\bibitem [{\citenamefont {Deja}\ \emph {et~al.}(2020)\citenamefont {Deja},
  \citenamefont {Trzci{\'{n}}ski},\ and\ \citenamefont
  {Graczykowski}}]{PhysicsGAN5}%
  \BibitemOpen
  \bibfield  {author} {\bibinfo {author} {\bibfnamefont {K.}~\bibnamefont
  {Deja}}, \bibinfo {author} {\bibfnamefont {T.}~\bibnamefont
  {Trzci{\'{n}}ski}},\ and\ \bibinfo {author} {\bibfnamefont
  {{\L}.}~\bibnamefont {Graczykowski}},\ }\bibfield  {title} {\bibinfo {title}
  {{Generative Models for Fast Cluster Simulations in the TPC for the ALICE
  Experiment}},\ }in\ \href@noop {} {\emph {\bibinfo {booktitle} {Information
  Technology, Systems Research, and Computational Physics}}}\ (\bibinfo
  {publisher} {Springer International Publishing},\ \bibinfo {address} {Cham},\
  \bibinfo {year} {(2020)})\ pp.\ \bibinfo {pages} {267--280}\BibitemShut
  {NoStop}%
\bibitem [{\citenamefont {Carminati}\ \emph {et~al.}(2017)\citenamefont
  {Carminati}, \citenamefont {Gulrukh~Khattak}, \citenamefont {Amir~Farbin},
  \citenamefont {Wei}, \citenamefont {Zhang}, \citenamefont {Pacela},
  \citenamefont {Vallecorsafac}, \citenamefont {Spiropulu},\ and\ \citenamefont
  {Vlimant}}]{PhysicsGAN6}%
  \BibitemOpen
  \bibfield  {author} {\bibinfo {author} {\bibfnamefont {F.}~\bibnamefont
  {Carminati}}, \bibinfo {author} {\bibfnamefont {M.~P.}\ \bibnamefont
  {Gulrukh~Khattak}}, \bibinfo {author} {\bibfnamefont {B.~H.}\ \bibnamefont
  {Amir~Farbin}}, \bibinfo {author} {\bibfnamefont {W.}~\bibnamefont {Wei}},
  \bibinfo {author} {\bibfnamefont {M.}~\bibnamefont {Zhang}}, \bibinfo
  {author} {\bibfnamefont {V.~B.}\ \bibnamefont {Pacela}}, \bibinfo {author}
  {\bibfnamefont {S.}~\bibnamefont {Vallecorsafac}}, \bibinfo {author}
  {\bibfnamefont {M.}~\bibnamefont {Spiropulu}},\ and\ \bibinfo {author}
  {\bibfnamefont {J.-R.}\ \bibnamefont {Vlimant}},\ }\bibfield  {title}
  {\bibinfo {title} {{Calorimetry with Deep Learning: Particle Classification,
  Energy Regression, and Simulation for High-Energy Physics}},\ }\href
  {https://dl4physicalsciences.github.io/files/nips_dlps_2017_15.pdf}
  {\bibfield  {journal} {\bibinfo  {journal} {Deep Learning for Physical
  Sciences, Workshop at the 31st Conference on Neural Information Processing
  Systems (NeurIPS)}\ } (\bibinfo {year} {2017})}\BibitemShut {NoStop}%
\bibitem [{\citenamefont {Shah}\ \emph {et~al.}(2019)\citenamefont {Shah},
  \citenamefont {Joshi}, \citenamefont {Ghosal}, \citenamefont {Pokuri},
  \citenamefont {Sarkar}, \citenamefont {Ganapathysubramanian},\ and\
  \citenamefont {Hegde}}]{PhysicsGAN7}%
  \BibitemOpen
  \bibfield  {author} {\bibinfo {author} {\bibfnamefont {V.}~\bibnamefont
  {Shah}}, \bibinfo {author} {\bibfnamefont {A.}~\bibnamefont {Joshi}},
  \bibinfo {author} {\bibfnamefont {S.}~\bibnamefont {Ghosal}}, \bibinfo
  {author} {\bibfnamefont {B.~S.~S.}\ \bibnamefont {Pokuri}}, \bibinfo {author}
  {\bibfnamefont {S.}~\bibnamefont {Sarkar}}, \bibinfo {author} {\bibfnamefont
  {B.}~\bibnamefont {Ganapathysubramanian}},\ and\ \bibinfo {author}
  {\bibfnamefont {C.}~\bibnamefont {Hegde}},\ }\bibfield  {title} {\bibinfo
  {title} {{Encoding Invariances in Deep Generative Models}},\ }\href
  {http://arxiv.org/abs/1906.01626} {\bibfield  {journal} {\bibinfo  {journal}
  {CoRR}\ } (\bibinfo {year} {2019})},\ \Eprint
  {https://arxiv.org/abs/1906.01626} {arXiv:1906.01626} \BibitemShut {NoStop}%
\bibitem [{\citenamefont {Cranmer}\ \emph {et~al.}(2018)\citenamefont
  {Cranmer}, \citenamefont {Gadatsch}, \citenamefont {Ghosh}, \citenamefont
  {Golling}, \citenamefont {Gilles~Louppe}, \citenamefont {Salamani},\ and\
  \citenamefont {on~behalf of~the ATLAS~Collaboration}}]{PhysicsGAN8}%
  \BibitemOpen
  \bibfield  {author} {\bibinfo {author} {\bibfnamefont {K.}~\bibnamefont
  {Cranmer}}, \bibinfo {author} {\bibfnamefont {S.}~\bibnamefont {Gadatsch}},
  \bibinfo {author} {\bibfnamefont {A.}~\bibnamefont {Ghosh}}, \bibinfo
  {author} {\bibfnamefont {T.}~\bibnamefont {Golling}}, \bibinfo {author}
  {\bibfnamefont {D.~R.}\ \bibnamefont {Gilles~Louppe}}, \bibinfo {author}
  {\bibfnamefont {D.}~\bibnamefont {Salamani}},\ and\ \bibinfo {author}
  {\bibfnamefont {G.~S.}\ \bibnamefont {on~behalf of~the
  ATLAS~Collaboration}},\ }\bibfield  {title} {\bibinfo {title} {{Deep
  generative models for fast shower simulation in ATLAS}},\ }\href
  {http://bayesiandeeplearning.org/2018/papers/24.pdf} {\bibfield  {journal}
  {\bibinfo  {journal} {Bayesian Deep Learning, Workshop at the 32nd Conference
  on Neural Information Processing Systems (NeurIPS)}\ } (\bibinfo {year}
  {2018})},\ \bibinfo {note}
  {\url{http://bayesiandeeplearning.org/2018/papers/24.pdf}}\BibitemShut
  {NoStop}%
\bibitem [{\citenamefont {Hashemi}\ \emph {et~al.}(2019)\citenamefont
  {Hashemi}, \citenamefont {Amin}, \citenamefont {Datta}, \citenamefont
  {Olivito},\ and\ \citenamefont {Pierini}}]{PhysicsGAN9}%
  \BibitemOpen
  \bibfield  {author} {\bibinfo {author} {\bibfnamefont {B.}~\bibnamefont
  {Hashemi}}, \bibinfo {author} {\bibfnamefont {N.}~\bibnamefont {Amin}},
  \bibinfo {author} {\bibfnamefont {K.}~\bibnamefont {Datta}}, \bibinfo
  {author} {\bibfnamefont {D.}~\bibnamefont {Olivito}},\ and\ \bibinfo {author}
  {\bibfnamefont {M.}~\bibnamefont {Pierini}},\ }\bibfield  {title} {\bibinfo
  {title} {{LHC analysis-specific datasets with Generative Adversarial
  Networks}},\ }\href@noop {} {\bibfield  {journal} {\bibinfo  {journal}
  {CoRR}\ } (\bibinfo {year} {2019})},\ \Eprint
  {https://arxiv.org/abs/1901.05282} {arXiv:1901.05282} \BibitemShut {NoStop}%
\bibitem [{\citenamefont {Otten}\ \emph {et~al.}(2019)\citenamefont {Otten},
  \citenamefont {Caron}, \citenamefont {de~Swart}, \citenamefont {van
  Beekveld}, \citenamefont {Hendriks}, \citenamefont {van Leeuwen},
  \citenamefont {Podareanu}, \citenamefont {de~Austri},\ and\ \citenamefont
  {Verheyen}}]{PhysicsGAN10}%
  \BibitemOpen
  \bibfield  {author} {\bibinfo {author} {\bibfnamefont {S.}~\bibnamefont
  {Otten}}, \bibinfo {author} {\bibfnamefont {S.}~\bibnamefont {Caron}},
  \bibinfo {author} {\bibfnamefont {W.}~\bibnamefont {de~Swart}}, \bibinfo
  {author} {\bibfnamefont {M.}~\bibnamefont {van Beekveld}}, \bibinfo {author}
  {\bibfnamefont {L.}~\bibnamefont {Hendriks}}, \bibinfo {author}
  {\bibfnamefont {C.}~\bibnamefont {van Leeuwen}}, \bibinfo {author}
  {\bibfnamefont {D.}~\bibnamefont {Podareanu}}, \bibinfo {author}
  {\bibfnamefont {R.~R.}\ \bibnamefont {de~Austri}},\ and\ \bibinfo {author}
  {\bibfnamefont {R.}~\bibnamefont {Verheyen}},\ }\bibfield  {title} {\bibinfo
  {title} {{Event Generation and Statistical Sampling for Physics with Deep
  Generative Models and a Density Information Buffer}},\ }\href@noop {}
  {\bibfield  {journal} {\bibinfo  {journal} {CoRR}\ } (\bibinfo {year}
  {2019})},\ \Eprint {https://arxiv.org/abs/1901.00875} {arXiv:1901.00875}
  \BibitemShut {NoStop}%
\bibitem [{\citenamefont {Cogan}\ \emph {et~al.}(2015)\citenamefont {Cogan},
  \citenamefont {Kagan}, \citenamefont {Strauss},\ and\ \citenamefont
  {Schwarztman}}]{sparsityInHEP}%
  \BibitemOpen
  \bibfield  {author} {\bibinfo {author} {\bibfnamefont {J.}~\bibnamefont
  {Cogan}}, \bibinfo {author} {\bibfnamefont {M.}~\bibnamefont {Kagan}},
  \bibinfo {author} {\bibfnamefont {E.}~\bibnamefont {Strauss}},\ and\ \bibinfo
  {author} {\bibfnamefont {A.}~\bibnamefont {Schwarztman}},\ }\bibfield
  {title} {\bibinfo {title} {{Jet-Images: Computer Vision Inspired Techniques
  for Jet Tagging}},\ }\href {https://doi.org/10.1007/JHEP02(2015)118}
  {\bibfield  {journal} {\bibinfo  {journal} {JHEP}\ }\textbf {\bibinfo
  {volume} {02}}},\ \Eprint {https://arxiv.org/abs/1407.5675} {arXiv:1407.5675
  [hep-ph]} \BibitemShut {NoStop}%
\bibitem [{\citenamefont {Paganini}\ \emph {et~al.}(2018)\citenamefont
  {Paganini}, \citenamefont {de~Oliveira},\ and\ \citenamefont
  {Nachman}}]{CALOGAN}%
  \BibitemOpen
  \bibfield  {author} {\bibinfo {author} {\bibfnamefont {M.}~\bibnamefont
  {Paganini}}, \bibinfo {author} {\bibfnamefont {L.}~\bibnamefont
  {de~Oliveira}},\ and\ \bibinfo {author} {\bibfnamefont {B.}~\bibnamefont
  {Nachman}},\ }\bibfield  {title} {\bibinfo {title} {{CaloGAN: Simulating 3D
  High Energy Particle Showers in Multi-Layer Electromagnetic Calorimeters with
  Generative Adversarial Networks}},\ }\href {http://arxiv.org/abs/1712.10321}
  {\bibfield  {journal} {\bibinfo  {journal} {Phys. Rev. D}\ }\textbf {\bibinfo
  {volume} {97}} (\bibinfo {year} {2018})},\ \Eprint
  {https://arxiv.org/abs/1712.10321} {arXiv:1712.10321} \BibitemShut {NoStop}%
\bibitem [{\citenamefont {Chintala}(2016)}]{sparseGradientsGANWorkshopNIPS}%
  \BibitemOpen
  \bibfield  {author} {\bibinfo {author} {\bibfnamefont {S.}~\bibnamefont
  {Chintala}},\ }\bibfield  {title} {\bibinfo {title} {{How to train a GAN?}},\
  }in\ \href@noop {} {\emph {\bibinfo {booktitle} {Workshop on Generative
  Adversarial Networks}}}\ (\bibinfo {year} {2016})\BibitemShut {NoStop}%
\bibitem [{\citenamefont {Bottou}(2010)}]{sgd}%
  \BibitemOpen
  \bibfield  {author} {\bibinfo {author} {\bibfnamefont {L.}~\bibnamefont
  {Bottou}},\ }\bibfield  {title} {\bibinfo {title} {Large-scale machine
  learning with stochastic gradient descent},\ }in\ \href@noop {} {\emph
  {\bibinfo {booktitle} {COMPSTAT}}}\ (\bibinfo {year} {(2010)})\ \bibinfo
  {note}
  {\url{https://leon.bottou.org/publications/pdf/compstat-2010.pdf}}\BibitemShut
  {NoStop}%
\bibitem [{\citenamefont {Kingma}\ and\ \citenamefont {Ba}(2014)}]{adam}%
  \BibitemOpen
  \bibfield  {author} {\bibinfo {author} {\bibfnamefont {D.~P.}\ \bibnamefont
  {Kingma}}\ and\ \bibinfo {author} {\bibfnamefont {J.}~\bibnamefont {Ba}},\
  }\bibfield  {title} {\bibinfo {title} {{Adam: A Method for Stochastic
  Optimization}},\ }in\ \href {http://arxiv.org/abs/1412.6980} {\emph {\bibinfo
  {booktitle} {{3rd International Conference on Learning Representations,
  {ICLR}}}}}\ (\bibinfo {year} {(2014)})\ \Eprint
  {https://arxiv.org/abs/1412.6980} {arXiv:1412.6980} \BibitemShut {NoStop}%
\bibitem [{\citenamefont {Arjovsky}\ and\ \citenamefont
  {Bottou}(2017)}]{convergence1}%
  \BibitemOpen
  \bibfield  {author} {\bibinfo {author} {\bibfnamefont {M.}~\bibnamefont
  {Arjovsky}}\ and\ \bibinfo {author} {\bibfnamefont {L.}~\bibnamefont
  {Bottou}},\ }\bibfield  {title} {\bibinfo {title} {{Towards Principled
  Methods for Training Generative Adversarial Networks}},\ }in\ \href
  {https://openreview.net/forum?id=Hk4\_qw5xe} {\emph {\bibinfo {booktitle}
  {5th International Conference on Learning Representations, {ICLR} 2017,
  Toulon, France, April 24-26, 2017, Conference Track Proceedings}}}\ (\bibinfo
  {year} {(2017)})\ \Eprint {https://arxiv.org/abs/1701.04862}
  {arXiv:1701.04862} \BibitemShut {NoStop}%
\bibitem [{\citenamefont {Nagarajan}\ and\ \citenamefont
  {Kolter}(2017)}]{modecollapse}%
  \BibitemOpen
  \bibfield  {author} {\bibinfo {author} {\bibfnamefont {V.}~\bibnamefont
  {Nagarajan}}\ and\ \bibinfo {author} {\bibfnamefont {J.~Z.}\ \bibnamefont
  {Kolter}},\ }\bibfield  {title} {\bibinfo {title} {{Gradient descent GAN
  optimization is locally stable}},\ }in\ \href
  {http://papers.nips.cc/paper/7142-gradient-descent-gan-optimization-is-locally-stable.pdf}
  {\emph {\bibinfo {booktitle} {Advances in Neural Information Processing
  Systems 30}}}\ (\bibinfo  {publisher} {Curran Associates, Inc.},\ \bibinfo
  {year} {(2017)})\ pp.\ \bibinfo {pages} {5585--5595},\ \Eprint
  {https://arxiv.org/abs/1706.04156} {arXiv:1706.04156} \BibitemShut {NoStop}%
\bibitem [{\citenamefont {Radford}\ \emph {et~al.}(2015)\citenamefont
  {Radford}, \citenamefont {Metz},\ and\ \citenamefont
  {Chintala}}]{Radford2015UnsupervisedRL}%
  \BibitemOpen
  \bibfield  {author} {\bibinfo {author} {\bibfnamefont {A.}~\bibnamefont
  {Radford}}, \bibinfo {author} {\bibfnamefont {L.}~\bibnamefont {Metz}},\ and\
  \bibinfo {author} {\bibfnamefont {S.}~\bibnamefont {Chintala}},\ }\bibfield
  {title} {\bibinfo {title} {{Unsupervised Representation Learning with Deep
  Convolutional Generative Adversarial Networks}},\ }\href@noop {} {\bibfield
  {journal} {\bibinfo  {journal} {CoRR}\ } (\bibinfo {year} {2015})},\ \Eprint
  {https://arxiv.org/abs/1511.06434} {arXiv:1511.06434} \BibitemShut {NoStop}%
\bibitem [{\citenamefont {Salimans}\ \emph {et~al.}(2017)\citenamefont
  {Salimans}, \citenamefont {Karpathy}, \citenamefont {Chen},\ and\
  \citenamefont {Kingma}}]{pixelcnn++}%
  \BibitemOpen
  \bibfield  {author} {\bibinfo {author} {\bibfnamefont {T.}~\bibnamefont
  {Salimans}}, \bibinfo {author} {\bibfnamefont {A.}~\bibnamefont {Karpathy}},
  \bibinfo {author} {\bibfnamefont {X.}~\bibnamefont {Chen}},\ and\ \bibinfo
  {author} {\bibfnamefont {D.~P.}\ \bibnamefont {Kingma}},\ }\bibfield  {title}
  {\bibinfo {title} {{PixelCNN++: Improving the PixelCNN with Discretized
  Logistic Mixture Likelihood and Other Modifications}},\ }\href
  {http://arxiv.org/abs/1701.05517} {\bibfield  {journal} {\bibinfo  {journal}
  {CoRR}\ } (\bibinfo {year} {2017})},\ \Eprint
  {https://arxiv.org/abs/1701.05517} {arXiv:1701.05517} \BibitemShut {NoStop}%
\bibitem [{\citenamefont {Almeida}\ \emph {et~al.}(2015)\citenamefont
  {Almeida}, \citenamefont {Backović}, \citenamefont {Cliche}, \citenamefont
  {Lee},\ and\ \citenamefont {Perelstein}}]{Almeida:2015jua}%
  \BibitemOpen
  \bibfield  {author} {\bibinfo {author} {\bibfnamefont {L.~G.}\ \bibnamefont
  {Almeida}}, \bibinfo {author} {\bibfnamefont {M.}~\bibnamefont {Backović}},
  \bibinfo {author} {\bibfnamefont {M.}~\bibnamefont {Cliche}}, \bibinfo
  {author} {\bibfnamefont {S.~J.}\ \bibnamefont {Lee}},\ and\ \bibinfo {author}
  {\bibfnamefont {M.}~\bibnamefont {Perelstein}},\ }\bibfield  {title}
  {\bibinfo {title} {{Playing Tag with ANN: Boosted Top Identification with
  Pattern Recognition}},\ }\href {https://doi.org/10.1007/JHEP07(2015)086}
  {\bibfield  {journal} {\bibinfo  {journal} {JHEP}\ }\textbf {\bibinfo
  {volume} {07}},\ \bibinfo {pages} {086}},\ \Eprint
  {https://arxiv.org/abs/1501.05968} {arXiv:1501.05968 [hep-ph]} \BibitemShut
  {NoStop}%
\bibitem [{\citenamefont {de~Oliveira}\ \emph {et~al.}(2016)\citenamefont
  {de~Oliveira}, \citenamefont {Kagan}, \citenamefont {Mackey}, \citenamefont
  {Nachman},\ and\ \citenamefont
  {Schwartzman}}]{de_Oliveira_jet_images_deep_learning_edition_2016}%
  \BibitemOpen
  \bibfield  {author} {\bibinfo {author} {\bibfnamefont {L.}~\bibnamefont
  {de~Oliveira}}, \bibinfo {author} {\bibfnamefont {M.}~\bibnamefont {Kagan}},
  \bibinfo {author} {\bibfnamefont {L.}~\bibnamefont {Mackey}}, \bibinfo
  {author} {\bibfnamefont {B.}~\bibnamefont {Nachman}},\ and\ \bibinfo {author}
  {\bibfnamefont {A.}~\bibnamefont {Schwartzman}},\ }\bibfield  {title}
  {\bibinfo {title} {{Jet-Images — Deep Learning Edition}},\ }\bibfield
  {journal} {\bibinfo  {journal} {Journal of High Energy Physics}\ }\href
  {https://doi.org/10.1007/jhep07(2016)069} {10.1007/jhep07(2016)069} (\bibinfo
  {year} {2016}),\ \Eprint {https://arxiv.org/abs/1511.05190}
  {arXiv:1511.05190} \BibitemShut {NoStop}%
\bibitem [{\citenamefont {Barnard}\ \emph {et~al.}(2017)\citenamefont
  {Barnard}, \citenamefont {Dawe}, \citenamefont {Dolan},\ and\ \citenamefont
  {Rajcic}}]{Barnard:2016qma}%
  \BibitemOpen
  \bibfield  {author} {\bibinfo {author} {\bibfnamefont {J.}~\bibnamefont
  {Barnard}}, \bibinfo {author} {\bibfnamefont {E.~N.}\ \bibnamefont {Dawe}},
  \bibinfo {author} {\bibfnamefont {M.~J.}\ \bibnamefont {Dolan}},\ and\
  \bibinfo {author} {\bibfnamefont {N.}~\bibnamefont {Rajcic}},\ }\bibfield
  {title} {\bibinfo {title} {{Parton Shower Uncertainties in Jet Substructure
  Analyses with Deep Neural Networks}},\ }\href
  {https://doi.org/10.1103/PhysRevD.95.014018} {\bibfield  {journal} {\bibinfo
  {journal} {Phys. Rev.}\ }\textbf {\bibinfo {volume} {D95}},\ \bibinfo {pages}
  {014018} (\bibinfo {year} {2017})},\ \Eprint
  {https://arxiv.org/abs/1609.00607} {arXiv:1609.00607 [hep-ph]} \BibitemShut
  {NoStop}%
\bibitem [{\citenamefont {Komiske}\ \emph {et~al.}(2017)\citenamefont
  {Komiske}, \citenamefont {Metodiev},\ and\ \citenamefont
  {Schwartz}}]{Komiske:2016rsd}%
  \BibitemOpen
  \bibfield  {author} {\bibinfo {author} {\bibfnamefont {P.~T.}\ \bibnamefont
  {Komiske}}, \bibinfo {author} {\bibfnamefont {E.~M.}\ \bibnamefont
  {Metodiev}},\ and\ \bibinfo {author} {\bibfnamefont {M.~D.}\ \bibnamefont
  {Schwartz}},\ }\bibfield  {title} {\bibinfo {title} {{Deep learning in color:
  towards automated quark/gluon jet discrimination}},\ }\href
  {https://doi.org/10.1007/JHEP01(2017)110} {\bibfield  {journal} {\bibinfo
  {journal} {JHEP}\ }\textbf {\bibinfo {volume} {01}},\ \bibinfo {pages}
  {110}},\ \Eprint {https://arxiv.org/abs/1612.01551} {arXiv:1612.01551
  [hep-ph]} \BibitemShut {NoStop}%
\bibitem [{\citenamefont {Sjostrand}\ \emph {et~al.}(2006)\citenamefont
  {Sjostrand}, \citenamefont {Mrenna},\ and\ \citenamefont {Skands}}]{pythia}%
  \BibitemOpen
  \bibfield  {author} {\bibinfo {author} {\bibfnamefont {T.}~\bibnamefont
  {Sjostrand}}, \bibinfo {author} {\bibfnamefont {S.}~\bibnamefont {Mrenna}},\
  and\ \bibinfo {author} {\bibfnamefont {P.~Z.}\ \bibnamefont {Skands}},\
  }\bibfield  {title} {\bibinfo {title} {{PYTHIA 6.4 Physics and Manual}},\
  }\href {https://doi.org/10.1088/1126-6708/2006/05/026} {\bibfield  {journal}
  {\bibinfo  {journal} {JHEP}\ }\textbf {\bibinfo {volume} {0605}},\ \bibinfo
  {pages} {026}},\ \Eprint {https://arxiv.org/abs/hep-ph/0603175}
  {arXiv:hep-ph/0603175 [hep-ph]} \BibitemShut {NoStop}%
\bibitem [{\citenamefont {Alwall}\ \emph {et~al.}(2014)\citenamefont {Alwall},
  \citenamefont {Frederix}, \citenamefont {Frixione}, \citenamefont {Hirschi},
  \citenamefont {Maltoni}, \citenamefont {Mattelaer}, \citenamefont {Shao},
  \citenamefont {Stelzer}, \citenamefont {Torrielli},\ and\ \citenamefont
  {Zaro}}]{madgraph}%
  \BibitemOpen
  \bibfield  {author} {\bibinfo {author} {\bibfnamefont {J.}~\bibnamefont
  {Alwall}}, \bibinfo {author} {\bibfnamefont {R.}~\bibnamefont {Frederix}},
  \bibinfo {author} {\bibfnamefont {S.}~\bibnamefont {Frixione}}, \bibinfo
  {author} {\bibfnamefont {V.}~\bibnamefont {Hirschi}}, \bibinfo {author}
  {\bibfnamefont {F.}~\bibnamefont {Maltoni}}, \bibinfo {author} {\bibfnamefont
  {O.}~\bibnamefont {Mattelaer}}, \bibinfo {author} {\bibfnamefont {H.~S.}\
  \bibnamefont {Shao}}, \bibinfo {author} {\bibfnamefont {T.}~\bibnamefont
  {Stelzer}}, \bibinfo {author} {\bibfnamefont {P.}~\bibnamefont {Torrielli}},\
  and\ \bibinfo {author} {\bibfnamefont {M.}~\bibnamefont {Zaro}},\ }\bibfield
  {title} {\bibinfo {title} {{The automated computation of tree-level and
  next-to-leading order differential cross sections, and their matching to
  parton shower simulations}},\ }\href
  {https://doi.org/10.1007/JHEP07(2014)079} {\bibfield  {journal} {\bibinfo
  {journal} {JHEP}\ }\textbf {\bibinfo {volume} {07}},\ \bibinfo {pages}
  {079}},\ \Eprint {https://arxiv.org/abs/1405.0301} {arXiv:1405.0301 [hep-ph]}
  \BibitemShut {NoStop}%
\bibitem [{\citenamefont {de~Favereau}\ \emph {et~al.}(2014)\citenamefont
  {de~Favereau} \emph {et~al.}}]{delphes}%
  \BibitemOpen
  \bibfield  {author} {\bibinfo {author} {\bibfnamefont {J.}~\bibnamefont
  {de~Favereau}} \emph {et~al.} (\bibinfo {collaboration} {DELPHES 3}),\
  }\bibfield  {title} {\bibinfo {title} {{DELPHES 3, A modular framework for
  fast simulation of a generic collider experiment}},\ }\href
  {https://doi.org/10.1007/JHEP02(2014)057} {\bibfield  {journal} {\bibinfo
  {journal} {JHEP}\ }\textbf {\bibinfo {volume} {1402}},\ \bibinfo {pages}
  {057}},\ \Eprint {https://arxiv.org/abs/1307.6346} {arXiv:1307.6346 [hep-ex]}
  \BibitemShut {NoStop}%
\bibitem [{\citenamefont {Larochelle}\ and\ \citenamefont
  {Murray}(2011)}]{NADE}%
  \BibitemOpen
  \bibfield  {author} {\bibinfo {author} {\bibfnamefont {H.}~\bibnamefont
  {Larochelle}}\ and\ \bibinfo {author} {\bibfnamefont {I.}~\bibnamefont
  {Murray}},\ }\bibfield  {title} {\bibinfo {title} {{The Neural Autoregressive
  Distribution Estimator}},\ }in\ \href
  {http://proceedings.mlr.press/v15/larochelle11a.html} {\emph {\bibinfo
  {booktitle} {Proceedings of the Fourteenth International Conference on
  Artificial Intelligence and Statistics}}},\ \bibinfo {series} {Proceedings of
  Machine Learning Research}, Vol.~\bibinfo {volume} {15}\ (\bibinfo
  {publisher} {PMLR},\ \bibinfo {address} {Fort Lauderdale, FL, USA},\ \bibinfo
  {year} {(2011)})\ pp.\ \bibinfo {pages} {29--37},\ \bibinfo {note}
  {\url{http://proceedings.mlr.press/v15/larochelle11a/larochelle11a.pdf}}\BibitemShut
  {NoStop}%
\bibitem [{\citenamefont {Germain}\ \emph {et~al.}(2015)\citenamefont
  {Germain}, \citenamefont {Gregor}, \citenamefont {Murray},\ and\
  \citenamefont {Larochelle}}]{MADE}%
  \BibitemOpen
  \bibfield  {author} {\bibinfo {author} {\bibfnamefont {M.}~\bibnamefont
  {Germain}}, \bibinfo {author} {\bibfnamefont {K.}~\bibnamefont {Gregor}},
  \bibinfo {author} {\bibfnamefont {I.}~\bibnamefont {Murray}},\ and\ \bibinfo
  {author} {\bibfnamefont {H.}~\bibnamefont {Larochelle}},\ }\bibfield  {title}
  {\bibinfo {title} {{{MADE:} Masked Autoencoder for Distribution
  Estimation}},\ }\href {http://arxiv.org/abs/1502.03509} {\bibfield  {journal}
  {\bibinfo  {journal} {CoRR}\ }\textbf {\bibinfo {volume} {abs/1502.03509}}
  (\bibinfo {year} {2015})},\ \Eprint {https://arxiv.org/abs/1502.03509}
  {arXiv:1502.03509} \BibitemShut {NoStop}%
\bibitem [{\citenamefont {Uria}\ \emph {et~al.}(2013)\citenamefont {Uria},
  \citenamefont {Murray},\ and\ \citenamefont {Larochelle}}]{RNADE}%
  \BibitemOpen
  \bibfield  {author} {\bibinfo {author} {\bibfnamefont {B.}~\bibnamefont
  {Uria}}, \bibinfo {author} {\bibfnamefont {I.}~\bibnamefont {Murray}},\ and\
  \bibinfo {author} {\bibfnamefont {H.}~\bibnamefont {Larochelle}},\ }\bibfield
   {title} {\bibinfo {title} {{RNADE: The real-valued neural autoregressive
  density-estimator}},\ }in\ \href
  {http://papers.nips.cc/paper/5060-rnade-the-real-valued-neural-autoregressive-density-estimator.pdf}
  {\emph {\bibinfo {booktitle} {Advances in Neural Information Processing
  Systems 26}}}\ (\bibinfo  {publisher} {Curran Associates, Inc.},\ \bibinfo
  {year} {(2013)})\ pp.\ \bibinfo {pages} {2175--2183},\ \Eprint
  {https://arxiv.org/abs/1306.0186} {arXiv:1306.0186} \BibitemShut {NoStop}%
\bibitem [{\citenamefont {Huang}\ \emph {et~al.}(2018)\citenamefont {Huang},
  \citenamefont {Krueger}, \citenamefont {Lacoste},\ and\ \citenamefont
  {Courville}}]{Neural_Autoregressive_Flows}%
  \BibitemOpen
  \bibfield  {author} {\bibinfo {author} {\bibfnamefont {C.-W.}\ \bibnamefont
  {Huang}}, \bibinfo {author} {\bibfnamefont {D.}~\bibnamefont {Krueger}},
  \bibinfo {author} {\bibfnamefont {A.}~\bibnamefont {Lacoste}},\ and\ \bibinfo
  {author} {\bibfnamefont {A.}~\bibnamefont {Courville}},\ }\bibfield  {title}
  {\bibinfo {title} {{Neural Autoregressive Flows}},\ }in\ \href
  {http://proceedings.mlr.press/v80/huang18d.html} {\emph {\bibinfo {booktitle}
  {Proceedings of the 35th International Conference on Machine Learning}}},\
  \bibinfo {series} {Proceedings of Machine Learning Research}, Vol.~\bibinfo
  {volume} {80}\ (\bibinfo  {publisher} {PMLR},\ \bibinfo {address}
  {Stockholmsmässan, Stockholm Sweden},\ \bibinfo {year} {(2018)})\ pp.\
  \bibinfo {pages} {2078--2087},\ \bibinfo {note}
  {\url{http://proceedings.mlr.press/v80/huang18d/huang18d.pdf}}\BibitemShut
  {NoStop}%
\bibitem [{\citenamefont {Gregor}\ \emph {et~al.}(2014)\citenamefont {Gregor},
  \citenamefont {Danihelka}, \citenamefont {Mnih}, \citenamefont {Blundell},\
  and\ \citenamefont {Wierstra}}]{DARNs}%
  \BibitemOpen
  \bibfield  {author} {\bibinfo {author} {\bibfnamefont {K.}~\bibnamefont
  {Gregor}}, \bibinfo {author} {\bibfnamefont {I.}~\bibnamefont {Danihelka}},
  \bibinfo {author} {\bibfnamefont {A.}~\bibnamefont {Mnih}}, \bibinfo {author}
  {\bibfnamefont {C.}~\bibnamefont {Blundell}},\ and\ \bibinfo {author}
  {\bibfnamefont {D.}~\bibnamefont {Wierstra}},\ }\bibfield  {title} {\bibinfo
  {title} {{Deep AutoRegressive Networks}},\ }in\ \href
  {http://proceedings.mlr.press/v32/gregor14.html} {\emph {\bibinfo {booktitle}
  {Proceedings of the 31st International Conference on Machine Learning}}},\
  \bibinfo {series} {Proceedings of Machine Learning Research}, Vol.~\bibinfo
  {volume} {32}\ (\bibinfo  {publisher} {PMLR},\ \bibinfo {address} {Bejing,
  China},\ \bibinfo {year} {(2014)})\ pp.\ \bibinfo {pages} {1242--1250},\
  \Eprint {https://arxiv.org/abs/1310.8499} {arXiv:1310.8499} \BibitemShut
  {NoStop}%
\bibitem [{\citenamefont {Deng}\ \emph {et~al.}(2009)\citenamefont {Deng},
  \citenamefont {Dong}, \citenamefont {Socher}, \citenamefont {Li},
  \citenamefont {Li},\ and\ \citenamefont {Fei-Fei}}]{imagenet_cvpr09}%
  \BibitemOpen
  \bibfield  {author} {\bibinfo {author} {\bibfnamefont {J.}~\bibnamefont
  {Deng}}, \bibinfo {author} {\bibfnamefont {W.}~\bibnamefont {Dong}}, \bibinfo
  {author} {\bibfnamefont {R.}~\bibnamefont {Socher}}, \bibinfo {author}
  {\bibfnamefont {L.-J.}\ \bibnamefont {Li}}, \bibinfo {author} {\bibfnamefont
  {K.}~\bibnamefont {Li}},\ and\ \bibinfo {author} {\bibfnamefont
  {L.}~\bibnamefont {Fei-Fei}},\ }\bibfield  {title} {\bibinfo {title}
  {{ImageNet: A Large-Scale Hierarchical Image Database}},\ }in\ \href@noop {}
  {\emph {\bibinfo {booktitle} {Conference on Computer Vision and Pattern
  Recognition}}}\ (\bibinfo {year} {(2009)})\ \bibinfo {note}
  {\url{http://www.image-net.org/papers/imagenet\_cvpr09}}\BibitemShut
  {NoStop}%
\bibitem [{\citenamefont {Rubner}\ \emph {et~al.}(2000)\citenamefont {Rubner},
  \citenamefont {Tomasi},\ and\ \citenamefont {Guibas}}]{Wasserstein_distance}%
  \BibitemOpen
  \bibfield  {author} {\bibinfo {author} {\bibfnamefont {Y.}~\bibnamefont
  {Rubner}}, \bibinfo {author} {\bibfnamefont {C.}~\bibnamefont {Tomasi}},\
  and\ \bibinfo {author} {\bibfnamefont {L.~J.}\ \bibnamefont {Guibas}},\
  }\bibfield  {title} {\bibinfo {title} {{The Earth Mover's Distance as a
  Metric for Image Retrieval}},\ }\href@noop {} {\bibfield  {journal} {\bibinfo
   {journal} {International Journal of Computer Vision}\ }\textbf {\bibinfo
  {volume} {40}} (\bibinfo {year} {2000})},\ \bibinfo {note}
  {\url{https://link.springer.com/article/10.1023/A:1026543900054}}\BibitemShut
  {NoStop}%
\bibitem [{\citenamefont {Baldi}\ \emph {et~al.}(2016)\citenamefont {Baldi},
  \citenamefont {Bauer}, \citenamefont {Eng}, \citenamefont {Sadowski},\ and\
  \citenamefont {Whiteson}}]{Baldi:2016fql}%
  \BibitemOpen
  \bibfield  {author} {\bibinfo {author} {\bibfnamefont {P.}~\bibnamefont
  {Baldi}}, \bibinfo {author} {\bibfnamefont {K.}~\bibnamefont {Bauer}},
  \bibinfo {author} {\bibfnamefont {C.}~\bibnamefont {Eng}}, \bibinfo {author}
  {\bibfnamefont {P.}~\bibnamefont {Sadowski}},\ and\ \bibinfo {author}
  {\bibfnamefont {D.}~\bibnamefont {Whiteson}},\ }\bibfield  {title} {\bibinfo
  {title} {{Jet Substructure Classification in High-Energy Physics with Deep
  Neural Networks}},\ }\bibfield  {journal} {\bibinfo  {journal} {Phys. Rev.}\
  }\textbf {\bibinfo {volume} {D93}},\ \href
  {https://doi.org/10.1103/PhysRevD.93.094034} {10.1103/PhysRevD.93.094034}
  (\bibinfo {year} {2016}),\ \Eprint {https://arxiv.org/abs/1603.09349}
  {arXiv:1603.09349 [hep-ex]} \BibitemShut {NoStop}%
\bibitem [{\citenamefont {{He}}\ \emph {et~al.}(2016)\citenamefont {{He}},
  \citenamefont {{Zhang}}, \citenamefont {{Ren}},\ and\ \citenamefont
  {{Sun}}}]{resnet}%
  \BibitemOpen
  \bibfield  {author} {\bibinfo {author} {\bibfnamefont {K.}~\bibnamefont
  {{He}}}, \bibinfo {author} {\bibfnamefont {X.}~\bibnamefont {{Zhang}}},
  \bibinfo {author} {\bibfnamefont {S.}~\bibnamefont {{Ren}}},\ and\ \bibinfo
  {author} {\bibfnamefont {J.}~\bibnamefont {{Sun}}},\ }\bibfield  {title}
  {\bibinfo {title} {{Deep Residual Learning for Image Recognition}},\ }in\
  \href@noop {} {\emph {\bibinfo {booktitle} {IEEE Conference on Computer
  Vision and Pattern Recognition (CVPR)}}}\ (\bibinfo {year} {(2016)})\ pp.\
  \bibinfo {pages} {770--778},\ \Eprint {https://arxiv.org/abs/1512.03385}
  {arXiv:1512.03385} \BibitemShut {NoStop}%
\bibitem [{\citenamefont {Odena}\ \emph {et~al.}(2016)\citenamefont {Odena},
  \citenamefont {Dumoulin},\ and\ \citenamefont {Olah}}]{checkerboard}%
  \BibitemOpen
  \bibfield  {author} {\bibinfo {author} {\bibfnamefont {A.}~\bibnamefont
  {Odena}}, \bibinfo {author} {\bibfnamefont {V.}~\bibnamefont {Dumoulin}},\
  and\ \bibinfo {author} {\bibfnamefont {C.}~\bibnamefont {Olah}},\ }\bibfield
  {title} {\bibinfo {title} {Deconvolution and {Checkerboard} {Artifacts}},\
  }\bibfield  {journal} {\bibinfo  {journal} {Distill}\ }\textbf {\bibinfo
  {volume} {1}},\ \href {https://doi.org/10.23915/distill.00003}
  {10.23915/distill.00003} (\bibinfo {year} {2016}),\ \bibinfo {note}
  {\url{http://distill.pub/2016/deconv-checkerboard}}\BibitemShut {NoStop}%
\bibitem [{\citenamefont {Hendrycks}\ and\ \citenamefont
  {Gimpel}(2016)}]{GeLU}%
  \BibitemOpen
  \bibfield  {author} {\bibinfo {author} {\bibfnamefont {D.}~\bibnamefont
  {Hendrycks}}\ and\ \bibinfo {author} {\bibfnamefont {K.}~\bibnamefont
  {Gimpel}},\ }\bibfield  {title} {\bibinfo {title} {{Bridging Nonlinearities
  and Stochastic Regularizers with Gaussian Error Linear Units}},\ }\href
  {http://arxiv.org/abs/1606.08415} {\bibfield  {journal} {\bibinfo  {journal}
  {CoRR}\ }\textbf {\bibinfo {volume} {abs/1606.08415}} (\bibinfo {year}
  {2016})},\ \Eprint {https://arxiv.org/abs/1606.08415} {arXiv:1606.08415}
  \BibitemShut {NoStop}%
\bibitem [{\citenamefont {Hertel}\ \emph {et~al.}(2020)\citenamefont {Hertel},
  \citenamefont {Collado}, \citenamefont {Sadowski}, \citenamefont {Ott},\ and\
  \citenamefont {Baldi}}]{hertel2020sherpa}%
  \BibitemOpen
  \bibfield  {author} {\bibinfo {author} {\bibfnamefont {L.}~\bibnamefont
  {Hertel}}, \bibinfo {author} {\bibfnamefont {J.}~\bibnamefont {Collado}},
  \bibinfo {author} {\bibfnamefont {P.}~\bibnamefont {Sadowski}}, \bibinfo
  {author} {\bibfnamefont {J.}~\bibnamefont {Ott}},\ and\ \bibinfo {author}
  {\bibfnamefont {P.}~\bibnamefont {Baldi}},\ }\bibfield  {title} {\bibinfo
  {title} {{Sherpa: Robust Hyperparameter Optimization for Machine Learning}},\
  }\href@noop {} {\bibfield  {journal} {\bibinfo  {journal} {SoftwareX}\ }
  (\bibinfo {year} {2020})},\ \bibinfo {note} {in press. Software available at:
  https://github.com/sherpa-ai/sherpa},\ \Eprint
  {https://arxiv.org/abs/2005.04048} {arXiv:2005.04048 [cs.LG]} \BibitemShut
  {NoStop}%
\bibitem [{\citenamefont {Paszke}\ \emph {et~al.}(2017)\citenamefont {Paszke},
  \citenamefont {Gross}, \citenamefont {Chintala}, \citenamefont {Chanan},
  \citenamefont {Yang}, \citenamefont {DeVito}, \citenamefont {Lin},
  \citenamefont {Desmaison}, \citenamefont {Antiga},\ and\ \citenamefont
  {Lerer}}]{pytorch}%
  \BibitemOpen
  \bibfield  {author} {\bibinfo {author} {\bibfnamefont {A.}~\bibnamefont
  {Paszke}}, \bibinfo {author} {\bibfnamefont {S.}~\bibnamefont {Gross}},
  \bibinfo {author} {\bibfnamefont {S.}~\bibnamefont {Chintala}}, \bibinfo
  {author} {\bibfnamefont {G.}~\bibnamefont {Chanan}}, \bibinfo {author}
  {\bibfnamefont {E.}~\bibnamefont {Yang}}, \bibinfo {author} {\bibfnamefont
  {Z.}~\bibnamefont {DeVito}}, \bibinfo {author} {\bibfnamefont
  {Z.}~\bibnamefont {Lin}}, \bibinfo {author} {\bibfnamefont {A.}~\bibnamefont
  {Desmaison}}, \bibinfo {author} {\bibfnamefont {L.}~\bibnamefont {Antiga}},\
  and\ \bibinfo {author} {\bibfnamefont {A.}~\bibnamefont {Lerer}},\ }\bibfield
   {title} {\bibinfo {title} {{Automatic Differentiation in PyTorch}},\ }in\
  \href {https://openreview.net/forum?id=BJJsrmfCZ} {\emph {\bibinfo
  {booktitle} {NIPS 2017 Workshop on Autodiff}}}\ (\bibinfo {year} {2017})\
  \bibinfo {note} {\url{https://openreview.net/forum?id=BJJsrmfCZ}}\BibitemShut
  {NoStop}%
\end{thebibliography}%

\end{document}